# Nuclear Data Needs and Capabilities for Applications

*May 27-29, 2015*

*Lawrence Berkeley National Laboratory,*

*Berkeley, CA USA*

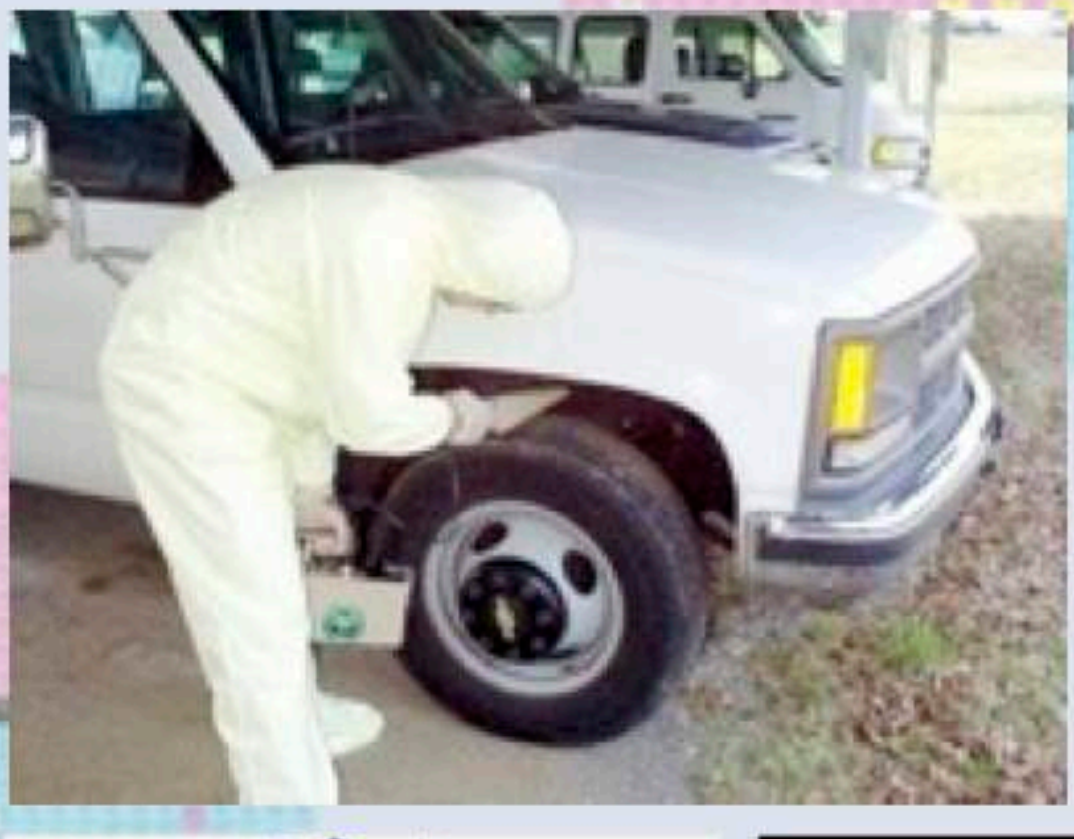

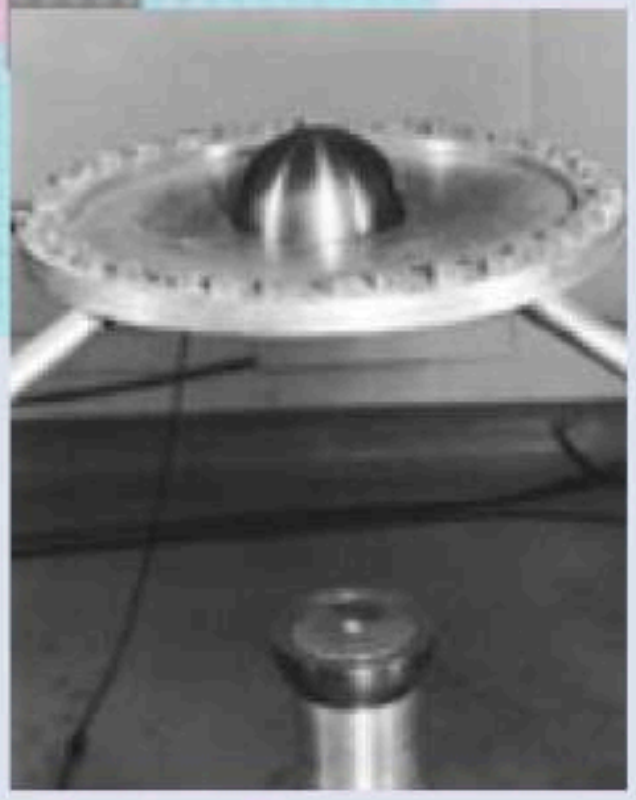

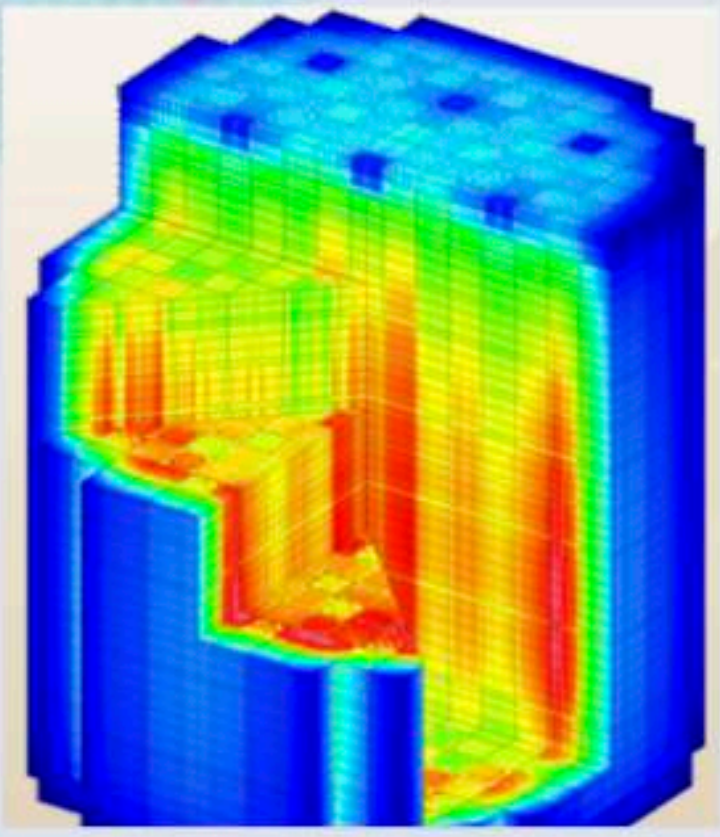

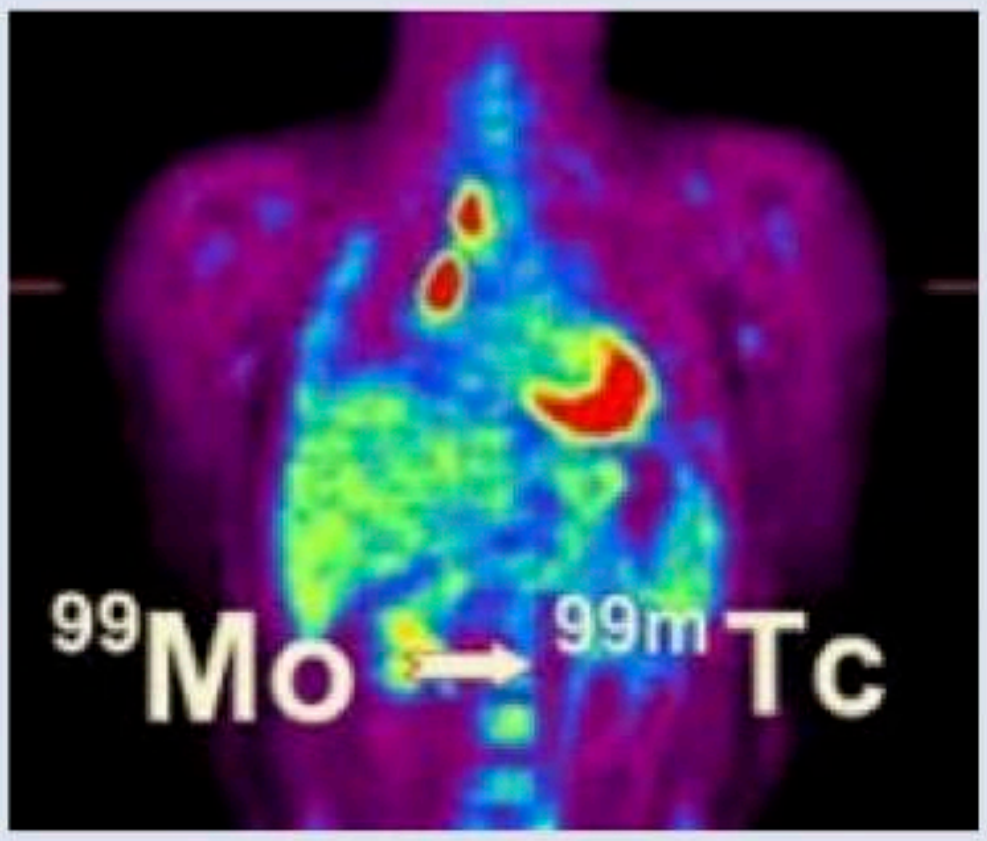

U.S. DEPARTMENT OF ENERGY | Office of Science

# Table of Contents

## Table of Contents

# Executive Summary

A detailed understanding of Nuclear Physics is important for many areas of science and technology. The nucleus is a complicated, strongly interacting, many-body dynamical system, whose accurate description requires a precise treatment of the combined effects of 3 of the 4 fundamental forces in the Standard Model (strong, weak, and electromagnetic). As a mesoscopic system composed of up to ≈ 300 particles, the nucleus exhibits a wide range of collective phenomena, including vibrational and rotational motion, and nuclear fission. Knowledge of the properties of nuclei is crucial for the study of many topics in pure research, such as the use of weak decays to search for evidence of new physics "Beyond the Standard Model." In addition to providing a unique laboratory for scientific research, nuclei are also important for a wide range of applications that are critical to society, including the generation of energy, medical applications, and national (and international) security.

Quantitative predictions regarding the rich and complex phenomena that take place in the nucleus require the use of a wide array of models and theoretical approaches, almost all of which lack predictive capabilities of sufficient accuracy to meet the needs of the applications communities. For example, although more than a century has passed since the discovery of the neutron, we still lack the ability to predict the excitation energy or the lifetime of the first excited state of most nuclei to within ±20% accuracy! As a result, nuclear physics applications continue to be strongly dependent on the measurement, publication, compilation, and evaluation of *experimental* nuclear data.

The United States Nuclear Data Program (USNDP) of the Department of Energy, Office of Science, Office of Nuclear Physics (DOE NP) is the primary custodian of nuclear data in the US, and compiles, evaluates and archives nuclear reaction and structure data for use in both basic and applied nuclear science and engineering. The USNDP also serves as an interface to the international nuclear data community, notably the International Atomic Energy Agency (IAEA) and the Organization for Economic Co-operation and Development's (OECD) Nuclear Energy Agency (NEA).

In July 2014, DOE NP carried out a review of the US Nuclear Data Program. This led to several recommendations, including that the USNDP should "*devise effective and transparent mechanisms to solicit input and feedback from all stakeholders on nuclear data needs and priorities.*" The review also recommended that USNDP pursue experimental activities of relevance to nuclear data; the revised 2014 Mission Statement accordingly states that the USNDP uses "*targeted experimental studies*" to address gaps in nuclear data.

In support of these recommendations, DOE NP requested that USNDP personnel organize a Workshop on Nuclear Data Needs and Capabilities for Applications (NDNCA). This Workshop was held at Lawrence Berkeley National Laboratory (LBNL) on 27-29 May 2015. The goal of the NDNCA Workshop was the compilation nuclear data needs across a wide spectrum of applied nuclear science, and to provide a summary of associated capabilities (accelerators, reactors, spectrometers, *etc*.) available for the required measurements. The first two days of the workshop consisted of 25 plenary talks by speakers from 16 different institutions, on nuclear energy (NE), national security (NS), isotope production (IP), and industrial applications (IA). There were also shorter "capabilities" talks that described the experimental facilities and instrumentation available for the measurement of nuclear data. This was followed by a third day of topic-specific "breakout" sessions and a final closeout session. The agenda and copies of these talks are available online at http://bang.berkeley.edu/ events/NDNCA/agenda and a copy of the agenda is included in this whitepaper as well. The importance of nuclear data to both basic and applied nuclear science was reflected in the fact that while the impetus for the workshop arose from the 2014 USNDP review, joint sponsorship for the workshop was provided by the Nuclear Science and Security Consortium, a UC-Berkeley based organization funded by the National Nuclear Security Administration (NNSA).

# A Path Forward

A principal goal of the NDNCA Workshop was to produce this whitepaper, which summarizes the data needs of the participants and of others in the community who were not in attendance. This whitepaper is more than just a bulleted list of needs; several items reappeared in multiple discussions, and are highlighted in the Cross-cutting Needs section below. References [1-9] in the section titled Cross-cutting Needs section provide additional lists of some additional needs that were collected prior to this meeting. There is substantial overlap between these data needs and the needs outlined in the current whitepaper, and we will highlight these overlaps within this document. While the main focus of the meeting was to identify capabilities and experimental needs (especially overlapping needs), other theoretical and workflow/process needs were also identified. It is hoped that this whitepaper will provide useful guidance for DOE NP and partnering DOE offices in their planning exercises. The authors also view it as a useful reference for the nuclear data community's future strategic planning.

Although many of the talks in the Workshop were focused on specific needs, there were several non-specific themes that were repeatedly emphasized, both in the talks and in discussions within the breakout groups. One of these themes was that the immense progress in computational and analysis capabilities, especially in the case of national security, have led to the discovery of significant, and in some cases dramatic, deficiencies in USNDP databases. Unfortunately, the "time constant" for addressing deficiencies in the USNDP evaluation process is often too slow relative to the needs of the programs that use this data. In the worst case this delay could conceivably lead to catastrophic failures, since certain applications rely on simulations using the best currently available nuclear data for their predictions of system performance. A recurring theme of the meeting was that addressing these nuclear data problems would necessitate an increased effort on the part of the USNDP to incorporate input from external applications groups, as well as the support from non-USNDP programs, through additional experiments and evaluation activities. In summary, the assessment of this Workshop was that *although the USNDP can help address the problem of making the most accurate nuclear data available to the users in a timely fashion, it cannot solve this problem alone.*

The USNDP can however serve as a "central clearing house" that compiles nuclear data needs from the entire application space, and provides a framework for assessing the relative priority of each need. One model of such a prioritization framework is the High Priority List (HPRL) used by the nuclear energy community to assess nuclear data needs, and assign a relative importance to each [1]. Nuclear Energy Agency (NEA) member countries then use this list to decide how to apportion resources to address the needs. The USNDP could coordinate the formation of a "Super-HPRL," in which the data needs of the entire US application space, in nuclear energy, would be compiled regularly, reviewed, and assigned priorities. The resources needed for such a "super-HPRL" would be modest, presumably not much more than the resources expended by the International Atomic Energy Agency (IAEA) in its Coordinated Research Proposal (CRP) process. Although the IAEA provides only modest funding for their CRP, this does offer a proven model. Similarly, this new USNDP-coordinated effort would help provide expert, program-neutral input to funding agencies that would help in assigning priority to specific needs, taking into account the resources available to the various government agencies.

# White Paper Outline

The organization of this whitepaper is as follows. First we will detail the cross-cutting data needs that were highlighted by more than one application area. The needs described here are:

- Dosimetry Standards
- A Deeper Understanding of Fission
- Decay Data and Gamma Branching Ratios
- Targeted Covariance Reduction For Neutron Transport
- Expanded Integral Validation
- $^{252}Cf$ Production
- Nuclear Reactor Antineutrinos

Following this, we outline the contributions of the speakers in the areas of Isotope Production, Nuclear Security, and Nuclear Energy, and note other specific needs as they are encountered.
We also note that although the National Security section is shorter than those of the other areas, this is not due to a relative lack of needs, but is instead the result of classification issues that precluded presenting a detailed justification in many cases. Many of the National Security needs have a strong overlap with those reported in the Nuclear Energy section. Occasionally in these discussions we will highlight a related issue, capability or accomplishment.

The whitepaper also includes four appendices, A-D. The first provides a summary of all the nuclear data needs presented at this workshop by application area, in a matrix format. This matrix summarizes the contents of this whitepaper, and will hopefully serve as a convenient guide to needs that overlap multiple application domains. The second appendix is a series of lists of the specific data requested by several programs. The third appendix provides a historical perspective of nuclear data in the Nuclear Energy domain. The final, fourth appendix summarizes the experimental resources that are currently available to address the nuclear data needs identified in this Workshop.

# Whitepaper Contributors

**Editors**
Bernstein, Lee (LLNL/LBNL/U.C. Berkeley)
Brown, David (BNL)
Hurst, Aaron (LBNL)
Kelly, John (TUNL)
Kondev, Filip (ANL)
McCutchan, Elizabeth (BNL)
Nesaraja, Caroline (ORNL)
Slaybaugh, Rachael (U.C. Berkeley)
Sonzogni, Alejandro (BNL)

**Plenary Speakers**
Cerjan, Charlie (LLNL)
Chadwick, Mark (LANL)
Chowdhury, Partha (U. Mass., Lowell)
Danon, Yaron (RPI)
Dean, David (ORNL)
Engle, Jonathan (LANL)
Gauld, Ian (ORNL)
Griffin, Patrick (SNL)
Hayes-Sterbenz, Anna (LANL)
Herman, Mike (BNL)
Liddick, Sean (MSU)
Nelson, Ron (LANL)
Peters, Nickie, (U. of Missouri)
Phair, Larry (LBNL)
Qaim, Syed (Forschungszentrum Jülich)
Quiter, Brian (LBNL)
Rykaczewski, Krzysztof (ORNL)
Savard, Guy (ANL)
Sleaford, Brad (LLNL)

**Plenary Speakers (cont.)**
Smith, Suzanne (BNL)
Stanculescu, Andrew (ANL/INL)
Tornow, Werner (TUNL)
Vujic, Jasmina (U.C. Berkeley)
White, Morgan (LANL)

**National Security**
Bleuel, Darren (LLNL)
Keillor, Martin (PNNL)
Ludewigt, Bernhard (LBNL)
Matthews, Eric (U.C. Berkeley)
Tonchev, Anton (LLNL)
Vogt, Ramona (LLNL)
Younes, Walid (LLNL)

**Nuclear Energy**
Dwyer, Dan (LBNL)
Grimes, Steve (Ohio U.)
Palmiotti, Pino (ANL/INL)
Pigni, Marco (ORNL)
Bradley Rearden (ORNL)

**Isotope Production & Other Applications**
Hogle, Susan (ORNL)
Perfetti, Chris (ORNL)
Julie Ezold (ORNL)

# Acknowledgements

This Workshop was sponsored jointly by the Lawrence Berkeley National Laboratory and the Nuclear Science and Security Consortium at U.C. Berkeley. The work performed by staff from several laboratories was sponsored by the US DOE Office of Nuclear Physics under Contracts DE-AC02-06CH11357 (Argonne National Laboratory), DE-AC02-05CB11231 (Lawrence Berkeley National Laboratory), DE-AC02-98CH10886 with Brookhaven Science Associates, LLC (Brookhaven National Laboratory), and DE-AC05-76OR00022 with UNIRIB (Oak Ridge National Laboratory). The work by staff from Lawrence Livermore National Laboratory was sponsored by the US DOE under Contract DE-AC52-07NA27344. The work performed by staff from Duke University and Triangle Universities Nuclear Laboratory was supported by the NNSA Stewardship Science Academic Alliances Program under Contracts DE-FG52-09NA294965 and DE-FG52-09NA29448.

# Participants & Conference Photo

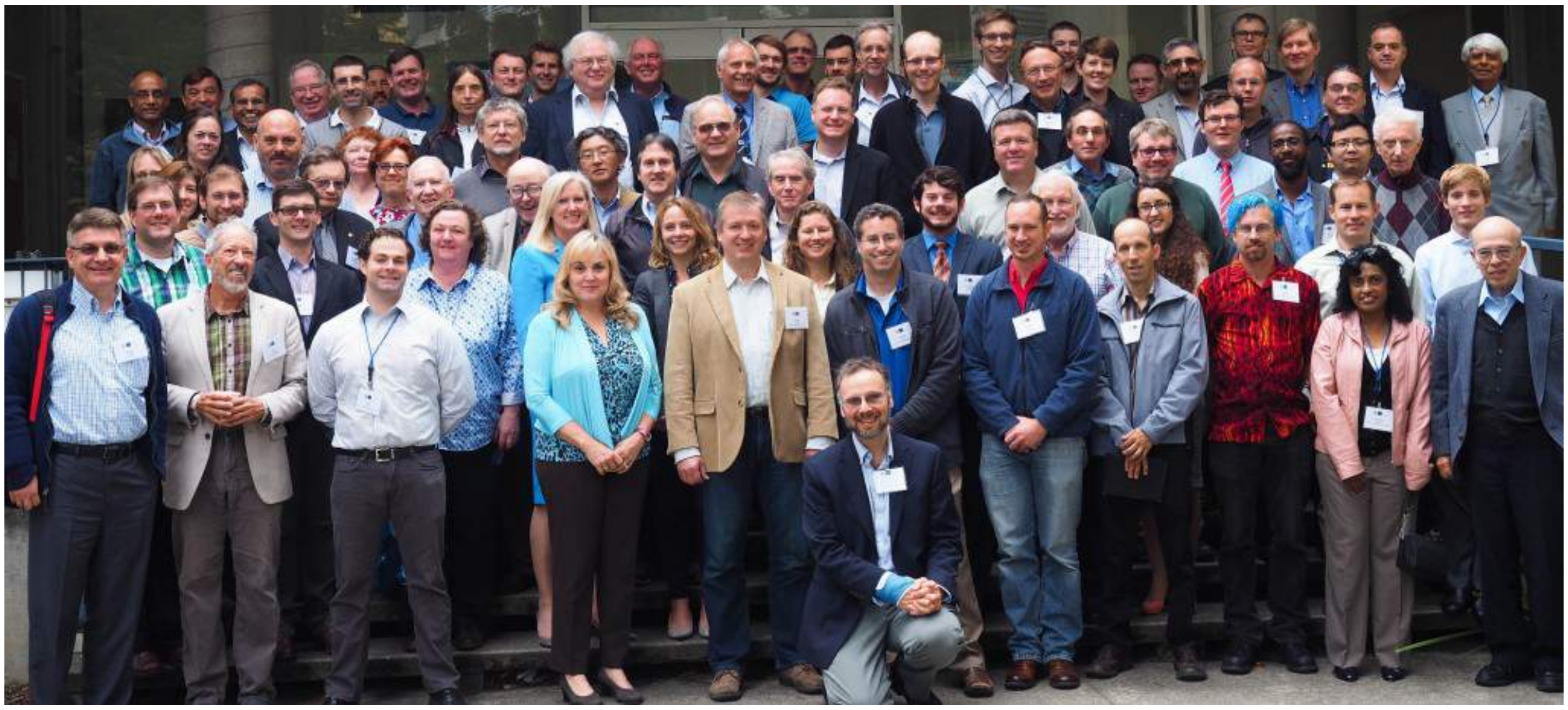

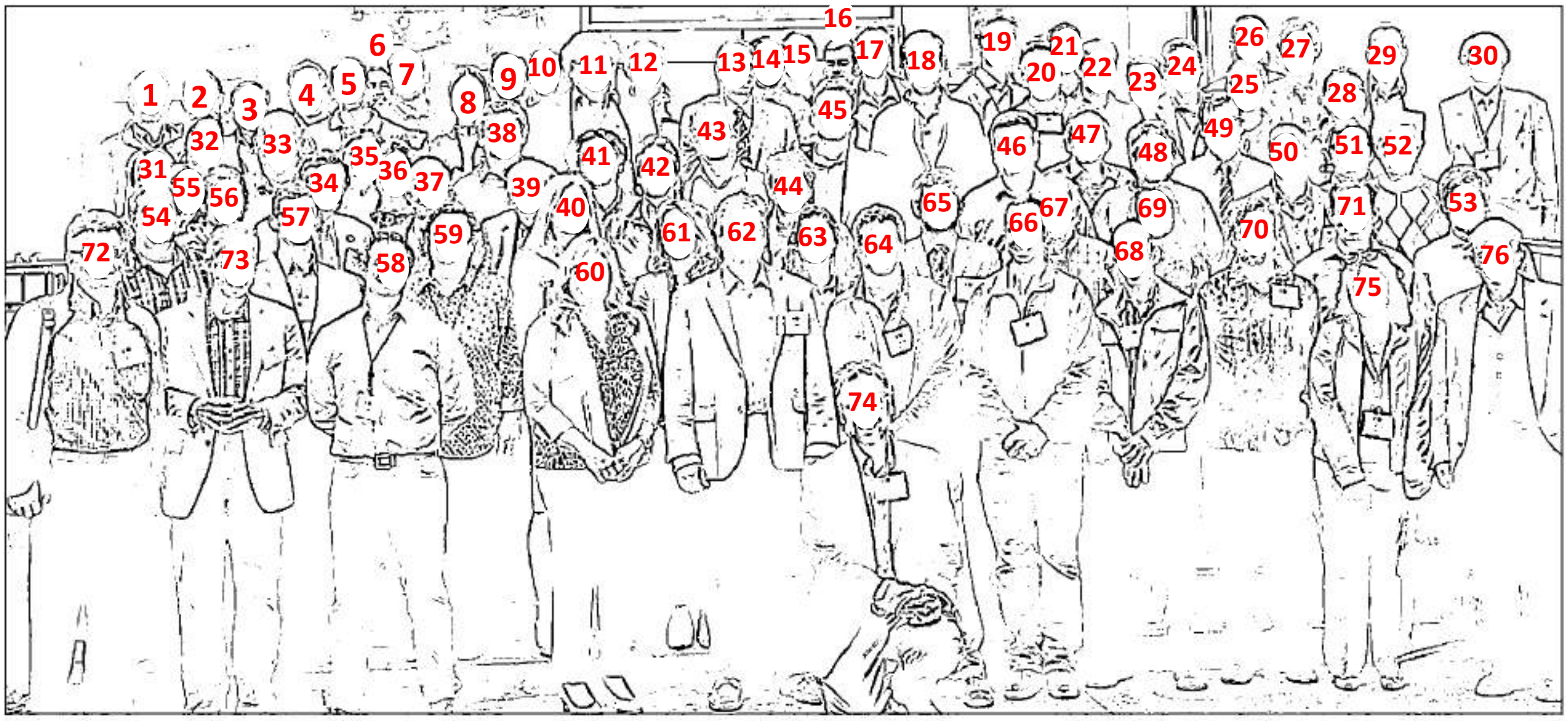

Abramovic, Ivana (Eindhoven/U.C. Berkeley) – 55
Azmy, Yousry (NCSU) - 24
Bahran, Rian (LANL)
Barnes, Ted (DOE NP) - 11
Basunia, Shamsuzzoha (LBNL) -3
Batchelder, Jon (ORAU/ORNL) - 28
Bauder, William (ANL/U. Notre Dame) - 10
Bernstein, Lee (LLNL/LBNL/U.C. Berkeley) - 74
Bevins, James (U.C. Berkeley)
Bleuel, Darren (LBNL) - 70
Brown, David (BNL)
Browne, Edgardo (BNL/LBNL) - 76
Cerjan, Charles (LLNL) - 44
Chadwick, Mark (LANL) - 17
Champine, Brian (U.C. Berkeley) - 66
Chen, Jun (MSU, NSCL) - 51
Chowdhury, Partha (U. Mass. Lowell) -1
Cooper, Reynold (LBNL) - 10
Danon, Yaron (RPI) - 29
Dean, David (ORNL)
Dunn, Michael (ORNL)
Dwyer, Daniel (LBNL) - 55
Engle, Jonathan (LANL) - 57
Ezold, Julie (ORNL)
Fallon, Paul (LBNL) - 25
Faye, Sherry (U.C. Berkeley) – 32
Firestone, Richard (U.C. Berkeley) -12
Forbes, Michael (Wash. State U.) - 49
Frey, Wesley (U.C. Davis, MNRC)
Gauld, Ian (ORNL) - 26
Goldblum, Bethany (NSSC/UCB) - 61
Griffin, Patrick (SNL) -4
Grimes, Steven (Ohio U.) - 39
Hallman, Timothy (DOE NP)
Halverson, Thomas (U.C. Berkeley) - 21
Hayes-Sterbenz, Anna (LANL) - 59
Henry, Gene (DOE ret./LLNL ret.) - 2
Herman, Michal (BNL) -13
Hogle, Susan (ORNL) - 69
Huff, Katy (UCB) - 22
Hurst, Aaron (LBNL) -23
Ingraham, Laura (DTRA) - 35
Jacak, Barbara (LBNL) - 36
Johnson, Mike (LBNL)
Kawano, Toshihiko (LANL) - 41
Keillor, Martin (PNNL) - 46
Kelley, John (NC State U./TUNL) - 48
Kenlow, Dorothy (LBNL)
Kirsch, Leo (UCB) - 19
Klein, Barry (U.C. Davis) - 73
Kolos, Karolina (U. Tenn.)
Kondev, Filip (ANL)
Kornell, Jim (STL/NSTec) - 73
Laplace, Thibault (U.C. Berkeley) - 16
Liddick, Sean (MSU, NSCL) - 53
Ludewigt, Bernhardt (LBNL) -15
Luo, Jianheng (U.C. Berkeley)
Matthews, Eric (U.C. Berkeley) - 53
Matters, David (AF Inst. Tech.) - 71
Mattoon, Caleb (LLNL) -18
McCutchan, Elizabeth (BNL) - 31
Moon, Namdoo (DHS/DNDO)
Nelson, Ronald (LANL) - 27
Nesaraja, Caroline (ORNL) - 75
Ortega, Mario (U. New Mexico)
Pardo, Richard (ANL)
Pehl, Dick (LBNL) - 52
Perfetti, Christopher (ORNL) - 65
Peters, Nickie (U. Missouri, MURR) - 50
Phair, Larry (LBNL)
Pigni, Marco (ORNL) - 5
Qaim, Syed (Forschungszentrum Jülich) - 30
Quiter, Brian (LBNL) – 57
Randrup, Jorgen - 38
Rearden, Bradley (ORNL) - 45
Rogers, Andrew (U. Mass., Lowell) - 9
Romano, Catherine (DOE NNSA, NA-22) - 40
Rykaczewski, Krzysztof (ORNL) - 34
Savard, Guy (ANL) - 62
Schroeder, Lee (LBNL/NSD/TechSource) - 67
Scielzo, Nicholas (LLNL) - 64
Shaughnessy, Dawn (LLNL)
Slaybaugh, Rachael (U.C. Berkeley) - 63
Sleaford, Brad (LLNL) - 42
Smith, Suzanne (BNL) - 60
Soltz, Ron (LLNL) - 68
Sonzogni, Alejandro (BNL) - 33
Stanculescu, Alexander (ANL/INL) - 20
Suzuki, Erika (NSSC)
Symons, James (LBNL)
Thompson, Ian (LLNL) - 43
Tonchev, Anton (LLNL) - 72
Tornow, Werner (Duke U./TUNL) - 37
Valentine, John (LBNL) - 6
Van Bibber, Karl (U.C. Berkeley)
Versluis, Robert (DOE NE, NE-42) - 73
Vogt, Ramona (LLNL/U.C. Davis) - 8
Vujic, Jasmina (U.C. Berkeley)
Wang, Tzu-Fang (LLNL)
White, Morgan (LANL) – 7
Wiens, Andreas (LBNL) -14
Younes, Walid (LLNL) - 47

# NDNCA Agenda

*Copies of the talks can be found on the web at bang.berkeley.edu/events/NDNCA/agenda*

## Wednesday, May 27 — Day 1

### Session 0: Welcome
### (Chair: Lee Bernstein, LBNL, UC Berkeley and LLNL)

| | |
|---|---|
| 8:00 - 8:30am | Registration and Refreshments |
| 8:30 - 9:00am | Welcome and Charge<br>*Lee Bernstein, UC Berkeley & LBNL*<br><br>*Karl Van Bibber, UC Berkeley Department of Nuclear Engineering*<br><br>*Tim Hallman, Office of Science, DOE*<br><br>*Ted Barnes, Nuclear Data & Nuclear Theory Computing, DOE* |
| 9:00 – 9:40am | Overview of Nuclear Data<br>*Mike W. Herman, Brookhaven National Laboratory* |

### Session 1: Nuclear Data Needs for Nuclear Energy
### (Chair: Shamsuzzoha Basunia, LBNL)

| | |
|---|---|
| 9:40 – 10:20am | Several Illustrative Examples of Nuclear Data Needs for Nuclear Energy Systems<br>*Jasmina Vujic, Nuclear Science and Security Consortium* |
| 10:20 – 10:35am | Break |
| 10:35 – 11:15am | Nuclear Data Needs in Nuclear Energy Application<br>*Alexander Stanculescu, Idaho National Laboratory* |
| 11:15 – 11:45am | Nuclear data uncertainty quantification for applications in energy, security, and isotope production<br>*Ian Gauld, Oak Ridge National Laboratory* |

11:45 – 12:00pm Discussion

## Session 2a: Energy and Other Applications
## (Alejandro Sonzogni, BNL)

2:00 – 2:35pm Nuclear Data Needs for Understanding Material Damage
*Patrick Griffin, SNL*

2:35 –3:05pm Challenges and Successes in Application of the Evaluated Nuclear Data Libraries for the Missouri University Research Reactor Core Irradiation Simulations
*Nickie Peters*

3:05 – 3:15pm Discussion

## Session 2b: Capabilities Part 1
## (Chair: P. Fallon, LBNL)

3:15 – 3:35pm Lawrence Berkeley National Laboratory Facility Review
*Larry Phair, LBNL*

3:35 – 3:55pm Los Alamos National Laboratory Facility Review
*Ronald Nelson, LANL*

3:55 – 4:10pm Capabilities
*Guy Savard, ANL*

4:10 – 4:40pm Break and Discussion

## Session 2c: Capabilities Part 2
## (Chair: Elizabeth McCutchan, BNL)

4:40 – 5:00pm Associate for Research at University Nuclear Accelerators Facilities Review
*Partha Chowdhury, ARUNA*

5:00 – 5:20pm Rensselaer Polytechnic Institute Facility Review
*Yaron Danon, RPI*

5:20 – 5:40pm Oak Ridge National Laboratory Facilities Review
*Krzysztof Rykaczewski, ORNL*

5:40 – 6:00pm Discussion

## Thursday, May 28 Day 2

## Session 3a: National Security Part 1
**(Chair: Aaron Hurst, LBNL)**

8:30 - 9:10am Needs for Neutron Reactions on Actinides
*Mark Chadwick,* LANL

9:10 –9:50am Nuclear Data's Hidden Dysfunctia: Applications Don't Actually Depend on Structure, Do They
*Morgan White, LANL*

9:50 – 10:20am Gamma Spectroscopic Data for Non-proliferation Applications
*Brad Sleaford, LLNL*

10:20 – 10:35am Discussion

10:35 – 10:50am Break

## Session 3b: National Security Part 2
**(Chair: Bethany Goldblum, UCB)**

10:50 – 11:20am Neutron and Charged Particle Reaction Data Needs for NIF Implosion Experiments
Charles Cerjan, LLNL

11:20 – 11:50am NRF Applications – An Unplanned Examination of Nuclear Data for 1-5 MeV Photons
*Brian Quiter, LBNL*

11:50 – 12:00pm Discussion

12:00 – 1:45pm Lunch

1:45 – 2:25pm Fission Product Yields for Neutrino Physics and Non-proliferation
*Anna Hayes-Sterbenz, LANL*

## Session 4: Medical Isotope Production
### (Chair: Caroline Nesaraja, ORNL)

| | |
|---|---|
| 2:25 –3:05pm | Nuclear Data for Medical Radionuclide Production: Present Status and Future Needs<br>*Syed Qaim, Research Centre Juelich and University of Cologne, Germany* |
| 3:05 – 3:35pm | Nuclear Data, Nuclear Theory, and Isotopes<br>*David Dean, ORNL* |
| 3:35 – 3:50pm | Discussion |
| 3:50 – 4:00pm | Break |
| 4:00 – 4:30pm | Nuclear Reaction and Decay Data for Medium Energy Radionuclide Production<br>*Jonathan Engle, LANL* |
| 4:30 – 5:00pm | Radioisotope Research and Production at Brookhaven Linac Isotope Producer<br>*Suzanne Smith, BNL* |
| 5:00 – 5:10pm | Discussion |

## Session 5: Capabilities Part 3
### (Chair: John Kelley)

| | |
|---|---|
| 5:10 – 5:30pm | Michigan State University Facilities Review<br>*Sean Liddick, MSU* |
| 5:30 – 5:50pm | Triangle Universities Nuclear Laboratory Facilities Review<br>*Werner Tornow, DUKE/HIGS* |
| 5:50 – 6:00pm | Discussion and Charge to breakout groups |
| 6:00pm | Adjourn |

## Friday, May 29 — Day 3

## Session 6: Breakout Sessions

9:00 – 12:00pm (note later start)

1) National Security Breakout Session (Chair: Bethany Goldblum)

2) Isotope Production/Other Breakout Session (Co-chairs: Caroline Nesaraja, Elizabeth McCutchan, Alejandro Sonzogni)

3) Nuclear Energy Breakout Session (Chair: Rachel Slaybaugh)
Location: Building 74, Room 104

## Session 7: Comments from PMs/Closeout
### (Chair: Michal Herman, BNL)

# Prioritizing Needs

We emphasize that the topics discussed here necessarily represent an incomplete list of the needs of nuclear data users, and in any case the importance of these needs will change with time as missions evolve. In view of the very extensive needs for nuclear data, including many that were not discussed in this workshop, it is evident that the USNDP cannot address more than a selected fraction of high-priority needs. It is accordingly very important to develop a procedure for prioritizing this work. It will also be important for USNDP to incorporate input from external applications groups, and to collaborate with non-USNDP programs in addressing nuclear data needs.

*A priori* one might assume that a sensitivity study using modern uncertainty quantification (UQ) techniques would provide an unbiased, and therefore preferable, method for establishing nuclear data measurement priorities. An idealized workflow representing this is shown in Figure 1 below. Here, covariance/uncertainty data from experiment and the evaluation process are folded into a simulation of an application. With careful study, one can determine which application metrics are sensitive to which experimental/theoretical inputs. One can then use these sensitivities to prioritize experiments to reduce the underlying covariance/uncertainty. As we will discuss below, such studies do have an important role in establishing relative priorities within a given application area. That being said, sensitivity studies are ultimately application-driven, and are specialized to a specific application, or class of applications; they do not reflect the needs of the community at large. Another concern is that it is often not possible to vary all quantities that a particular application depends on, in which case a complete sensitivity study is not feasible.

An alternate and likely more appropriate approach to prioritizing nuclear data needs is through expert consensus. This approach is followed by the Nuclear Data High Priority Request List (HPRL), which is coordinated by the Nuclear Energy Agency [1]. The HPRL is a long-running project that is focused on the data needs of nuclear energy, and documents and quantifies target accuracies for each identified need, and ranks the needs according to the consensus of Subgroup C. **This type of approach should be considered for the USNDP, in collaboration with CSEWG and other US nuclear data users.**

How might this work in practice? The USNDP could serve as a “central clearing house,” compiling nuclear data needs from the entire application space, and providing a framework for assessing the relative priority of each item. The resources needed for such a “super-HPRL” would be modest, probably not much more than the resources expended by the International Atomic Energy Agency (IAEA) in its coordinated research proposal (CRP) process. Although the IAEA provides only modest funding for the CRP process, it also provides an organizational structure and an unbiased viewpoint. Similarly, a new, coordinated USNDP-sponsored HPRL type effort could provide expert, program-neutral input to funding agencies, which would be very useful in assigning priorities to specific nuclear data needs, taking into account the resources available to the various government agencies.

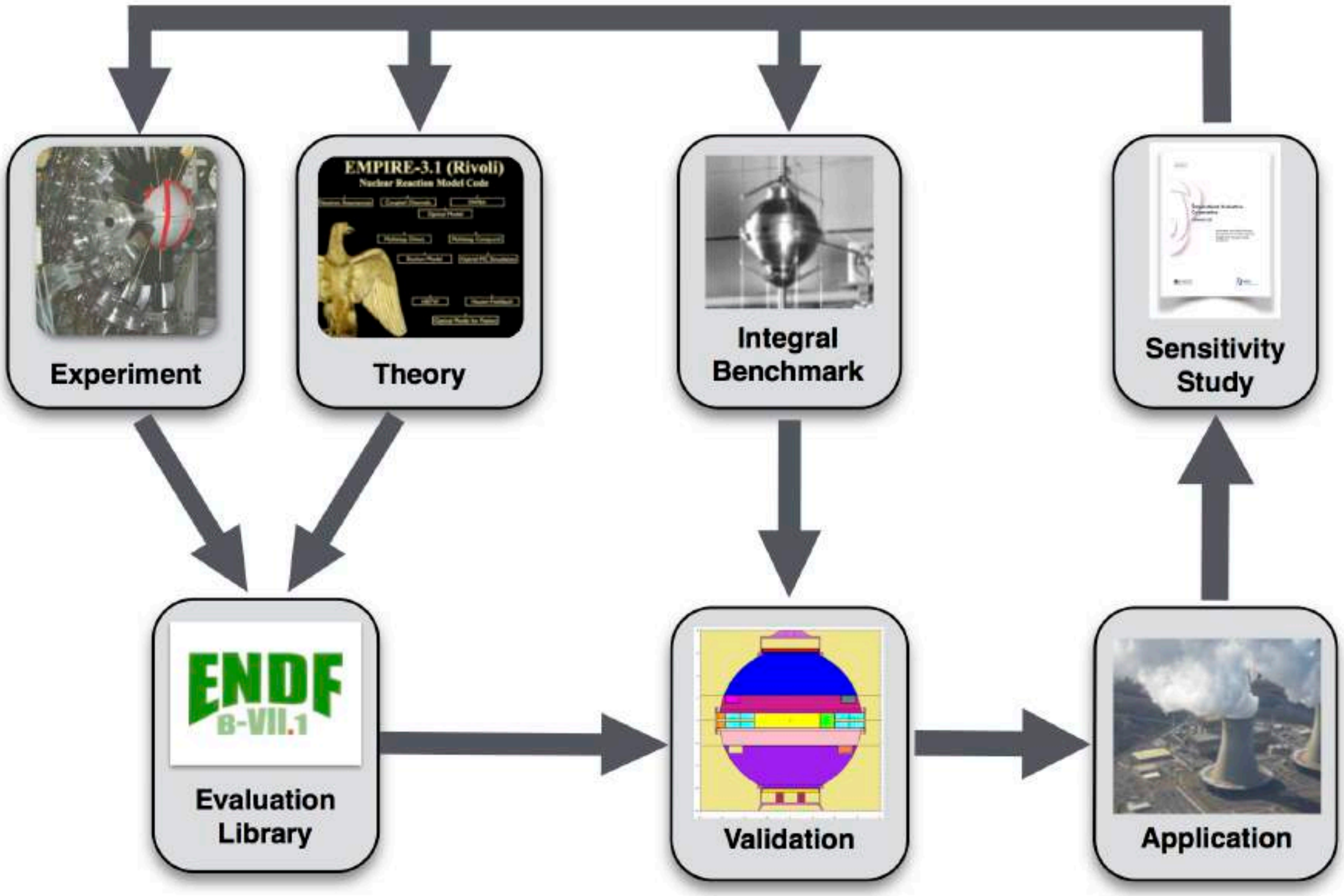

**Figure 1.** Covariance data plays a valuable role in the program planning process, as illustrated by this figure. An evaluator fitting theoretical calculations to experimental data, e.g. from the EXFOR library, constructs evaluated reaction data. These evaluated data are then used to simulate the performance of integral benchmarks such as critical assemblies. If the agreement is acceptable, the data are then used in applications. Sensitivity assessments of the benchmark and application performance are then used to define the target uncertainties needed in the nuclear data, employing the covariances in current data libraries. UQ tools such as DAKOTA [10] and TSUNAMI [11] are used in this process. A set of prioritized data needs can then be developed from these sensitivities, and the appropriate measurements performed. These new measurements can then be incorporated in the evaluations, and a new assessment can be carried out. (Figure from reference [12] of the Cross Cutting Needs section.)

# Cross-cutting Needs

In preparing this whitepaper it quickly became evident that several needs reappear in different contexts. These include:

1. Dosimetry standards
2. A deeper understanding of nuclear fission
3. Decay data and gamma branching ratios
4. Targeted covariance reduction in neutron transport
5. Expanded integral validation
6. $^{252}$Cf production
7. Antineutrinos from nuclear reactors

These needs and others discussed later in the paper are summarized in Appendices A and B. The order in which topics are presented below does not reflect any preference on the part of the authors or meeting participants. In the text below and in succeeding sections, needs or recommendations are highlighted in bold face.

## Dosimetry Standards

***(Nuclear Energy, National Security, Isotope Production, Industrial Applications, Safeguards)***

One of the most commonly used experimental techniques in nuclear physics is activation, in which a sample is irradiated to induce transitions to unstable nuclei, whose subsequent decays can be detected and quantified. This technique can be used for example to measure the fluence of a beam, or, in conjunction with a witness foil, as a monitor reaction. A monitor reaction uses a well-characterized reaction to reduce or eliminate systematic experimental errors. When used in this manner, the product radionuclide is chosen so that the decay radiation (usually emitted gammas) is distinct, and is emitted with a half-life commensurate with those of the other reaction products being studied.

The widespread use of the activation technique has motivated several efforts to establish dosimetry standards. The Neutron Standards Project is prototypical of these efforts. Another such effort supports the International Reactor Dosimetry and Fusion (IRDFF) library [13], which provides up to date source data and a wide range of monitor reaction data, including production cross sections, decay data, and recoil spectra.

There are many reactions that experimentalists would in principle like to employ as monitors, which are unfortunately not currently applicable due to poorly characterized decay products or poorly known production cross sections. In particular, **there is a need for standards for higher neutron energies (up to 60 MeV) to support studies of material damage from fusion simulators such as the International Fusion Material Irradiation Facility (IFMIF), accelerator driven systems (ADS), and spallation neutron sources (SNS)** [14-15]. To address this need, an IAEA Coordinated Research Project (CRP) [16] was initiated to improve the coverage of the IRDFF for higher-energy neutron standards. A large set of understudied monitor reactions was identified in reference [9], in which the authors followed an unorthodox approach in studying the EXFOR database using “Big Data” network theory methods. The full list of dosimetry reactions is given in Appendix B.

## A Deeper Understanding of Nuclear Fission

***(Nuclear Energy, National Security, Basic Research, Safeguards, Isotope Production)***

A first principles understanding of nuclear fission would likely be of great importance to nuclear physics applications, since this understanding could lead to a predictive fission model that provides reliable information about fission nuclides and fission products at energies not normally accessed by experiment. Improved fission models will provide fission product information required by multiple applications. Without a high-fidelity fission models, one may still infer systematics from the limited existing experimental fission studies, which may introduce large uncertainties to calculated values in regimes that have not been addressed experimentally.

There are two major classes of fission models that are in development by USNDP members and collaborators. The first addresses the scission process itself, and seeks to develop a more fundamental understanding of fission as it proceeds through the scission point, tackling the difficult many-body problem starting either from protons, neutrons, and an effective interaction between them [5-8], or from a liquid-drop picture with shell corrections [7-4]. This type of model, which addresses pre-scission physics, would benefit from data that directly probes fission dynamics, including fission time scales and pre-scission photon emission, which can affect the final neutron multiplicity. The second focuses on the description of post-scission emission of prompt and/or delayed neutrons, photons, and other particles [9-12]. These models can be either deterministic or stochastic; stochastic models of (predominantly) prompt emission can address a wider range of observables, but require more input data. A potential concern is that many database files related to actinide fission are not actually based on measured data, but instead incorporate results from deterministic models based only on fission systematics.

For both pre- and post-scission models, input data are required for validation, and post-scission models also use input data to fix parameters. A quantity of particular importance for both classes of models, but which contains large uncertainties, is the yield of fission fragments (before prompt emission) and fission products (after prompt emission). Fission fragment yields are important inputs for post-scission models, and also serve to validate pre-scission models. The fragments themselves become products after de-excitation through beta decays and prompt neutron and photon emission. Fission products are the sources of delayed photon emission, as well as neutrons, electrons and neutrinos (through beta decays). These delayed decays lead to both decay heat and decay radiation, which are important in reactors as well as in the fission byproducts of spent nuclear fuel. They also provide useful signatures for various detection schemes.

Although neutron-induced fission data are a high priority for many programs, the relevant data are usually only available for thermal, fast, and high-energy (14 MeV) neutrons. Careful measurements of fission yields are needed for more isotopes and more incident energies. Photofission is also of interest for applications, but there is very little information available about the fission fragment yields needed as input in this case. Fission yields of metastable states are also important.

Fission product yields are required for post-detonation forensics. The blocked cesium and iodine fission products can be used to determine whether the fuel is uranium- or plutonium-based, and provide an indication of the incident neutron energy in neutron-induced fission, in particular whether 14 MeV neutrons were involved. For this particular application, the 130I and 135I thermal, fast and 14 MeV fission yields need to be measured. Codes such as FIER, CASCADES, and ORIGEN use fission product yields to simulate the isotopic inventory of fission products and their decay signatures. However, in ENDF, the independent fission product yields (those following prompt particle emission) are inconsistent with the cumulative fission product yields (those following all the fission product beta decays) and the ENDF decay sub-library. This discrepancy must be resolved at some point. Furthermore, users would like to have full sets of covariance data for the fission product yields of all fissionable materials, for uncertainty quantification.

A second important input for post-scission models is the total kinetic energy (TKE) of the fragments, a quantity that has often only been measured at thermal energies, or for spontaneous fission. There have been recent measurements of the TKE as a function of incident neutron energy, but these are averaged over the fragment mass, and do not give an indication of how TKE(AH), usually presented as a function of the heavy fragment mass (AH), changes with energy. Differential measurements for more isotopes and incident energies are critical for model tuning and validation. In addition, such measurements do not yet exist for photofission.

An accurate description of Neutron and Photon Yields and Spectra is a goal of many fission models. Highly-excited fission fragments typically de-excite by neutron and gamma emission. The prompt gamma spectrum and multiplicity are crucial inputs in determining local heating post-fission, and are not well known. Prompt gammas also induce radiation damage surrounding a fission event, and as such must be accounted for when computing dosage. These data are needed not just for 235U, 238U, and 239Pu, but for all minor actinides, and for neutron and gammas as projectiles. Photon branching ratios are also required. The prompt-fission neutron spectra of actinides were the subject of a recent IAEA Coordinated Research Project. Much progress was made, but it was clear that further work was needed, not only on the major actinides needed for security, but also on the minor ones needed for nuclear power, isotope production and forensics. Stochastic post-fission models are only beginning to be able to address evaluations of neutron and photon spectra. To broaden their reach, more differential data, such as neutron, $\gamma$-ray energy and multiplicity as a function of incident neutron energy from thermal to >20 MeV, are needed for validation.

Unfortunately, most recent experiments have tended to measure only one quantity well, such as the fission cross section or the prompt neutron or photon spectrum, with no accompanying measurement of the fragments. An ideal experiment for addressing prompt post-scission physics would be the 'Mother of All Fission Experiments,' in which the fission fragments, prompt neutrons, and prompt photons are all measured in the same setup at the same time, for a range of actinides, and for a range of energies from thermal to > 20 MeV.

# Decay Data and Gamma Branching Ratios

***(Nuclear Energy, National Security, Isotope Production, Safeguards, Industrial Applications)***

The fission of a heavy nucleus such as $^{238}$U can produce significant amounts of more than 800 different types of radioactive fission fragments, which then decay back to stability by emitting (primarily) photons, electrons, neutrons, and antineutrinos. Given a sufficiently good decay database, the absolute yield and spectrum of each type of radiation can be calculated. As these fission spectra have many observable consequences, this decay data is correspondingly of great importance for a wide range of applications. The electromagnetic and light-particle energy released by the fission fragments, the so-called decay heat, is essential for a precise modeling of refueling and reprocessing strategies for nuclear reactor fuel and materials. Reactor operation and control relies heavily on knowing the flux of beta-delayed neutrons or antineutrinos. In fundamental physics, the estimated antineutrino flux is currently being used to understand the properties of neutrino oscillations, and to search for evidence of physics beyond the standard model in the weak interaction more generally. The same antineutrino flux is of interest in nonproliferation studies, as it is sensitive to the mix of actinides being burned. For each of the applications discussed above there have been several IAEA consultants' meetings to identify exactly which nuclides most urgently require better nuclear data. Here we provide references to those reports in which specific high-priority nuclei are identified through sensitivity analyses. For decay heat, a series of IAEA investigations led to a priority list of nuclides [29], which required new measurements to better understand the decay heat, generated by a reactor. This decay-heat priority list encouraged several new measurements, which are summarized in a new IAEA study [30] that includes an updated high-priority list. This new study also considers which nuclei require new measurements to better model the antineutrino spectrum generated by a reactor, and provides an additional high-priority list. Finally, a summary of the most important delayed neutron precursor nuclides for reactor kinetics studies, based on a sensitivity study of delayed neutrons, was presented in another IAEA coordinated research project [31].

The importance of decay data goes beyond fission-related applications. In the field of medical isotopes, a precise understanding of the radiation emitted by radionuclides is needed to determine the total dose received by the patient, the specific dose to targeted tissue, the cost of infrastructure in production facilities (*i.e.* shielding requirements), the background in imaging technologies, *etc*. A measurement of a single quantity can have a huge impact on the production and supply chain of a particular isotope. As an example, a change in the absolute intensity of the 776-keV transition in $^{82}$Rb decay (used to determine the dose of this frequently used cardiac PET isotope) from 13% to 15% [32] had major implications for the suppliers of the $^{82}$Sr/$^{82}$Rb generator. This higher value was recently verified by C. Gross *et al.* [33]. **The IAEA has investigated the decay data needs of certain medical isotopes, and provides recommendations for new measurements and evaluations [34-37].**

Conventional Non-Destructive Assay (NDA) methods for fresh fuel assay, forensics, and irradiated fuel characterization also rely on the properties of the radiation emitted following beta decay [38]. Traditional NDA methods utilize the absolute gamma-ray intensity, whereas newer techniques will need additional information on the cascade nature of the gamma rays for coincidence and spectral techniques, thereby yielding higher sensitivity and unique isotopic identification.

# Targeted Covariance Reduction for Neutron Transport

***(Nuclear Energy, National Security, Safeguards, Isotope Production)***

The applications discussed at the workshop largely involve design and analysis through modeling and simulation. To enable effective prediction, design, and analysis we need well-quantified uncertainties, so that we can clearly characterize safe and economical operational areas, detailed isotope production estimates, and calculate key quantities of relevance to national security.

As one example, isotope production processes can involve multiple neutron captures, so there are many contributing uncertainties, and it can be difficult to establish which reaction rates are most significant. Furthermore, uncertainties in the individual actinide absorption cross sections themselves can obscure production capabilities; the resonance capture cross section is significant in many key actinides, but the data may be poor, and analysis options are resource intensive.

High-fidelity sensitivity and uncertainty analysis tools, which are applicable to these problems, have been developed [39-43]. These tools are very general, and can assess the contribution of various (but not all) important parameters, including energy-dependent cross sections, angular distribution, reactivity coefficients, criticality, reaction rates, reaction rate ratios, *etc.* The development of these sensitivity and uncertainty analysis tools is continuing. These tools can be used for uncertainty quantification, target uncertainty assessment, the validation of simulation tools, and for nuclear data/parameters assimilation using integral experiments. These analytical tools are one key component in the ideal, information-rich workflow discussed above, and illustrated in Figure 1. However, sensitivity studies require accurate covariance estimates, in addition to accurate input values. **A key cross-cutting need that was expressed throughout the workshop was for more accurate covariance data, so that we can clearly identify which isotopes most need improved data or additional measurements.** This includes not just higher quality covariances, but covariances on all possible input data. Improving these data could enable the analysis of the contribution of the energy-dependent cross sections to isotope yields, which would be a very powerful tool for identifying and prioritizing the nuclear data needs of the all of these communities.

A series of sophisticated sensitivity studies [6,13-14] revealed the following needs:

- **Large uncertainties in many neutron-induced actinide reaction cross sections:** Actinide cross sections are the most common need for modeling nuclear systems. **In this workshop the data on neutron capture cross sections for $^{235}$U, $^{238}$U and $^{239}$Pu were repeatedly called out as a source of concern**. It is not enough to have a high precision cross section; one must also provide covariance data for Uncertainty Quantification (UQ) applications. The CIELO pilot project aims to resolve some of this need by providing standards level evaluations of these isotopes. However, the strong competition with fission in all of these isotopes means that the (n,$\gamma$) cross section cannot be evaluated accurately in the absence of experimental capture data in the energy range of interest. **In addition, issues with the $^{237}$Np(n,f) cross section in the energy region from 1-100 keV were mentioned.**

## Highlight 1: Compensating Errors in the Jezebel $k_{eff}$

Eric Bauge [39] reported on an analysis where components of the Bruyères-le-Châtel (BRC) $^{239}$Pu evaluation were replaced with those from ENDF/B-VII.1. At each step in the replacement process, $k_{eff}$ of the Jezebel critical assembly was computed. While both the BRC and ENDF/B-VII.1 give the same $k_{eff}$ for Jezebel, they do so for very different reasons. This replacement study shows how different parts of the evaluation substantially shift the reactivity of Jezebel. We do not know if either evaluation is “correct” but both get the “correct” answer.

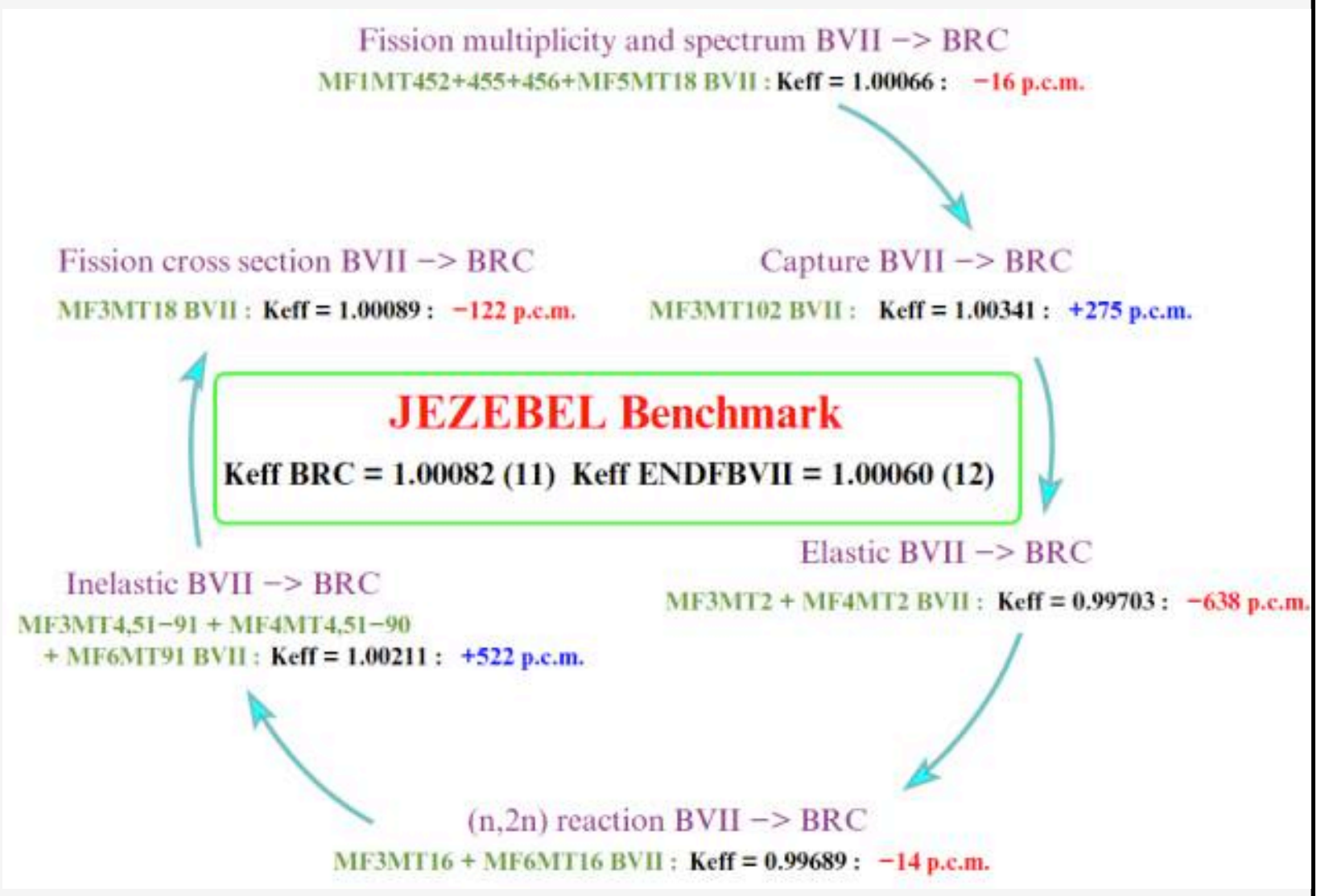

- **(n,n’) and Cross Sections and Angular Distributions**: Another recurring need was for accurate modeling of neutron elastic and inelastic scattering, not just on actinides, but also on structural materials. Both the cross sections and outgoing angular distributions are needed. These data are important in small systems in which neutron leakage plays an outsized role. To a large degree, WPEC Subgroup 35 and recent advances are addressing this problem in the EMPIRE [44] and CoH [45] reaction code systems. This has already fed back into the CIELO evaluation for $^{238}$U and $^{56}$Fe. Nonetheless **we still need integral benchmark data for validating codes and evaluations.** An RPI group [46] has been investigating the measurements of semi-integral data for testing and has developed a test for $^{56}$Fe and $^{238}$U, but **additional tests are needed, especially for $^{235}$U.**

# Expanded Integral Validation

***(Nuclear Energy, National Security)***

It is expected that the providers of nuclear data are responsible for ensuring the quality of the nuclear data they generate: the National Nuclear Data Center is responsible for simple format and physics testing of nuclear reaction data files in the ENDF library on behalf of the Cross Section Working Group (CSEWG), and the US Nuclear Data Program performs similar tests on the nuclear structure data files in the ENSDF library. Beyond this testing, more advanced benchmarking is done, comparing results from simulations done for example with ENDF files to results from high-fidelity integral experiments. These experiments typically are critical assemblies or other simple benchmark problems that can be simulated with modest computer resources, but test the underlying nuclear data in a rigorous and targeted way. During the meeting, **the need for more semi-integral and differential experiments that are driven by application and science needs was raised. In particular, an understanding of the new Ohio University [47] and older LLNL pulsed-sphere experiments is needed, to separate the various effects and achieve an understanding of some of the basic phenomena. Semi-integral data can help diagnose shortcomings in elastic and inelastic neutron scattering [48]. There is also a serious lack of integral tests for incident charged-particle reaction data.** Note: Highlight box #2 discusses a familiar problem in the study of integral benchmarks, specifically that of compensating errors.

# $^{252}$Cf Production

***(Nuclear Energy, National Security, Isotope Production, Industrial Applications)***

The production of $^{252}$Cf is essential for many applications, and the data needs to optimize $^{252}$Cf production would have a simultaneous cross-cutting effect on these applications. These include applications in the energy industry (nuclear fuel quality control, reactor startup sources, coal analyzers, oil exploration); construction (mineral and cement analyzers, corrosion inspection); and security (handheld contraband detectors, fission source, monitoring HEU down-blending, identifying unexploded ordnance, landmine detection). ORNL is the world's leader in the production of $^{252}$Cf. $^{252}$Cf and other heavy isotopes are produced by successive neutron captures on mixed actinide targets containing Cm/Am/Pu (curium feedstock) at the High Flux Isotope Reactor (HFIR). During this process, about 95% of the initial heavy target nuclei undergo fission into lighter nuclides. These losses are highlighted in Figure 2 below.

In addition to consuming valuable curium feedstock, fission heating constrains both target design and chemical processing schedule flexibility. Target yields are further limited due to neutron absorption by $^{252}$Cf during production. The efficiency with which $^{252}$Cf is produced, incorporating both the isotope transmutation fractions and $^{252}$Cf retention, is strongly dependent on the incident neutron energy spectrum. By shifting the hardness of the spectrum, or by suppressing the flux in certain energy ranges, the ratio of beneficial to destructive neutron captures can be increased. Researchers are exploring the use of focused resonance shielding [49] through a variety of neutron flux filter materials to increase this ratio of (n,γ) captures in the curium feedstock relative to destructive captures. $^{103}$Rh is being considered as a potential filter material for $^{252}$Cf, and has undergone preliminary experimental evaluation. Sensitivity analyses are being performed to identify other possible filter materials as well as to assess other methods of flux optimization such as target shuffling and the use of alternative geometries. Because this optimization relies on the variation of neutron absorption ratios throughout the energy spectrum, accurate neutron cross sections for key isotopes in the $^{252}$Cf production chain are needed (see Appendix B).

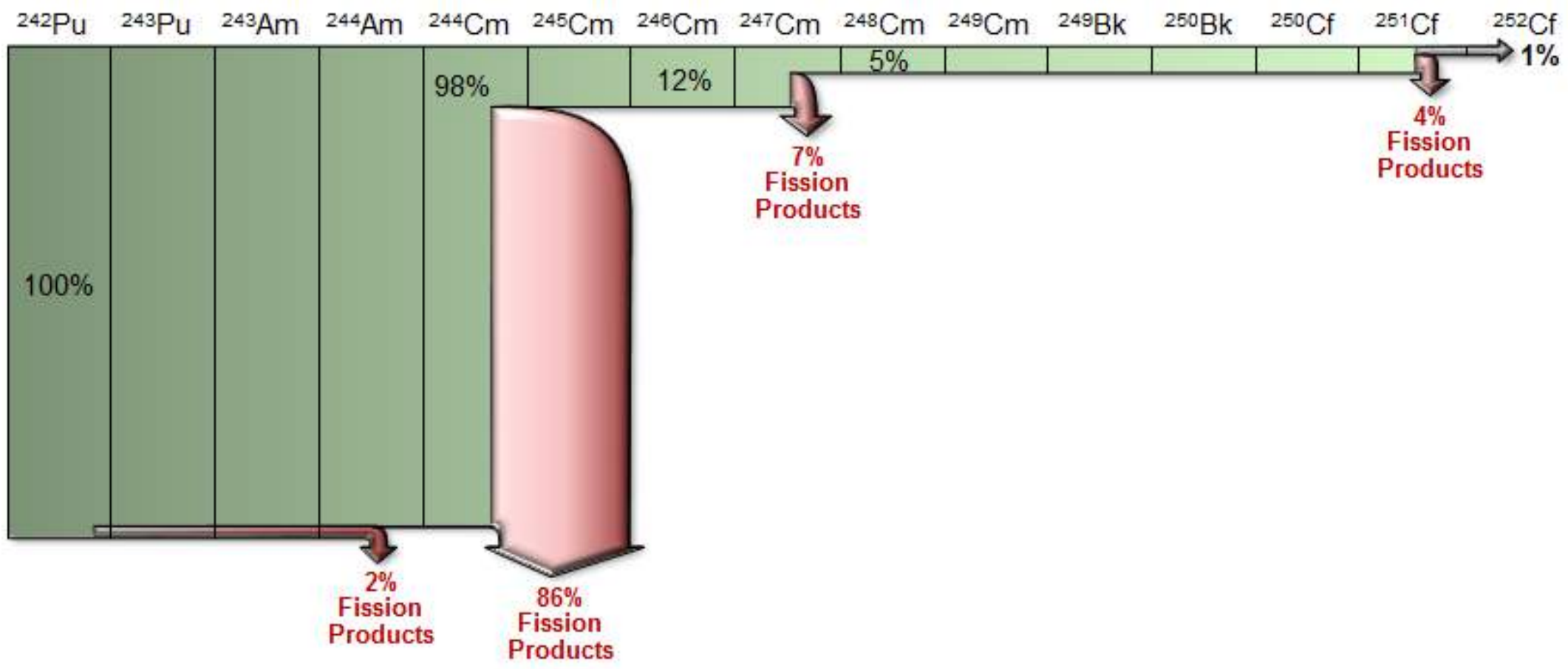

**Figure 2.** Production of $^{252}Cf$ through successive neutron captures, showing losses due to fission

## Antineutrinos from Nuclear Reactors

***(Nuclear Energy, National Security)***

Nuclear reactors are copious producers of antineutrinos, and generate about $10^{21}$ per second per GW from the $\beta^-$ decay of neutron-rich fission products. In the first experimental observation of neutrinos [50], this very large flux compensated for the extremely small neutrino cross sections. A white paper that was prepared following a recent meeting at BNL [51] details recent advances and current research in this field.

By placing detectors next to power reactors, in the last few years the neutrino oscillation parameter $\theta_{13}$ was precisely measured [52-54]. These experiments have also yielded precise antineutrino spectra. The energy-integrated antineutrino spectrum appears to be about 6% smaller than expected; this is known as the short-distance anomaly. In addition, a distortion in the spectrum at around 5 MeV has been seen.

An accurate calculation of the antineutrino spectrum emitted by a reactor requires knowledge of a) the reactor core fuel composition, in other words the different power contributions from $^{235}U$, $^{238}U$, $^{239}Pu$, and $^{241}Pu$, and b) the antineutrino spectrum that results from the neutron-induced fission of each of these nuclides. Fission fragment and decay data are essential to determine the latter.

Two methods are available to obtain the antineutrino spectra from $^{235,238}U$ and $^{239,241}Pu$. The summation method [55] uses fission yields and decay data to compute the contributions of each decay branch of the over 800 $\beta^-$-decaying fission fragments involved. The main advantage of this method is that a good understanding of the spectral features can be obtained [56-57]. However, in view of the incomplete decay data and imprecise fission yield data, there is ample scope for improvement in the accuracy of this method [58].

In the conversion method, one fits the corresponding measured electron spectra [59] with a number of artificial decay branches. Because the electron data has been accurately measured (except for $^{238}U$), the resulting predicted antineutrino spectra have smaller errors than those derived using the summation method. The principal drawback of the conversion method is that only about 30 decay branches can be used in fits to the electron spectra at present, due to the limited experimental resolution. This method is not independent of fission yield and decay data, as they are used to obtain the Fermi function effective charge of the decay branches as a function of the end-point energy [60].

To better understand the antineutrino anomaly and the 5 MeV spectral distortion we need the following improvements in nuclear data:

- **New measurements of the electron spectra following fission for the 4 main fuel component nuclides, $^{235}$U, $^{238}$U, $^{239}$Pu, and $^{241}$Pu, to confirm the Institut Laue-Langevin (Grenoble, France) data.**

- **New fission yield measurements, in particular for odd-Z, odd-N nuclides with two long-lived levels of low and high spin.**

- **Precise measurements of beta intensities for some 20-30 relevant nuclides.**

- **Precise measurement of the beta spectra for those same 20-30 nuclides, to test the suggestion that non-allowed shapes may cause distortions in the spectra that solve the anomaly problem** [61].

We note that there are cross-cutting benefits associated with these measurements; since the energy carried away by the neutrinos is correlated with the energy deposited by the electrons, and anti-correlated with that of the gammas, these new measurements will also result in more precise decay heat calculations [62-63]. Additionally, these new measurements would allow for more precise non-proliferation uses of antineutrino detectors [64]. Moreover, the Total Absorption Gamma Spectrometry (TAGS) technique used to obtain beta intensities can also produce precise values of gamma and neutron widths, which in turn can be used to calculate neutron capture cross sections for neutron-rich short-lived nuclides [63] which are relevant in nuclear astrophysics, and reactor fuel burnup and isotope production calculations.

The future of this field also includes new measurements of the neutrino spectra at short distances, (7-15 km) to better understand the anomaly, as well as medium distance experiments (47 - 53 km) that will aim to measure the $\theta_{12}$ parameter and the antineutrino mass hierarchy [66-68].

# References

1. OECD/NEA WPEC Nuclear Data High Priority Request List (HPRL) https://www.oecd-nea.org/dbdata/hprl/ (2015).
2. R. Bahran, S. Croft, J. Hutchinson, M. Smith, A. Sood, “A Survey of Nuclear Data Deficiencies Affecting Nuclear Non-Proliferation,” Proc. of the 2014 INMM Annual Meeting, Atlanta GA, LANL Report LA-UR-14-26531.
3. P. Santi, D. Vo *et al*. “The Role of Nuclear Data in Advanced Safeguards,” Proc. Of Global 2007: Advanced Nuclear Fuel Cycles and Systems, Boise, ID pp. 1670-1678 (2007).
4. D. McNabb, “Nuclear Data Needs for Homeland Security,” LLNL Report UCRL-MI-207715 (2005).
5. T. Yoshida *et al*., “Assessment of Fission Product Decay Data for Decay Heat Calculations,” OECD/NEA WPEC Subgroup 25, ISBN 978-92-64-99034-0 (2007).
6. A. Plompen, “IAEA Report on Long-term Needs for Nuclear Data Development,” INDC(NDS)-0601 (2012).
7. A.L. Nichols, S.M. Qaim, R. Capote Noy, “IAEA Intermediate-term Nuclear Data Needs for Medical Applications,” INDC(NDS)-0596 (2015).
8. Proceedings of the Workshop on "Decay Spectroscopy at CARIBU: Advanced Fuel Cycle Applications, Nuclear Structure and Astrophysics," http://www.ne.anl.gov/capabilities/nd/AFC-Apr11/index.shtml.
9. J.A. Hirdt and D.A. Brown, “Identifying Understudied Nuclear Reactions by Text-mining the EXFOR Experimental Nuclear Reaction Library,” to be published in Nuclear Data Sheets, Jan. 2016.
10. Dakota 6.2, https://dakota.sandia.gov/ (2015).

11. B.T. Rearden, D.E. Mueller, S.M. Bowman, R.D. Busch, and S.J. Emerson, "TSUNAMI Primer: A Primer for Sensitivity/Uncertainty Calculations with SCALE," ORNL/TM-2009/027, ORNL, Oak Ridge, Tenn., (2009).
12. D. Brown *et al.,* J. Phys. G: Nucl. Part. Phys. 42, 034020 (2015).
13. R. Capote, K.I. Zolotarev, V.G. Pronyaev, A. Trkov, Journal of ASTM International (JAI) - Volume 9, Issue 4, April 2012, JAI104119; E.M. Zsolnay, R. Capote, H.K. Nolthenius, and A. Trkov, Technical report INDC(NDS)-0616, IAEA, Vienna (2012).
14. A. Trkov, L.R. Greenwood, S.P. Simakov, "Testing and Improving the International Reactor Dosimetry and Fusion File (IRDFF)," IAEA Report INDC(NDS)-0639 (2013); K.I. Zolotarev, "Evaluation of excitation functions for $^{28}$Si(n,p)$^{28}$Al, $^{31}$P(n,p)$^{31}$Si, and $^{113}$In(n,$\gamma$)$^{114m}$In reactions," IAEA Report INDC(NDS)-0668 (2014); https://www-nds.iaea.org/IRDFFtest/.
15. IAEA 616, 616INDC(NDS)-05, "Summary Description of the New International Reactor Dosimetry and Fusion File (IRDFF release 1.0)" (2012).
16. IAEA 682, 682INDC(NDS)-05, "Summary Report of the Second Research Coordination Meeting on Testing and Improving the International Reactor Dosimetry and Fusion File (IRDFF)" (2015).
17. Jørgen Randrup and Peter Möller, "Brownian Shape Motion on Five-Dimensional Potential-Energy Surfaces:Nuclear Fission-Fragment Mass Distributions", Phys. Rev. Lett. 106, 132503 (2011).
18. Jørgen Randrup and Peter Möller, "Energy dependence of fission-fragment mass distributions from strongly damped shape evolution", Phys. Rev. C 88, 064606 (2013).
19. E. G. Ryabov, A. V. Karpov, P. N. Nadtochy, and G. D. Adeev, "Application of a temperature-dependent liquid-drop model to dynamical Langevin calculations of fission-fragment distributions of excited nuclei", Phys. Rev. C 78, 044614 (2008).
20. M. Borunov, P. N. Nadtochy, G. D. Adeev, "Nuclear scission and fission-fragment kinetic-energy distribution: Study within three-dimensional Langevin dynamics", Nucl. Phys. A799, 56 (2008).
21. J. F. Berger, M. Girod, D. Gogny, "Microscopic analysis of collective dynamics in low energy fission", Nucl. Phys. A428, 23 (1984).
22. H. Goutte, J. F. Berger, P. Casoli, and D. Gogny, "Microscopic approach of fission dynamics applied to fragment kinetic energy and mass distributions in 238U", Phys. Rev. C 71, 024316 (2005).
23. W. Younes, and D. Gogny, "Nuclear Scission and Quantum Localization", Phys. Rev. Lett. 107, 132501 (2011).
24. Younes, W., Gogny, D., and Schunck, N., "A microscopic theory of low energy fission: Fragment properties", Proc. of the Fifth International Conference on ICFN5, Sanibel Island, Florida, USA, 4-10 Nov. 2012, p. 605, World Scientific, Eds. J.H. Hamilton and A.V. Ramayya (2013).
25. S. Lemaire, P. Talou, T. Kawano, M. B. Chadwick, and D. G. Madland, "Monte Carlo approach to sequential neutron emission from fission fragments", Phys. Rev. C **72**, 024601 (2005).
26. Jørgen Randrup and Ramona Vogt, "Calculation of fission observables through event-by-event simulation", Phys. Rev. C **80**, 024601 (2009).
27. P. Talou, T. Kawano, I. Stetcu, "Prompt fission neutrons and gamma rays in a Monte-Carlo Hauser-Feshbach formalism", Physics Procedia 47, 39 (2013).
28. R. Vogt, J. Randrup, "Event-by-event modeling of prompt neutrons and photons from neutron-induced and spontaneous fission with FREYA", Physics Procedia 47, 82 (2013).
29. IAEA 499, INDC(NDS)-0499, "Summary Report of Second Consultants' Meeting on Beta Decay and Decay Heat" (2006).
30. IAEA 676, INDC(NDS)-0676, "Total Absorption Gamma-ray Spectroscopy for Decay Heat Calculations and Other Applications" (2014).
31. IAEA 643, INDC(NDC)-0643, "Summary Report of First Consultants' Meeting on Development of a Reference Database for Beta-delayed Neutron Emission" (2013).
32. J.K. Tuli, Nuclear Data Sheets 98, 209 (2003).
33. C. Gross, Phys. Rev. C 85, 024319 (2012).
34. IAEA 535, INDC(NDS)-0535, "Consultants' Meeting on High-precision beta-intensity measurements and evaluations for specific PET radioisotopes" (2008).
35. IAEA 591, INDC(NDS)-0591, "Consultants' Meeting on Improvements in charged-particle monitor reactions and nuclear data for medical isotope production" (2011).

36. IAEA 596, INDC(NDS)-0596, “Technical Meeting on Intermediate-term Nuclear Data Needs for Medical Applications: Cross Sections and Decay Data” (2011).
37. IAEA 675, INDC(NDS)-0675, “Second Research Coordination Meeting on Nuclear Data for Charged-particle Monitor Reactions and Medical Isotope Production” (2015).
38. S. Croft and S.J. Tobin, “A Technical Review of Non-Destructive Assay Research for the Characterization of Spent Nuclear Fuel Assemblies Being Conducted Under the US DOE NGSI,” LANL Report LA-UR-10-08045 (2011).
39. E. Bauge *et al*., “Coherent investigation of nuclear data at CEA DAM: Theoretical models, experiments and evaluated data,” Eur. Phys. J. A 48:113 (2012).
40. C.M. Perfetti and B.T. Rearden, “Continuous-Energy Monte Carlo Methods for calculating Generalized Response Sensitivities using TSUNAMI-3D,” in Proc. of the 2014 International Conference on the Physics of Reactors (PHYSOR 2014), Kyoto, Japan, Sept. 28 – Oct. 3, 2014.
41. B.T. Rearden, M.L. Williams and J.E. Horwedel, "Advances in the TSUNAMI Sensitivity and Uncertainty Analysis Codes Beyond SCALE 5," Trans. Am. Nucl. Soc. 92, 760-762 (2005).
42. M. Salvatores *et al*., “Uncertainty and target accuracy assessment for innovative systems using recent covariance data evaluations,” Nuclear Energy Agency Report NEA/WPEC-26 (2008).
43. M. Salvatores *et al*., “Methods and issues for the combined use of integral experiments and covariance data: results of a NEA international collaborative study,” Nucl. Data Sheets 118 38 (2014).
44. M. Herman *et al*., "EMPIRE: Nuclear Reaction Model Code System for Data Evaluation", Nucl. Data Sheets, 108, 2655-2715 (2007).
45. T. Kawano, CoH3 Hauser-Feshbach code (2015).
46. Y. Danon, personal communication (2015).
47. Grimes, et al., “A study of the Absorption Cross Sec/on of Iron using Two Techniques”, this meeting.
48. Y. Danon, et al. NEMEA-7 Workshop Proceedings, OECD-NEA Report Number NEA/NSC/DOC(2014)13, (2014).
49. S. Hogle, G.I. Maldonado, and C.W. Alexander, “Increasing transcurium production efficiency through directed resonance shielding,” Annals of Nuclear Energy, 60, 267-273 (2013).
50. F. Reines, C.L. Cowan, F.B. Harrison, A.D. McGuire, and H.W. Kruse, “Detection of the free antineutrino,” Phys. Rev. 117, 159 (1960).
51. C. Adams *et al*., arXiv:1503.06637.
52. Y. Abe *et al*., Phys. Rev. Lett. 108, 131801 (2012).
53. F.P. An *et al*., Phys. Rev. Lett. 108, 171803 (2012).
54. J.K. Ahn *et al*., Phys. Rev. Lett. 108, 191802 (2012).
55. P. Vogel, G.K. Schenter, F.M. Mann, and R.E. Schenter, Phys. Rev. C 24, 1543 (1981).
56. D.A. Dwyer and T.J. Langford, Phys. Rev. Lett. 114, 012502 (2015).
57. A.A. Sonzogni, T.D. Johnson, and E.A. McCutchan, Phys. Rev. C 91, 011301(R) (2015).
58. M. Fallot *et al*., Phys. Rev. Lett. 109, 202504 (2012).
59. F. von Feilitzsch, A.A. Hahn, and K. Schreckenbach, Phys. Lett. B 118, 162 (1982); K. Schreckenbach, G. Colvin, W. Gelletly, and F. von Feilitzsch, Phys. Lett. B 160, 325 (1985); A.A. Hahn *et al*., Phys. Lett. B 218, 365 (1989); N. Haag *et al*., Phys. Rev. Lett. 112, 122501 (2014).
60. P. Huber, Phys. Rev. C 84, 024617 (2011).
61. A.C. Hayes, J.L. Friar, G.T. Garvey, G. Jungman, and G. Jonkmans, Phys. Rev. Lett. 112, 202501 (2014).
62. A. Algora *et al*., Phys. Rev. Lett. 105, 202501 (2010).
63. A. Fijalkowska *et al*., Nucl. Data Sheets 120, 26 (2014).
64. E. Christensen, P. Huber, P. Jaffke, and T.E. Shea, Phys. Rev. Lett. 113, 042503 (2014).
65. J. Tain *et al*., arXiv: 1505.0549 (2015).
66. J. Ashenfelter *et al*., arXiv: 1309.7647 (2013).
67. Yu-Feng Li, arXiv: 1402.6143 (2014).
68. Soo-Bong Kim, arXiv: 1412.2199 (2014).

# Specific Needs for Isotope Production

Over 20 million nuclear medicine procedures are performed each year in the United States [1]. Nuclear data are essential for both the production and the proper application of these radionuclides. The main goals of current research are to develop the production capabilities for new radionuclides, remove discrepancies in existing data, and find alternative production routes for established radionuclides. Here we will describe a representative set of current nuclear data needs for medical isotopes. We note however that the field of nuclear medicine is continually evolving, so an ongoing nuclear data research effort in this field is required to address changing trends in medical applications. We also note that the use of a particular isotope in a medical application may be driven by availability; research into the production of an isotope may therefore be driven by the prospects for applications of that isotope in a clinical setting.

Knowledge of nuclear excitation functions, which describe reaction probabilities as a function of the incident particle energy, is central to the isotope production effort. These functions are obviously necessary to determine the yields of the nuclei of interest. In addition they are required for the prediction of the amount of undesired “contaminant” materials produced, and are therefore crucial guides in the choice of target materials. In many cases, particularly for the wide range of possible contaminants, the excitation functions have never been measured. In some cases in which excitation functions have been measured, significant differences exist between the measurements. New measurements are required to resolve these discrepancies. As discussed below, there is also a need to have a set of excitation functions known to high precision for use as monitor reactions. Finally, accurate excitation functions are required for the verification, validation, and development of theoretical nuclear reaction model codes.

Cross sections are generally studied through the activation foil method, in which a well-characterized thin foil is irradiated at a single beam energy, and the produced radioactive residuals are quantified through off-line alpha, beta, gamma, or electron spectroscopy. In this approach, control of uncertainties depends strongly on the accurate characterization of the target material, precise measurements of the incident particle flux, and the spectroscopic assaying techniques used. Access to high quality nuclear data can be useful in reducing the latter two sources of uncertainty.

The incident particle flux is commonly measured using a monitor reaction. For charged-particle induced reactions, relatively few monitor reactions that can be applied in the low to intermediate energy regimes of 30-70 MeV have been accurately measured (*i.e*. to a few percent level). Above 70 MeV, monitor reactions whose experimental measurement is free from the potential influence of secondary neutron contributions to residual yields are not well characterized. **As both IPF and BLIP operate at proton energies in the 100 MeV regime, monitor reactions at these high energies are an obvious, high-priority nuclear data need for medical isotopes.**

Gamma-ray spectroscopy is the most common tool used in the identification and quantification of nuclei produced in a reaction. Traditionally, singles spectra are collected, and the gamma-ray intensity is measured as a function of time to correlate half-life with parent nuclide. This technique is particularly challenging at higher incident energies or with heavier target nuclei; as the number of reactions leading to unstable nuclides is greatly increased, so is the density of gamma-ray transitions. With little selectivity, products from contaminant reactions can compromise and degrade the quality and validity of the cross section determination. The field of gamma-ray spectroscopy has now matured to a level at which Compton-suppressed, gamma-gamma coincidence spectroscopy is the standard. Compton-suppression improves the peak to background ratio in the spectrum, thus giving higher sensitivity, which allows for more accurate peak area determinations. Gamma-gamma coincidence analysis enables one to uniquely identify the parent isotope. New measurements for isotope production R&D, or the quantification of decay properties, should ideally employ such techniques, to achieve higher quality and more reliable data. In some cases it will be necessary to combine Compton-suppressed gamma-gamma coincidence spectroscopy with chemical separations.

In the following we outline some current needs for medical isotope production. We again emphasize that this list is by no means exhaustive, and could change with the evolving needs of the medical isotope community. In addition to the need for more accurate cross sections for the isotopes of interest, it is also essential that the accompanying impurities be well characterized. Finally, we note that a recent series of IAEA studies has also identified nuclear data needs relating to medical isotope production [2,3,4].

- **Theranostic agents:** The future of personalized medicine lies with theranostic agents. Theranostic agents are identical in molecular structure to the chemically active agent, but incorporate an isotope that emits a gamma suitable for PET or SPECT imaging. These agents may also incorporate isotopes of the same element that have a therapeutic emission (*i.e.* Auger, alpha or beta) to radiotherapy applications. Such agents offer physicians the ability to perform dynamic visualization studies of the active agent's movement and localization at the target site. Understanding the agent's behavior allows the physician to optimize dosage, to enhance treatment and reduce the side effects of the drugs. In order for this procedure to be successful, the theranostic agents need to be produced with high-specific activity, and following purification, should contain only very small amounts of other isotopes of the same element. One specific example discussed in detail was the Platinum radioisotopes, which could be combined with standard platinum chemotherapeutic agents. $^{188}$Pt and $^{191}$Pt show potential for use in imaging chemotherapy, whereas $^{193m}$Pt, $^{195m}$Pt and $^{197}$Pt all have possible uses in radiotherapy. All require more cross-section measurements, to determine optimum production routes. A recent NSAC report [5] highlights additional pairs of isotopes that have potential uses as theranostic agents.

- **Intermediate-energy charged-particle reactions**: In the energy regime of 30 to 100 MeV and even beyond, which is most accessible to the major isotope production facilities BLIP and IPF, there are many unexplored reactions, which have strong cross section requirements. This includes all types of medical isotopes; non-standard $\beta^+$ emitters, SPECT radionuclides and their generator parents, as well as therapeutic isotopes [2,6]. Examples of specific isotopes and reactions are given in Appendix B. It was noted that a dedicated low-current, 200 MeV research beamline at BLIP could address many of these data needs, while providing a training ground for young nuclear scientists. Development of this capability at BLIP will complement the existing cross section measurement capability at Los Alamos, which offers the potential for the measurement of proton-induced reactions between 40 and 100 MeV, as well as at 200 and 800 MeV.

- **Alpha-emitting radioisotopes**: A major limitation encountered by studies of promising alpha-emitting radioisotopes for cancer therapy is their lack of availability. There are attempts now underway to address this problem in specific cases. Production of $^{225}$Ac through the reaction $^{232}$Th(p,x) is being actively pursued at both IPF and BLIP. Production through neutron irradiation of $^{226}$Ra is also being investigated at HFIR. Further research into new production methods and more efficient isolation methods is required. Production routes for $^{211}$At, $^{212}$Pb/$^{212}$Bi, $^{213}$Bi, $^{226}$Th and $^{227}$Th have all been identified as high priority topics by the most recent NSAC Long Range Plan on Isotope Production [5].

- **Auger and Coster-Kronig electron emitting nuclei:** Low-energy electron-targeted radiotherapy makes use of radionuclides which emit Auger and Coster-Kronig electrons. Characterization of the medium energy production of such nuclides, to allow the determination of achievable radioisotopic purities and yields, is lacking in many interesting cases. Accurate excitation functions that result in the no-carrier-added formation of $^{119}Sb$, $^{134}Ce/^{134}La$, $^{165}Tm/^{165}Er$, $^{71}Ge$, and $^{55}Co$ are needed.

- **Alternative pathways for $^{99m}Tc$ production:** $^{99m}Tc$ is perhaps the most well-known medical isotope, and is used in 80% of medical imaging procedures. There is serious concern that the production of this critical isotope through the usual fission process $^{235}U(n,f)$ is in jeopardy, due to reactor aging [7]. Furthermore, the use of reactors, especially those using highly enriched uranium (HEU) targets, carries a nuclear weapons proliferation risk; the identification of alternative pathways for $^{99m}Tc$ production is therefore a priority [8]. Various $^{99m}Tc$ production techniques have been suggested; these will require detailed new measurements to assess their feasibility for full-scale production. The specific reactions of current interest are listed in Appendix B. These require $^{99m}Tc$ production cross section measurements, as well as studies of the associated production of long-lived impurities [9].

## Highlight 2: ORNL Isotope Uncertainty Quantification

Since many isotope production processes involve multiple neutron captures and have the associated combined uncertainties, it can be difficult to establish which reaction rates are the most important limiting factors in production. ORNL is accordingly developing high fidelity sensitivity and uncertainty analysis tools [11-12] that are capable of assessing the importance of various parameters, including energy-dependent cross sections, in determining reaction rates and reaction rate ratios. This work relies on having accurate covariance estimates and heavy-actinide data. These tools could in future be incorporated ORIGEN (in development) to analyze the effects of energy dependent cross sections on isotope yields; this would be a very powerful computational tool for identifying and prioritizing the nuclear data needs of the isotope production community.

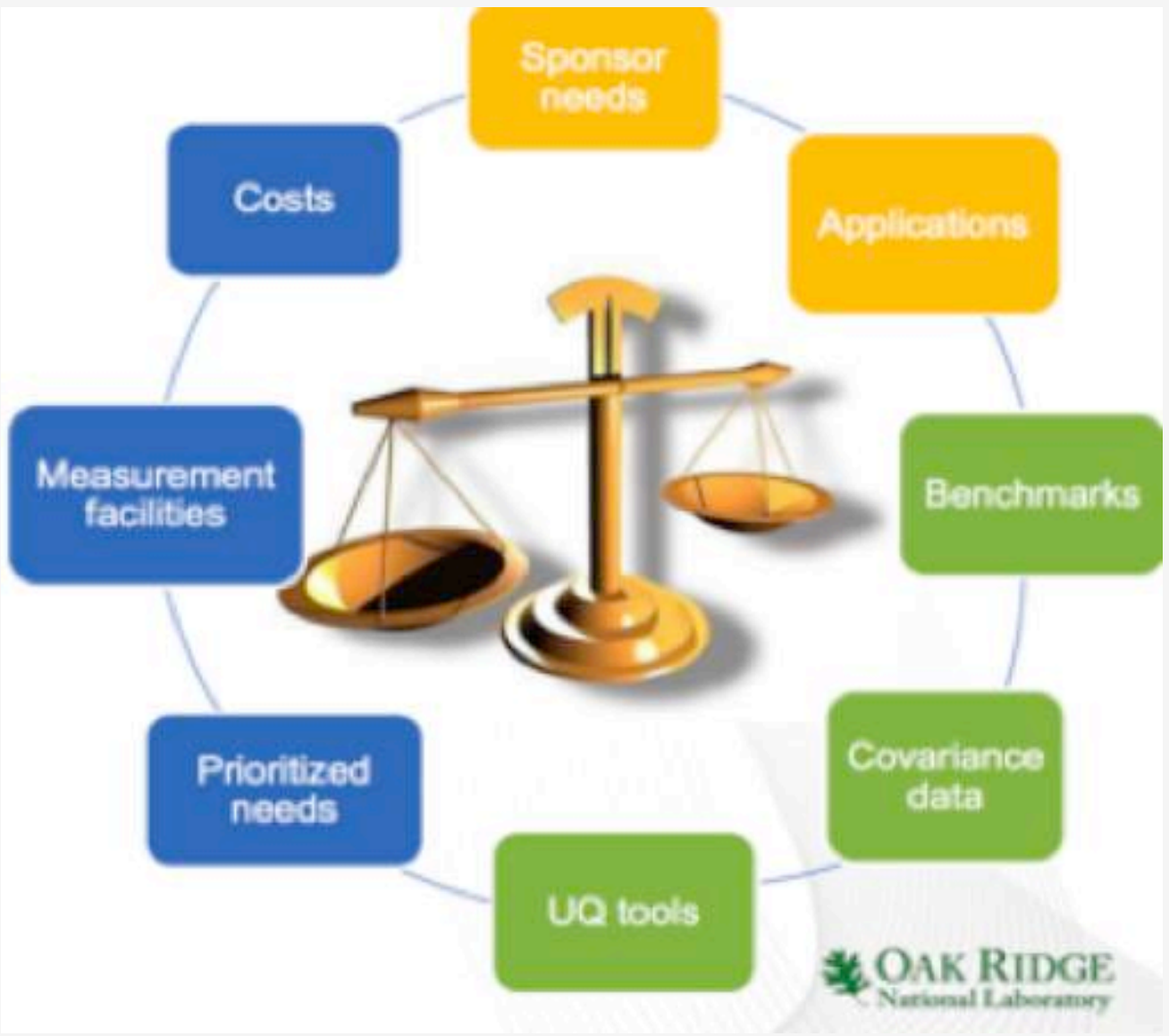

- **Fast-neutron induced reactions:** The production of medical radionuclides by fission neutrons is used routinely, and is often well characterized. Similarly, spallation or d/Be break-up neutrons could be advantageously utilized for radionuclide production [10], but there is very little data on these types of reactions. A number of therapeutic radionuclides can potentially be produced by spallation neutrons through (n,p) reactions. Examples of possible targets include $^{32}S$, $^{47}Ti$, $^{64,67}Zn$, $^{105}Pd$, $^{149}Sm$, $^{175}Lu$, and $^{177}Hf$. In addition, the production of alpha-emitting radionuclides such as $^{225}Ac$, $^{223}Ra$, and $^{227}Th$ using spallation neutrons has yet to be explored. Finally, $^{99}Mo$ could be produced by using spallation neutrons to induce fission in $^{232}Th$ or $^{238}U$.

- **Medical Isotope Production at MURR:** the Nuclear Regulatory Committee (NRC) oversees Reactor safety measures at MURR. One of these measures mandates that during irradiation the target and the host vessel must always remain at temperatures below half of their melting points. As a large fraction of the thermal energy deposited in these components is due to the absorption of locally produced gammas from radiative capture, it is accordingly very important to accurately model the amount of heating that results from this mechanism. However, the gamma yield production data for radiative capture in the relevant thermal – 10 MeV neutron energy regime is lacking in the evaluated neutron interaction files for the list of nuclides provided. (See Appendix B.) It will be very useful to carry out these measurements, since these data are essential for the accurate calculation of gamma heating required to ensure that MURR operates in compliance with NRC regulations.

- **$^{249}$Bk and $^{251}$Cf ; Production targets for Super Heavy Element (SHE) research:** The currently preferred SHE target materials are $^{249}$Bk and $^{251}$Cf. These are both byproducts of the usual $^{252}$Cf production process through multiple neutron captures, and have mass yields of 10% for $^{249}$Bk and ~2-5% for $^{251}$Cf, *albeit* with $^{252}$Cf contamination. Employing thermal neutron flux filters and short-term irradiation could increase the relative production of 249Bk. This procedure could be also employed to increase the relative $^{251}$Cf production and minimize $^{252}$Cf by extending the neutron flux filter out into the first few $^{251}$Cf resonances. Further neutron filtering, possibly with a staged approach, may be needed to produce significant quantities of $^{254}$Es and $^{255,257}$Fm as future SHE target materials. **To successfully design such filters, data on the significant neutron capture resonances for these heavy actinides are needed.**

- **$^{63}$Ni: Detectors for explosives and narcotics based on electron-capture technology:** $^{63}$Ni is produced through neutron absorption by highly enriched $^{62}$Ni; the target design and irradiation conditions are optimized to balance the total $^{63}$Ni yield and specific activity, and to minimizing the amount of expensive $^{62}$Ni target material consumed. Although neutron absorption by $^{62}$Ni is fairly well characterized, absorption by $^{63}$Ni, which ultimately limits both production and specific activity, is not. Thus, to optimize production of $^{63}$Ni, measurements of the cross sections for $^{63}$Ni(n,X) are needed.

- **$^{238}$Pu: Power source for satellites and NASA's deep space missions:**
$^{238}$Pu is used in Radioisotope Thermoelectric Generators (RTGs) and Radioisotope Heater Units (RHUs) to produce power for electronics and heat for environmental control in deep space missions [13-14]. $^{238}$Pu production in the US ended in the late 1980s with the shutdown of the Savannah River reactors. Since 1993 the domestic supply has consisted of purchases from Russia, which ended in 2009. DOE's goal to address this need is to produce 1.5 to 2 kg of $^{238}$Pu/year within the DOE complex, for example at HFIR, by 2018. Figure 3 shows the production sequence of $^{238}$Pu through neutron irradiation of a $^{237}$Np target, followed by beta decay. Simulations based on nuclear data (such as fission product evaluations and thermal cross section values) are used to estimate the heat generation rates for safety analyses of the $^{237}$Np targets during HFIR irradiation. Heat is generated primarily by fission of the $^{238}$Np and $^{239}$Pu formed during irradiation; validation of these fission rates is therefore critical to

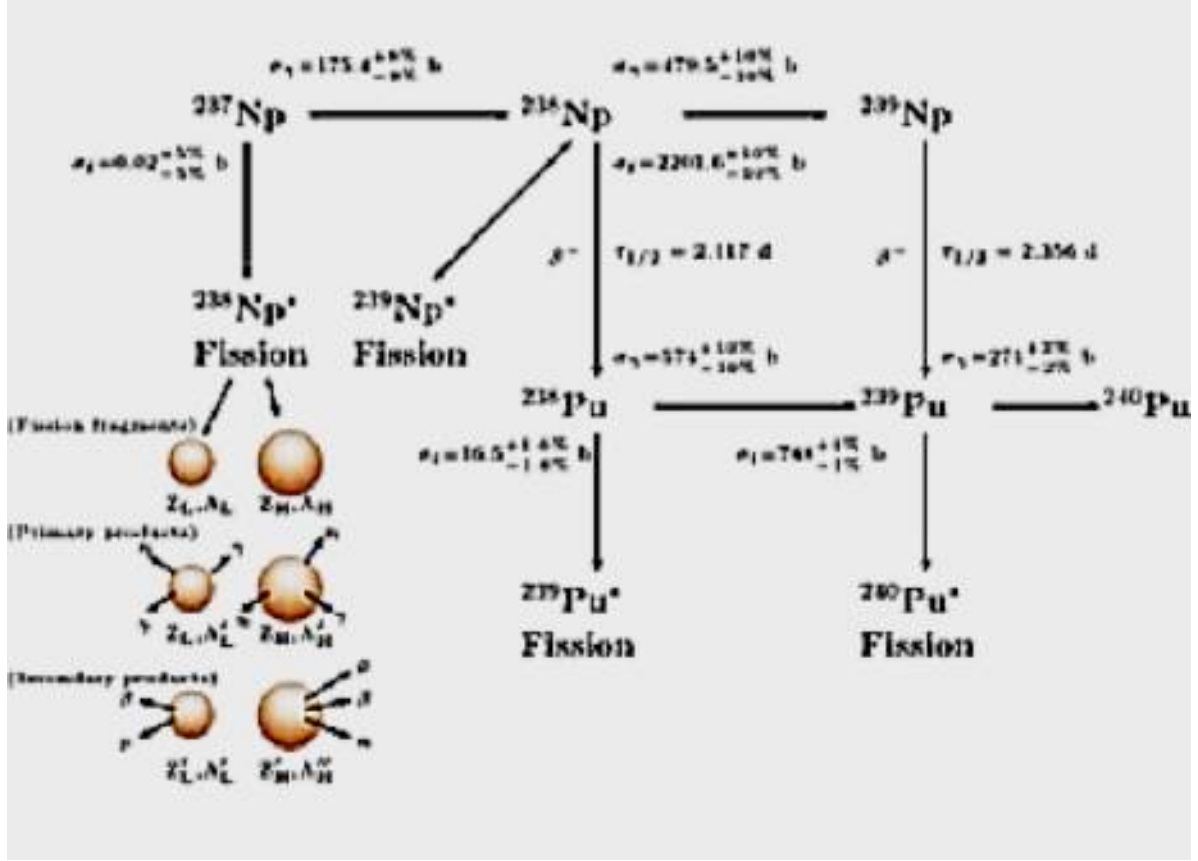

**Figure 3.** The reaction network involved in $^{238}$Pu production.

insure that these targets will operate within their safety limits. Some of the nuclear data used in these simulations, such as the $^{238}$Np fission product evaluation, have large uncertainties, which imply correspondingly large uncertainties in the target heating due to fission. New fission product measurements on $n_{th}$+$^{238}$Np are needed to assess the total fission product inventory, which will substantially reduce the level of uncertainty in target heating. Another concern is that the absence of measured yields for thermal fission may have been addressed by the substitution of fast fission data, which will clearly bias the fission heat calculations. The Appendix includes a summary of the nuclear data used for modeling and simulation of $^{238}$Pu production from $^{237}$Np targets; evidently there is considerable scope for improved measurements.

# References

1. F. Mettler *et al*., Radiology 253, 520 (2009).
2. IAEA 675, INDC(NDS)-0675, “Second Research Coordination Meeting on Nuclear Data for Charged-particle Monitor Reactions and Medical Isotope Production” (2015).
3. IAEA 596, INDC(NDS)-0596, “Technical Meeting on Intermediate-term Nuclear Data Needs for Medical Applications: Cross Sections and Decay Data” (2011).
4. IAEA 591, INDC(NDS)-0591, “Consultants' Meeting on Improvements in charged-particle monitor reactions and nuclear data for medical isotope production” (2011).
5. NSAC Isotope Subcommittee Report, “Meeting Isotope Needs and Capturing Opportunities for the Future: The 2015 Long Range Plan for the DOE NP Isotope Program,” July 20, 2015.
6. S.M. Qaim, Radiochimica Acta 100, 635 (2012).
7. T. Ruth, Nature 457, 536 (2009).
8. D. Updegraff and S.A. Hoedl, “Nuclear Medicine without Nuclear Reactors or Uranium Enrichment,” American Association for the Advancement of Science, June 13, 2013.
9. S.M. Qaim, J. Radioanal. Nucl. Chem. 305, 233 (2015).
10. K. Tadahiro *et al*., J. Phys. Soc. Japan 82, 034201 (2013).
11. C.M. Perfetti and B.T. Rearden, “Continuous-Energy Monte Carlo Methods for calculating Generalized Response Sensitivities using TSUNAMI-3D,” in Proc. of the 2014 International Conference on the Physics of Reactors (PHYSOR 2014), Kyoto, Japan, Sept. 28 – Oct. 3, 2014.
12. B.T. Rearden, M.L. Williams and J.E. Horwedel, "Advances in the TSUNAMI Sensitivity and Uncertainty Analysis Codes Beyond SCALE 5," Trans. Am. Nucl. Soc. 92, 760-762 (2005).
13. R. Wham, $^{238}$Pu Supply project - Technology Demonstration; http://web.ornl.gov/sci/aiche/presentations/2015-02-19AICHE-ANS-r2.pdf.
14. A. Witze, “Nuclear power: Desperately seeking plutonium.” Nature News Feature, 25 Nov. 2014.

# Specific Needs for National Security

As noted previously, there is considerable overlap between the nuclear data needs of National Security and Nuclear Energy. This duplication, coupled with the classification issues inherent in much National Security work, make this section somewhat less detailed than the sections that address other topical areas. There is also a long history of “needs documents” [1-3], and a well-supported series of experimental activities associated with the NNSA stockpile program, in the form of SSAA, campaigns, *etc.*, with overlaps with DNDO, NA-22, DTRA, and DHS. National Security applications nonetheless do involve some unique aspects of nuclear data, especially in the areas of detection and forensics, and neutronics and particle transport.

## Needs for Detection and Forensics

There are many applications in the areas of detection, forensics, and non-destructive assay in which one wishes to rapidly determine the isotopic composition of a sample, possibly in a high-background environment. Examples of such applications within the Next Generation Safeguards Initiative (NGSI) mission (NA-241) space are:

- Re-verification of material after a break in the chain of custody,
- Determination of criteria for the termination of safeguards at a geologic repository,
- Input accountability at reprocessing facilities,
- Enhanced containment during transshipment,
- Deterrence of diversion,
- Non-safeguards applications: burn-up credit, efficient facility operations, and heat-load determination in a repository.

Techniques that provide unique signatures rely on correlated data, such as γ-γ and γ-n coincidence data or time correlations from beta decay chains. Several techniques of this type were presented in this workshop: NRF, PGAA (the most well developed), the computer codes CASCADES and FIER. The coincidence and correlation data needed for these techniques and applications are derived from ENSDF, and provide yet another example of needs at the intersection of structure and reactions. This list of needs includes the following:

.

- The CASCADES tool developed at PNNL uses coincidence gammas to rapidly assay materials. **This tool needs up to date and complete gamma coincidence data.**

- The FIER tool developed at U.C. Berkeley models time-dependent gamma emission from decaying fission fragments. **This tool needs up to date and accurate decay data.**

- Nuclear Resonance Fluorescence (NRF) uses gamma rays to excite compound nuclei, which populates states with spin distributions, unlike other probes. The resulting de-excitation cascade or particle emission data provides another unique isotopic signature. **Data on the (γ,γ'), (γ,f), and (γ,n) interactions of major actinides and important fission products are needed.**

- Prompt Gamma Activation Analysis (PGAA) is an active interrogation technique in which thermal neutrons interact with a target, and excite a nucleus just at the neutron separation energy. The resulting gamma cascade is unique to the isotope, and offers another approach to assaying materials. The Evaluated Gamma Activation File (EGAF), a library of prompt gamma data, is crucial for this application. Data needs related to EGAF are given below:

- Efforts are currently underway at LBNL and LLNL to improve data for structural materials, such as **Fe and Zr has been identified as a future need (see Appendix A)**.

- **Recent work on $^{242}$Pu(n,γ) [9] finds a larger cross section than is currently adopted [10]. This discrepancy should be resolved.**

- **In addition, there are recent concerns regarding the accuracy of several other adopted cross sections, *e.g.* $^{186}$W(n,γ) [2] and $^{157}$Gd(n,γ) [4]**, demonstrating an urgency to confirm cross section measurements for isotopes critical to advanced reactor design and fuel cycle initiatives.

## Highlight 3: Neutron Capture Gammas

was initiated and led by LBNL through the USNDP, in collaboration with an International Atomic (IAEA Coordinated Research Project, to provide improved neutron-capture gamma-ray cross sections for over 260 isotopes from capture-gamma measurements using natural targets of all stable elements [1]. However, for many elements only data for the isotopes with the largest cross sections and/or abundances could be obtained with natural targets. Consequently, there are many gaps and inadequacies inherent in the EGAF database as currently constructed. This problem was highlighted recently in a study of tungsten isotopes, in collaboration with Global Security at LLNL; the isotopically-enriched cross section measurements [2] were found to differ considerably from those in the current EGAF database [1].

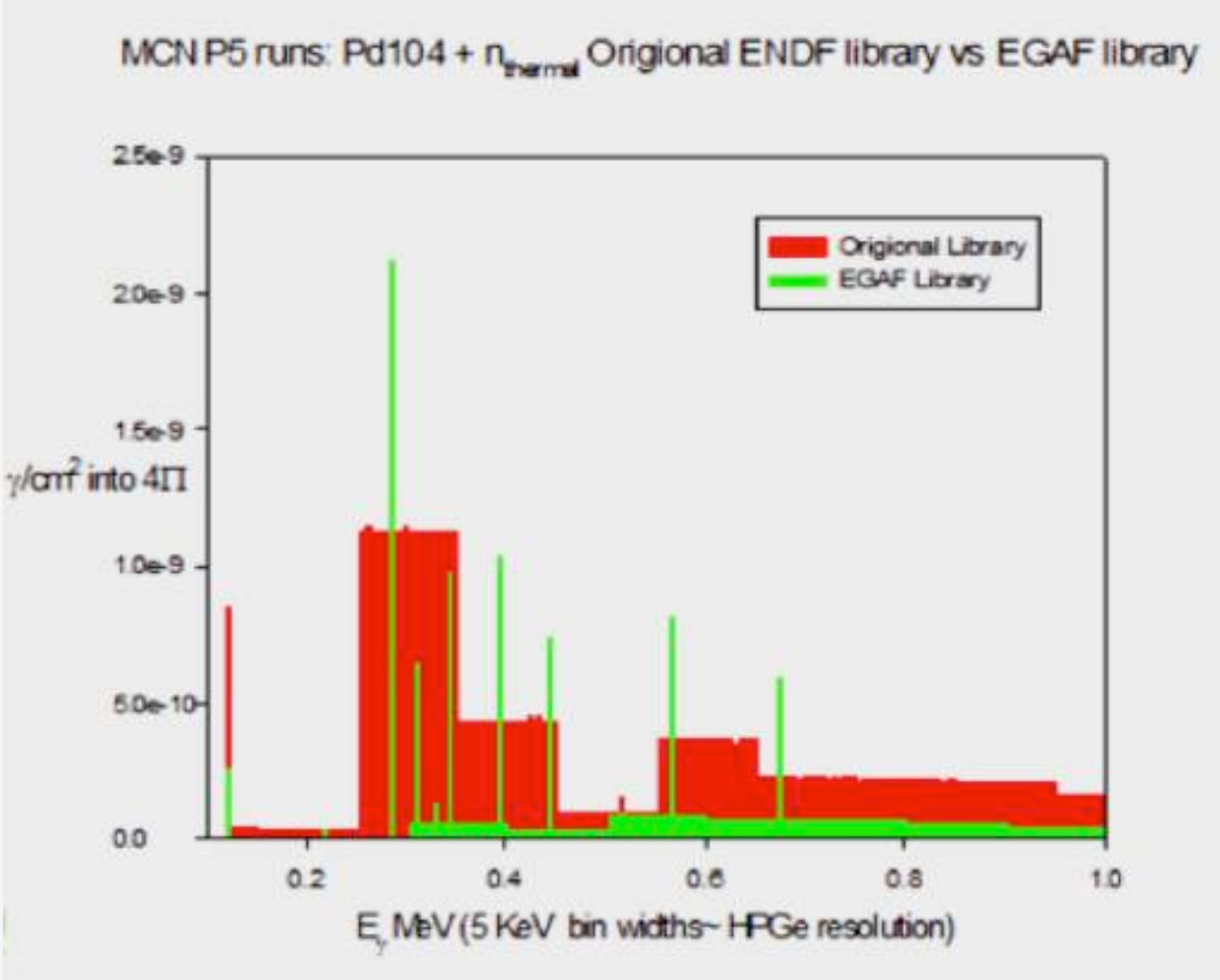

- The demonstrated success of the EGAF project (W isotopes [2], Gd isotopes [4], Pd isotopes [5], K isotopes [6], Na isotopes [7]) **still represents an ongoing need, especially for actinides (for which no data exits in ENDF!), fission products, and several other isotopes prioritized by the NA-22 office for nonproliferation applications (see Appendix A).** In addition, understanding the complete thermal capture-gamma spectra for these isotopes provides a natural segue into higher-energy neutron reactions where inelastic lines dominate the fast component of the spectrum. **The ongoing evaluation of the available (n,n'γ) data [8] will be essential to develop a future campaign of targeted (n,n'γ) measurements.**

# Needs for neutronics and other particle transport

- Most of the national security mission space requires a reliable, predictive capability for neutron transport in a large range of materials, and models of the resulting outputs. This need is greatest for reactions with incident neutron energies in the range 1-500 keV, which overlaps strongly with the prompt fission neutron spectrum.

- **The neutron energy range 1-500 keV requires many more measurements, since it falls in an experimental "gap" but covers a large fraction of the prompt fission multiplicity.** Criticality is very sensitive to data in this range. (See Figure 4 at right).

- **There is an unresolved disagreement regarding whether the average energy of thermal $^{235}$U PFNS is 2.03 MeV (ENDF) or 2.00 MeV (latest IAEA & Talou analysis). This average strongly influences criticality.**

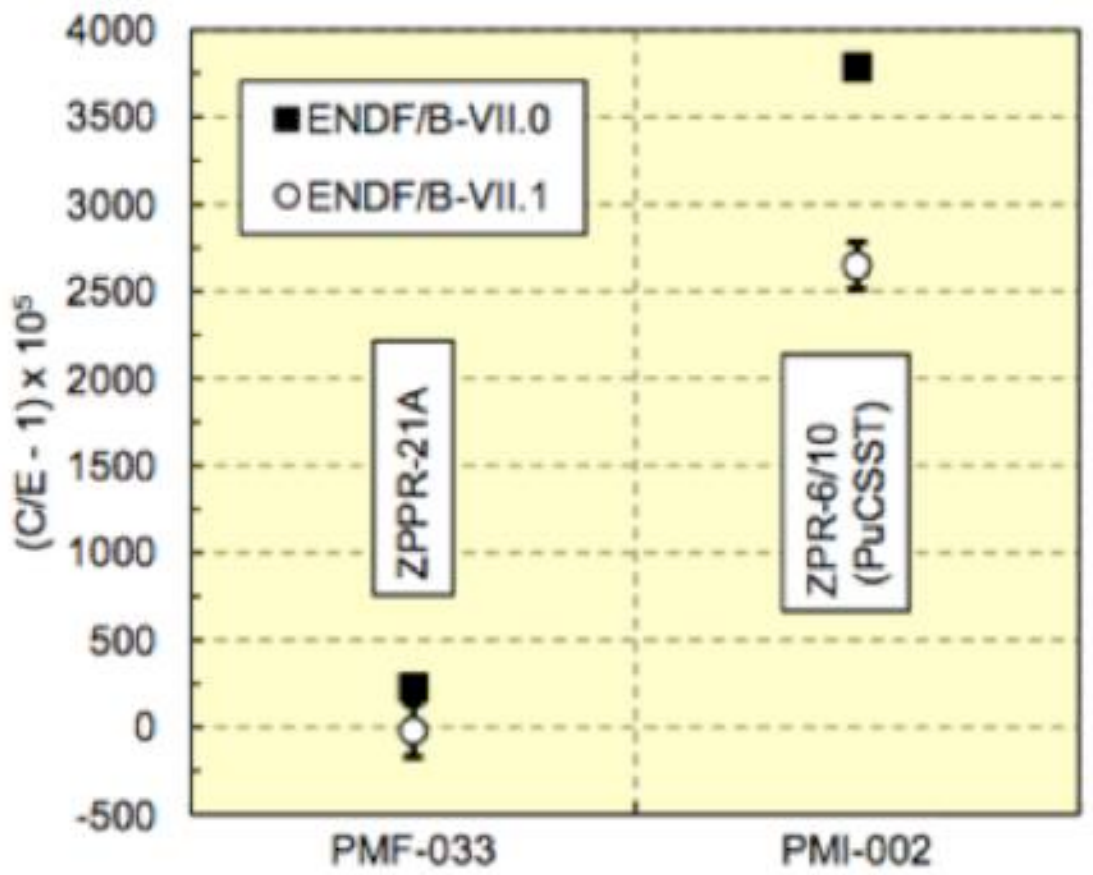

**Figure 4.** Difference between the calculated and experimental criticality parameters for two different assemblies highlighting the importance of "intermediate" energy neutrons (1-500 keV). (Reproduced from the talk by *Chadwick*.)

- **The thermal $^{239}$Pu PFNS is also insufficiently well known, and impacts solution criticals.** Figure 4 (at right) shows that the discrepancies in this "intermediate" region are ≈12% (PMI-002, right), whereas those of the simple fast PFNS are ≈ 6% (PMF-033, left).

- **$^{235}$U capture cross sections above 5 keV and below 500 keV are uncertain by ≈10%**

- **Neutron elastic and inelastic scattering on $^{235}$U and $^{239}$Pu (including angular dependence) are quite important, as mentioned in the section on cross-cutting needs [14]**

- **Prompt Fission Gamma Spectrum (PFGS) measurements are also critical (in addition to further PFNS studies) [15]**

- **Reactions on "long-lived" ≈100 ns states in populated after $^{235}$U(n,X) are important.** While it is clear that direct measurements of these reactions are not possible, improved nuclear data on both resolved and unresolved states would improve the fidelity of calculated cross sections.

- **Fission product yields for shielded fragments are needed.**

In nuclear forensics, key actinides are separated and isotopic ratios are measured by mass spectroscopy. The U and Pu chemistry makes this process difficult. Shielded fission fragments, such as $^{136}$Cs shown in Figure 5 on the right, offer a faster alternative means of identification. A shielded fragment is one that cannot be produced via beta decay of another fragment and must therefore be a prompt fission fragment.

- **Individual and cumulative fission fragment yields as a function of incident neutron energy for major actinides ($^{235}$U, $^{239}$Pu), and, to a lesser degree minor actinides, are required (see Appendix B).**

- **Nuclear physics data is essential to characterize ignition-relevant implosion experiments. A list of needs is provided in Appendix B.**

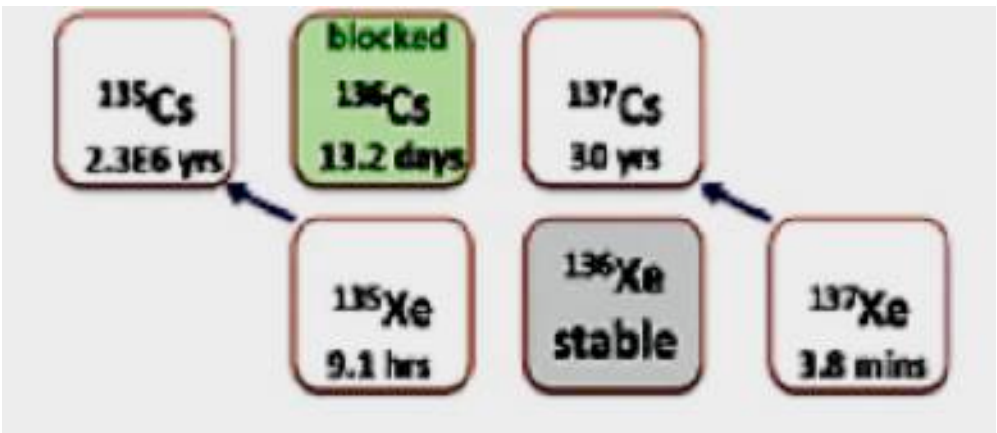

**Figure 5.** Shielded fission fragment yields near cesium and xenon used for forensic purposes. (Reproduced from the talk by *Hayes*.)

Furthermore, it is clear that reaction modeling will continue to play a major role in these applications since it is near-impossible to measure all needed (n,X) cross sections over the entire incident neutron energy range of interest for national security. While there is a fairly good understanding of the nuclear reaction models needed for neutrons, these models in turn require well-evaluated nuclear data to ensure their reliability. This includes:

- **Spin-parities, $J^{\pi}$, of low-lying states for improved modeling of (n,n'γ).**

- **Neutron resonance data for improved modeling of ($n,n_{el}$).**

- **Development of an improved targeted evaluation process for statistical nuclear properties, including level densities and radiative strength functions for nuclei near the valley of stability, of interest to national security applications.**

One last concept that appeared to be emerging at the workshop was the idea that there needed to be a new set of integral benchmark tests developed that would provide additional confidence in the cross sections in ENDF. The current approach to reaction evaluation, which optimizes cross sections using results from benchmark experiments, limits the use of this data to gain confidence in the resulting cross sections. This effort would clearly require significant coordination with the nuclear energy research community and will require more detailed study.

# References

1. R. Bahran, S. Croft, J. Hutchinson, M. Smith, A. Sood, "A Survey of Nuclear Data Deficiencies Affecting Nuclear Non-Proliferation," Proc. of the 2014 INMM Annual Meeting, Atlanta GA, LANL Report LA-UR-14-26531.
2. P. Santi, D. Vo *et al*. "The Role of Nuclear Data in Advanced Safeguards," Proc. of Global 2007: Advanced Nuclear Fuel Cycles and Systems, Boise, ID, pp. 1670-1678 (2007).
3. D. McNabb, "Nuclear Data Needs for Homeland Security", LLNL Report UCRL-MI-207715 (2005).
4. R.B. Firestone, "Database of Prompt Gamma Rays from Slow Neutron Capture for Elemental Analysis" (IAEA, Vienna, 2006), http://www.nds.iaea.org/pgaa/egaf.html.
5. A.M. Hurst *et al*., Phys. Rev. C 89, 014606 (2014).
6. M. Chadwick *et al*., Nucl. Data Sheets 112, 2887 (2011).
7. H.D. Choi *et al*., Nucl. Sci. Eng. 177, 219 (2014).
8. M. Kritčka *et al*., Phys. Rev. C 77, 054615 (2008).
9. R.B. Firestone *et al*., Phys. Rev. C 87, 024605 (2013).

10. R.B. Firestone *et al*., Phys. Rev. C 89, 014617 (2014).
11. A.M. Demidov *et al*., "Atlas of Gamma-Ray Spectra from the Inelastic Scattering of Reactor Fast Neutrons," Nuclear Research Institute, Baghdad, Iraq (1978).
12. C. Genreith, Ph.D. Thesis, "Partial Neutron Capture Cross Sections of Actinides using Cold Neutron Prompt Gamma Activation Analysis," FRM-II (2015).
13. S.F. Muhghabghab, Atlas of Neutron Resonances: Resonance Parameters and Thermal Cross Sections Z = 1-100, 5th ed. (Elsevier BV, New York, 2006).
14. A. Plompen, T. Kawano, and R. Capote Noy, "Summary Report of the Technical Meeting on Inelastic Scattering and Capture Cross-Section Data of Major Actinides in the Fast Neutron Region," IAEA Report INDC(NDS)-0597 (2012).
15. M. Jandel *et al*., Phys Rev Lett 109, (2012).

# Specific Needs for Nuclear Energy

The extensive, systematic worldwide nuclear data activities in support of energy applications have been coordinated and managed over the last few decades by the OECD/NEA. The OECD/NEA regularly publishes a comprehensive High Priority Request List (HPRL), which documents most of the nuclear data needs related to nuclear energy, and, most importantly, specifies their required accuracies. In the following, we will summarize the most relevant aspects of these nuclear data needs, and give a few specific examples.

Uncertainties have multiple, sometimes unexpected impacts on reactor design and on fuel cycle assessments. During recent decades, several especially notable cases have received considerable attention. The stringent design accuracies required to comply with safety and optimization requirements can only be met if very accurate nuclear data are used for a large number of isotopes, reaction types, and energy ranges.

Examples of potentially crucial uncertainties and their associated nuclear data needs can be found in several areas of nuclear system assessment. Some examples follow below.

## Highlight 4: The Need for Up-To-Date Nuclear Data

On March 10, 2004, Kakrapar Atomic Power Station Unit 1 (KAPS-1) in Gujarat, India, experienced an incident involving incapacitation of reactor regulating system, leading to an unintended rise in reactor power from 73%of full power to near 100% full power. The slow rise overpower transient could not be explained by the Design Manual based since it was based on the 27 group WIMS (Winfrith Improved Multi-group Scheme) 1971 library. The Indian Atomic Energy Regulatory Board shut down KAPS until incident was understood. The 2005 release of the WIMS nuclear library provided an explanation and brought the plant back into operation [6].

Similar nuclear data insufficiencies may also be, at least in part, responsible for unexplained behavior in the Canadian Maple reactors built for the production of $^{99m}$Tc. However, a definitive explanation was never determined, resulting in a forced closure of the reactors and a $1.6 billion lawsuit [7].

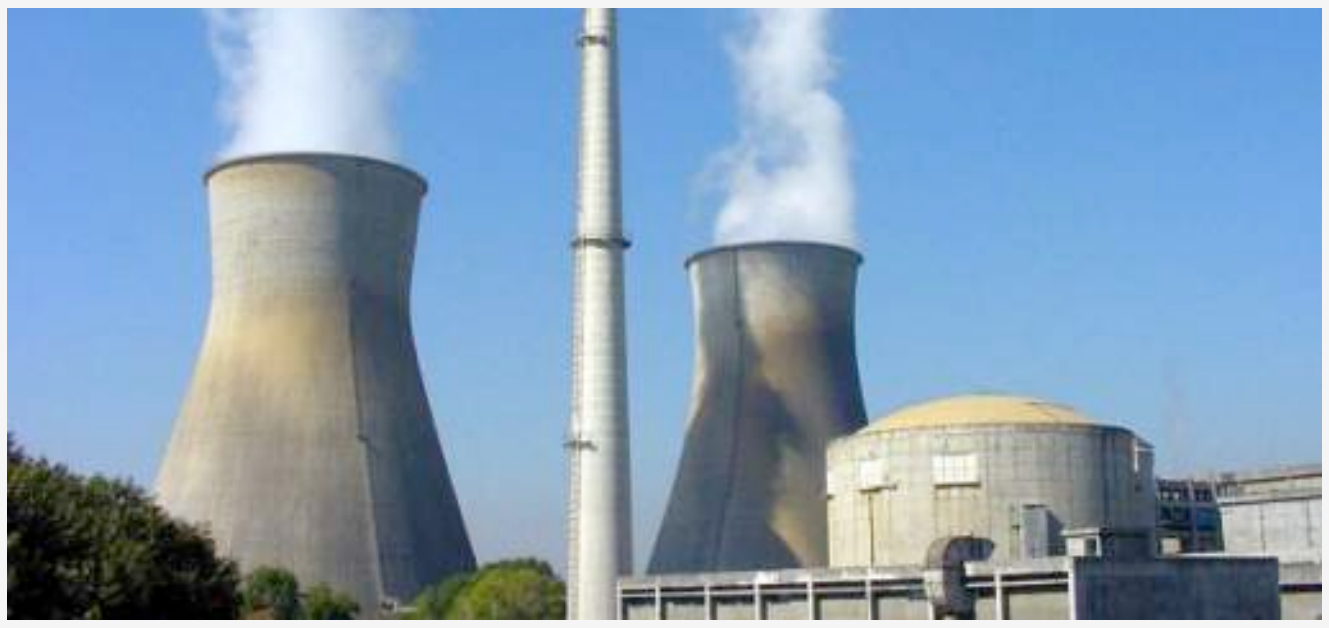

1. ***High burn-up systems will require additional and more accurate nuclear data.*** Increased burn-up scenarios will put greater emphasis on the quality of the higher actinide data and fission products evaluations. To better assess the neutron absorption rate of fission products, data on their cross sections, fission yields and radioactive decay properties need to be improved. Decay and fission yield data accordingly need to be critically assessed, and future evaluations should be accompanied by both uncertainty and covariance data. Absorption and fission cross sections, and even inelastic cross sections, of higher actinides (higher Pu isotopes and minor actinides) will play a much more crucial role in the future fuel cycle optimization studies foreseen by industry.

2. ***There is a need for more accurate and more complete covariance data in ENDF for all neutron reaction types for all nuclides (including cross correlations), fission product yields, gamma production, gamma interactions, and decay properties in fission products.*** This will facilitate more accurate uncertainty quantification and sensitivity analyses. We need these data to be accurate, or at least consistent, as they are used in calculations that provide licensing guidance.

3. ***We need to further reduce uncertainties in the neutron capture cross section data, particularly for minor actinides and Pu isotopes in the fast and epithermal energy ranges, to improve the modeling of advanced fuel cycles and fast-spectrum nuclear reactors.*** Specific integral experiments are required to validate this data:

    a. **New, science-driven integral experiments that can provide accurate information on separate physics effects.** For example, the next phase of MANTRA experiments can target fast and epithermal data for minor actinides, and reanalysis of past experiments available in the NEA DataBank could provide useful information. The application of sensitivity and uncertainty analyses to these semi-integral measurements could inform nuclear data evaluators of ways to improve ENDF data.
    b. **Follow up completed semi-integral experiments, to access existing information that could satisfy some of the needs discussed above.** For example, some pulsed iron-sphere experiments identified issues in several iron cross sections, which are important in nuclear engineering for operating reactors, spent fuel storage and transportation, and shielding design. If needed, follow with new, science-driven semi-integral experiments to generate targeted cross section data. Again, determine which data is most needed through sensitivity analyses and uncertainty quantification.

4. Regarding ***innovative systems under study at several institutions, such as the TerraPower concept, innovative molten salt concepts, and innovative and flexible breeder-burner fast reactor concepts***, typical examples of nuclear data dependent innovative design features are:
    a. Cores with low reactivity loss during the cycle: In this case the compensating effects of burn-out and built-in isotopes can strongly impact the safety case, since the control system has to accommodate significant margins. This requires a drastic reduction in certain nuclear data uncertainties which were not previously considered crucial.
    b. Cores with an increased inventory of minor actinides in the fuel: In these cases, both core criticality and the all-important safety-related reactivity coefficients are affected by large uncertainties in the data, due to our limited knowledge of the nuclear properties of minor actinide isotopes.
    c. Cores with no uranium blankets (*e.g.* to address non-proliferation concerns): Reflector effects are strongly dependent on anisotropic scattering effects, and the current nuclear data uncertainties can lead to very significant power distribution uncertainties for peripheral core fuel assemblies.
    d. Cores optimized to minimize coolant (Na) void coefficients: In safety case assessments it has been shown that current nuclear data uncertainties can result in the elimination of any potential benefits associated with such innovative core features.

5. ***In studies of innovative materials as structural or fuel components*****,** modern nuclear data evaluations and precision measurements of fast-neutron cross sections for structural materials and coolants are often missing or inadequate. For example, inelastic scattering cross sections are required for important system-dependent structural materials, coolants, and inert fuel elements. (The elements involved include Na, Mg, Si, Fe, Mo, Zr, Pb, and Bi.) As a specific example, an accurate determination of the sodium void coefficient of an SFR (Sodium Fast Reactor) requires improvements in the inelastic scattering cross sections for $^{23}$Na, as well as a complete covariance treatment. A careful reevaluation of uncertainties is definitely needed for materials associated with accident-tolerant fuels.

6. ***During and after the Fukushima events, a renewed in-depth assessment of the design and safety of spent nuclear fuel pools (SFP) at nuclear power plants was requested by all national regulators, to evaluate the expected fuel rod behavior as a function of cooling time in a realistic, modern PWR-BWR core stored in a SFP, assuming a sudden loss of cooling capacity.*** The total pool heat load can result from compensating effects, specifically opposing trends in the predicted decay heat for fresh versus highly burnt assemblies. Data uncertainties play a very important role in this case, especially for high burnup fuels.

7. ***Specific practical issues may be recognized as having been neglected or underestimated without convincing justification***. As one example, this is the case for the accumulated fluence at the tips of PWR control rods (CR). The fluence is required for both high-energy (E>1 MeV) and thermal neutrons, as these affect CR integrity through stresses and strains induced by coupled clad embrittlement and absorber swelling phenomena.

8. Optimization work on innovative features of the Canadian CANDU reactors has confirmed that ***the thermal elastic neutron scattering cross section for $^{16}O(n,n)^{16}O$ reported in some modern nuclear data libraries is too large relative to the best available experimental measurements.*** The reactivity impact of revising the $^{16}O$ scattering data was tested using the Replica Method. The uncertainty in the $^{16}O(n,n)^{16}O$ thermal cross section propagates into an uncertainty of the calculated $k_{eff}$ in thermal critical assemblies. For example, large reactivity differences of up to about 5−10 mk (500−1000 pcm) were observed using $^{16}O$ data files with different elastic scattering data. A similar discrepancy has been noted in graphite as well.

9. In studies of ***decay heat,*** uncertainty propagation is required in complex computational problems. This can be carried out by randomly perturbing the input data, using a probability distribution derived from the evaluated mean and standard deviation of each datum; a subsequent analysis gives the distribution of the quantity of interest about its mean. This basic approach however ignores correlations between the data. The use of correlations, including those associated with experiments, can have a significant effect on the final uncertainty assessment of the decay heat.

10. Future advanced nuclear reactors will require a better understanding of the fission process. Since four of the six impending Generation-IV reactors are fast ones, the high level of heat deposition requires an innovative core design. Approximately 10% of the deposited heat is due to gamma-ray energy, of which about 40% is due to prompt fission gamma-rays. Adequate modeling of heating in these cores requires estimated uncertainties of less than 7.5%. Use of the present evaluated nuclear data actually leads to an underestimate of gamma heating of the principal reactor isotopes $^{235}U$ and $^{239}Pu$ by up to 28%. ***We therefore need to considerably improve the accuracy of data on gamma production in fission events and in decay chains, as well as the associated KERMA (kinetic energy released in matter) data. Gamma production data are missing or are unbalanced for 155 of the 423 nuclides addressed in ENDF/B-VII.1, and 26 of the nuclides demonstrate negative KERMA data in at least one energy group of a commonly used 200-group structure.***

11. ***In a material irradiation test, it is essential to precisely characterize the irradiation field, considering not only the neutron fluence and displacements per atom (dpa), but also the temperature.*** Facilities and experiments for this type of study are essential for the development and validation of new fuels, and in view of the very limited availability of both, every effort should be made to extract as much reliable information as possible. The principal heat source in a material irradiation test is the gamma heating of surrounding materials, such as the stainless steels that constitute the irradiation rig, including the irradiation specimen and capsule. Accurate core and temperature calculation methods are required to predict the gamma heat rate and other key performance parameters for irradiation tests. To evaluate the spatial distribution of gamma heating in fast reactors, it is necessary to consider complete contributions to gamma intensities. Unfortunately, delayed gamma ray yield data for all actinides are not yet available in the standard evaluated nuclear data files.

12. Considerable effort has been expended in assessing and comparing various options for future fuel cycles (See for example the recent study by R. Wigeland *et al*.) ***The need to screen fuel cycle options requires that the appropriate uncertainties (including nuclear data uncertainties) be propagated in scenario codes. This is a fairly new nuclear data need, and no systematic approach is currently available.***

## Other Specific Needs for Nuclear Energy

Several other nuclear data needs were described in presentations at the NDNCA Workshop that were not directly driven by known data uncertainties.  These are:

- **One should determine of the accuracies required in β-delayed neutron data for reactor safety and criticality analyses. This should be followed by more and better measurements of β-delayed neutrons, branching ratios, and neutron energy spectra that achieve the necessary level of accuracy. These data are currently sparse and incomplete, and we need spectroscopic standards.** A small change in values could have a large impact on reactor accident analyses. VANDLE is an example of a detector that can help gather this data. These experiments can be directed towards isotopes of special interest for reactors, such as Br and I. In addition, new integral experiments for decay heat and β-delayed neutron energy spectra for the principal nuclear fuel components $^{235}$U, $^{238}$U, $^{239}$Pu, and $^{241}$Pu can generate the reliable data needed for reactors. We also need to verify the earlier measurements (typically from 40-50 years ago) with modern spectroscopy techniques.

- **Support the development of accurate neutral particle transport codes.** High-accuracy simulations are essential for identifying incorrect data; if something is modeled accurately and does not match experiment, we can then identify which data lead to the discrepancy. Correct methods are also essential for extracting data from integral and semi-integral experiments. For example, without the proper incorporation of self-shielding corrections in transport calculations, integral experimental results cannot be properly modeled.

- **Implement the deformed Hauser-Feshbach models in EMPIRE and TALYS**. These models include more complete physics information, and their predictions more closely match experimental results. This in turn generates better data for application users.

- **Better characterize fission product yield data.** Current inconsistencies in this data raises questions regarding how well we can characterize used nuclear fuel, which may impact the licensing of used fuel storage and transportation. For example, there are inconsistencies between cumulative versus independent yields.

- **Determine whether S(α, β) data are important for accurately modeling FLiBe as a coolant in advanced reactor designs such as the Fluoride Salt-cooled High-temperature Reactor (FHR).** If so, this data should be generated.

# References

1. T. Yoshida *et al*., "Assessment of Fission Product Decay Data for Decay Heat Calculations," OECD/NEA WPEC Subgroup 25, ISBN 978-92-64-99034-0 (2007).
2. OECD/NEA WPEC Nuclear Data High Priority Request List (HPRL), https://www.oecd-nea.org/dbdata/hprl/ (2015).
3. A. Plompen, "IAEA Report on Long-term Needs for Nuclear Data Development," INDC(NDS)-0601 (2012).
4. M. Salvatores *et al*., "Uncertainty and target accuracy assessment for innovative systems using recent covariance data evaluations." NEA Report NEA/WPEC-26 (2008).
5. M. Salvatores *et al*., "Methods and issues for the combined use of integral experiments and covariance data: Results of an NEA international collaborative study." Nucl. Data Sheets 118 38 (2014).
6. S. Ganesan, "New Reactor Concepts and New Nuclear Data Needed to Develop Them", AIP Conference Proceedings; Vol. 769 Issue 1, p1411 (2005).
7. Chris Whipple *et al.,*, "Medical Isotope Production without Highly Enriched Unraium" The National Academy Press. ISBN-978-0-309-13039-4 (2009).

# Closing Remarks & Thanks

The range of topics presented at the workshop and summarized in this whitepaper highlights the importance of nuclear data for a wide array of societal needs, including national and international security, aiding economic development through energy production, and medical applications, which can significantly improve our quality of life. There are few areas of research that can lay claim to being central to so many human endeavors. The extraordinary energy densities of nuclear materials, $10^3$ – $10^6$ times larger than in chemical systems, make them invaluable for many applications, while simultaneously presenting unique challenges for their responsible use. It is a clear statement of the importance of nuclear data that such a diverse body of experts could be gathered together to address these issues. The fact that the DOE's Office of Science/Nuclear Physics (NP) and National Nuclear Security Agency (NNSA) co-sponsored this workshop illustrates the broad importance of the material presented.

Although many of the participants in nuclear science are drawn to the subject by its intellectual challenges, this complexity also poses significant problems that the experimental, theoretical and evaluation communities must address in order to meet the demands of applications. Many of the nuclear data needs described here, such as an improved knowledge of the spectra of neutrons, β- and γ-rays emitted following nuclear fission, have remained unaddressed for decades due to the lack of committed resources or technical expertise required. In contrast, other nuclear data needs, such as the production cross sections of "emerging" radioisotopes for medical applications and the nascent interest in the reactor antineutrino spectrum, have arisen more recently. Moreover, many of the needs presented here, such as neutron dosimetry and covariance data, are continually evolving as new research and evaluation activities "push" uncertainties from one area to another. Clearly, the need for improved nuclear data will continue into the foreseeable future.

The need for work on nuclear data is widely appreciated within DOE. To this end, NP, NNSA and other organizations have supported a wide range of experimental capabilities in the US, whose purposes include performing nuclear data measurements in support of national nuclear data needs. At this workshop, presenters from across the country have described these capabilities, and we have attempted to document them in this whitepaper for the user community as a whole.

In addition to supporting experimental activities, NP also supports the US Nuclear Data Program (USNDP) as the national custodian of nuclear data. The primary USNDP databases are a core resource for both basic and applied nuclear science research, both nationally and internationally. In view of the expertise in nuclear data represented by the USNDP, it may be useful to consider mechanisms through which it can assist in planning experimental efforts to address future nuclear data needs.

Various possible "new directions" for the nuclear data effort have been considered in recent years, and were discussed in the course of this workshop. One possibility which we highlight here is to establish the "Super" High Priority List referred to in the beginning of this document, in which the data needs of the entire US application space would be compiled and updated regularly and assigned priorities. A periodic repetition of this workshop, perhaps on a biannual basis, would aid in this process. It was also noted that the ubiquitous role of nuclear data in society makes it well suited for study by a well-rounded expert panel, such as the National Academy of Sciences.

Finally, we acknowledge the assistance of the many people who made the workshop and this whitepaper possible. In particular we are grateful to Dorothy Kenlow and Erika Suzuki for their essential organizational skills, and we thank graduate students Ivana Abramovic, Leo Kirsch, Eric Matthews and others for their invaluable assistance.

# Appendix A: Matrix of Nuclear Data Needs

At the end of the workshop it was suggested that a matrix of nuclear data needs versus National Security applications would help to map the needs identified onto the offices and organizations that are charged with addressing specific application needs. However, it quickly became clear to the members of the writing committee that there were enough similarly cross-cutting nuclear data needs in the other areas to justify expanding this exercise to the entire application space represented in the workshop. (These being the four general areas of National Security/Defense Programs, Counter-Proliferation/Safeguards/Forensics, Nuclear Energy, and Isotope Production.) We have accordingly assembled matrices that present these cross-cutting nuclear data needs in all applications. Seven subsets of these four areas were found to have overlapping data needs, which are as follows:

- Matrix A.1: National Security, Counter-Proliferation, and Nuclear Energy
- Matrix A.2: National Security, Counter-Proliferation, and Isotope Production
- Matrix A.3: National Security, Nuclear Energy, and Isotope Production
- Matrix A.4: National Security and Counter-Proliferation
- Matrix A.5: Counter-Proliferation and Nuclear Energy
- Matrix A.6: Nuclear Energy and Isotope Production
- Matrix A.7: National Security and Isotope Production

These matrices are comprised of 2 columns:

1. **Nuclides and Issue:** A short summary of the nuclides involved and the topic requiring improved nuclear data.
2. **Nuclear Data Need:** The nuclear data quantity that requires improvement (cross section, gamma-spectrum *etc*.)

In addition to the needs that were of interest to more than one topic area there were 9 specific data needs that were identified either in the workshop or from one or more of the source documents listed in the references section on p.45 that are of interest to a single topical area. These are listed in a final three-column matrix as well (Matrix A.8):

- Matrix A.8: Single Area Nuclear Data Needs

The matrices provide tabular access to the cross-cutting nuclear data needs identified in this workshop and by other contemporary sources. Furthermore, it may also be useful in coordinating research and evaluation efforts and the preparation and evaluation of research proposals.

A partial list of the agencies that support these different applications would include:

- National Security/Defense Programs: NA-11, NA-51, NA-22, NA-24, DOD/DTRA
- Counter-Proliferation/Safeguards/Forensics: NA-22, NA-24, DHS/DNDO, DTRA
- Nuclear Energy: DOE/NE, Industry
- Isotope Production: DOE Isotope Program, DTRA, Industry, NA-22

In addition to the matrices presented in this Appendix, a 6-column version that includes “check marks” for all four topical areas was prepared as well and is in available as a separate file (NDNCA15_nuclear_data_needs_matrix.xlsx).

Several other documents were used in the preparation of these matrices in addition to the presentations and discussions during the workshop. A list of these source documents follows.

# References

1. M. B. Chadwick *et al*., “The CIELO Collaboration: Neutron Reactions on $^{1}$H, $^{16}$O, $^{56}$Fe, $^{235,238}$U, and $^{239}$Pu”, Nucl. Data Sheets 118, 1-25 (2014).
2. P. Santi, D. Vo, M. Todosow, A. Aronson, and H. Ludewig, “Nuclear Data Needs for Advanced Safeguards”, report prepared by Los Alamos National Laboratory and Brookhaven National Laboratory.
3. R. Bahran, S. Croft, J. Hutchinson, M. Smith, and A. Sood, “A Survey of Nuclear Data Deficiencies Affecting Nuclear Non-Proliferation”, Proc. 2014 INMM Ann. Meeting, Atlanta, GA; LA-UR-14-26531.
4. D. Updegraff and S.A. Hoedl, “Nuclear Medicine without Nuclear Reactors or Uranium Enrichment”, Center for Science, Technology, and Security Policy, American Association for the Advancement of Science, June 13, 2013.
5. OECD/NEA WPEC Nuclear Data High Priority Request List (HPRL) https://www.oecd-nea.org/dbdata/hprl/ (2015).

## Matrix A.1: National Security + Counter-Proliferation + Nuclear Energy

| Nuclides and Topic | Nuclear Data Need |
|---|---|
| **H, Li, Be, B, N, O, Mg, Al, Si, Ti, V, Cr, Fe, Ni, Cu, Ga, Zr, Nb, Mo, Eu, Gd, Ta, W, Ir, Pt, Au, Pb, Po, Ra, Th, U, Np, Pu, Am:** Isotopes of these elements have been prioritized by Nonproliferation and Homeland Security funding agencies: Improved data and corresponding evaluations are required to meet the demands of several applications of societal interest, including: transport modeling of unknown assemblies, NDA to enable reliable accounting for SNM, detection of contraband substances and explosives, radiation shielding design and characterization, and institutionalizing a “Safeguards by Design” approach in the development of clean, cost-effective, proliferation-resistant nuclear reactor facilities, enrichment, fuel-fabrication and reprocessing plants. Systematic experimental campaigns based on this set isotopes will greatly facilitate this need, and are described in turn. | Precise $\gamma$-ray energy data and their corresponding total and partial radiative-capture $(n,\gamma)$ cross sections, particularly for primary gamma rays, are needed for the EGAF library. New measurements for separated isotopes are especially required from thermal incident neutron energies to 20 MeV. These unique gamma-ray signatures are essential for ENDF to create complete and accurate libraries for nonproliferation applications predicated on credible high-fidelity data authentication. The actinides for which there are no primaries in ENDF are a particular concern. |
| as above | Improved inelastic scattering cross sections are needed over a wide range of neutron energies to provide data where none-to-little exists, and to meet targeted-accuracy application-driven uncertainty margins. New measurements of total inelastic and partial cross sections to individual levels are required. For many isotopes, there are considerable discrepancies between the evaluated data libraries and experimental information. |

| Nuclides and Topic | Nuclear Data Need |
|---|---|
| as above | There is a need for new and improved NRF data over proton energies of 1-5 MeV. Photonuclear elastic scattering cross-section data and electronic excitation cross section data are also required. |
| Neutron induced fission yields and cross sections are a cross cutting need for: prompt neutron spectroscopy; delayed gamma measurement for SNM identification; heat calculations for spent fuel storage; spent and fresh fuel assay; post-detonation forensics-based fallout analysis; reactor anti-neutrino source terms; isotope production calculations; new reactor design. Fission data of minor actinides are becoming more important as reactor fuels are highly burned | Improved cross section and prompt yield measurements are required as a function of incident neutron energy from thermal to ~20 MeV to provide improved correlated particle emission data from fission. Prompt fission neutron/gamma multiplicity and spectra as a function of fission fragment mass and TKE. The epithermal range of Pu-239 has large uncertainties, and the minor actinide fission cross sections and yields require improved data. Improved covariance data is required. In addition, fission fragment half-lives, peak gamma-ray energies and corresponding branching ratios are needed. |
| Numerous applications will be better served by targeted fission-data measurements, including: material characterization via neutron spectroscopy; spent fuel assay; post-detonation forensics-based fallout analysis; next generation safeguards. | New measurements are required in the thermal to fast region to provide improved correlated particle-emission data from fission corresponding to fission-product yields and covariances, prompt fission neutron spectra, half-lives, peak gamma-ray energies and corresponding branching ratios. |
| **$^{16}$O**: CIELO high-priority nucleus. Improved evaluated nuclear data needed to create accurate ENDF-formatted files for general purpose transport applications, *e.g.*, criticality, shielding, and activation. | Discrepancies of up to 30% in both measured and evaluated $^{16}O(n,\gamma)$ are problematic for fission applications. These discrepancies impact criticality predictions for reactors, and helium production rates. New measurements are needed in the 2.5-20 MeV region to reduce uncertainties to within 5-10%. |
| **$^{235}$U:** as above. | There are significant differences in evaluations in inelastic cross-section data from threshold to several MeV. These differences impact fast-criticality measurements. Differences exist in total and partial inelastic cross sections and angular distributions. New measurements and modeling are needed. |

| Nuclides and Topic | Nuclear Data Need |
|---|---|
| **$^{238}$U**: as above | Significant discrepancies between cross-section libraries for both elastic and inelastic scattering and angular distributions need to be addressed. Cross section differences are evident for total inelastic and partial cross sections to individual levels. This may have a severe impact on calculated criticality for fast systems. A global consensus on reactor sensitivity studies points to an urgent need for more accurate inelastic cross sections and angular distributions. |
| **$^{235}$U:** as above | Notable differences amongst the evaluations below 4 MeV for prompt average neutron multiplicity per fission. Libraries provide markedly different representations at 3 MeV, the average neutron energy causing fission in $^{238}$U in critical assemblies. This discrepancy has a clear impact on criticality calculations. |
| **$^{239}$Pu:** as above | Significant differences between evaluated data libraries for $^{239}$Pu in fast energy range for (n,inl). New measurements are needed to determine (n,inl) cross sections and theoretical work is needed by Hauser-Feshbach practitioners to better understand plutonium scattering reactions. |
| as above | Radiative-capture cross sections should be improved to meet the target accuracy requirements for advanced reactor systems. New measurements and evaluations are needed from 2 keV to 1.5 MeV to reduce the uncertainty to the 3-7% level (depending on the region and reactor considerations). |
| **$^{237}$Np, $^{233}$Pa**: Branching ratios for certain peak $\gamma$– and X-ray energy measurements differ by 5-15% with those in NuDat. Combining protactinium X-rays with uranium X-rays yields ratios approximately 30% higher than those in NuDat. | Further experimental decay-spectroscopy measurements needed for verification. |
| **$^{238,240\text{-}242}$Pu, $^{244}$Cm:** SF data are lacking to accurately model neutron characteristics of the advanced burner reactor fuel (ABR). | Improved data concerning the 1st, 2nd, and 3rd factorial moments of the SF neutron multiplicity distribution, and of the neutron-induced fission neutron multiplicity distribution are needed. |

| Nuclides and Topic | Nuclear Data Need |
|---|---|
| **$^{241}$Pu:** Decay-spectroscopy data are lacking for accurate fuel-cycle analyses. | Improved measurement of half-life required; contributes to $^{241}$Am in-growth. |
| **$^{242}$Pu:** Thermal neutron-capture radiative-capture (n, $\gamma$) cross section requires verification as several measurements are at odds with each other and the evaluated nuclear data libraries. This isotope contributes significantly to the mass inventory of spent fuel in the uranium nuclear fuel cycle and may have implications concerning the amount of spent fuel in burnup calculations. | Verification measurements of the total radiative thermal neutron-capture (n,$\gamma$) cross section are required. |
| **$^{232}$Th, $^{236}$U, $^{236,238,244}$Pu, $^{250}$Cm, $^{249}$Bk, $^{246,249,250,254}$Cf, $^{253}$Es, $^{244,246,254}$Fm, $^{252}$No**: Consensus values for average number of prompt neutrons (<v>) and prompt neutron multiplicity distributions (P) for these actinides are based on only 1 or 2 (max) SF measurements. The errors on <v> for Fm are currently 25%, while Es is reported without uncertainty. These quantities are important because it is unclear what actinides will be present in the ABR fuel after multiple cycles of the ABR. | New verification and precision measurements of P and <v> are highly desirable. |
| **$^{241}$Am, $^{243,245}$Cm:** Neutron multiplicity data does not currently exist for these actinides. Although these actinides have relatively small SF decay rates, and therefore negligible impact in modeling neutron emissions from ABR fuel, SF multiplicity data is essential for safeguarding fuel discharged from the ABR. | New verification and precision measurements of P and <v> are highly desirable. |
| **$^{243,254}$Cm, $^{237}$Np**: There are no neutron-induced fission neutron multiplicity data for these actinides, which will be present in ABR fuel. This hampers efforts to reliably model neutron emissions. | Measurements of neutron-induced-fission neutron multiplicity are required. |

| **Nuclides and Topic** | **Nuclear Data Need** |
|---|---|
| **$^{233,235,238}$U, $^{239,241}$Pu**: Very limited measured neutron-emission probability distribution data exist. This data is needed to accurately model neutron emissions in the burnup fuel and also has importance for NDA techniques. | Experimental measurements of neutron-emission probabilities are required as a function of incident neutron energy. |
| **Li, N, B, C, $^{16-18}$O , $^{19}$F, $^{23}$Na :** Nuclear materials that produce a much larger number of neutrons from ($\alpha$,n) reactions on light low-Z elements than from SF may cause biases in neutron-coincidence or multiplicity measurements due to the large number of accidental coincidences. Improved ($\alpha$,n) cross sections are essential for background modeling in multiplicity measurements for SNM characterization, enrichment verification, and Pu oxide characterization at reprocessing facilities. Not all ($\alpha$,n) cross sections for these nuclei have uncertainties associated with them. | New and improved measurements of ($\alpha$,n) cross sections as a function of incident alpha-particle energy from 0-10 MeV are essential. |
| **$^{244,248}$Cm**: Longer-lived alternatives for $^{252}$Cf as SF sources for neutron-detector characterization for nonproliferation applications. | Improvements in correlated-fission particle-emission data and ($\alpha$,n) yields required. |
| **$^{233}$U:** High-quality data is needed for the fission-to-capture ratio for $^{233}$U to facilitate thorium-based reactor design. | Measurements of neutron-induced fission and radiative capture are required to yield improved data for the (n,f)/(n,$\gamma$) ratio. |
| **Cd:** New capture-gamma data is needed for safeguards instruments that use Cd to get flux ratios (*e.g.* PNAR, SINRD). | Radiative-capture (n,$\gamma$) cross sections need improving from thermal to 20 MeV for all major Cd isotopes; enriched-sample measurements are required to provide desired cross-section information. |
| **Gd, Pb:** Total and partial cross sections from individual gamma transitions are necessary for applications in neutron radiography and prompt neutron gamma activation analysis. | Radiative-capture (n,$\gamma$) cross sections need improving from thermal to 20 MeV for all major Gd and Pb isotopes; enriched-sample measurements required. |

| Nuclides and Topic | Nuclear Data Need |
|---|---|
| **$^{92,94-96,98-100}$Mo:** New data are needed to assess HEU to LEU fuel conversion feasibility at the MURR facility and to predict recoverable capture energy for proposed U10Mo LEU matrix. | Total and partial radiative neutron-capture (n,γ) measurements from thermal to 10 MeV incident neutron energies in addition to photonuclear (γ,n) and neutron-induced (n,xn) cross-section measurements. |
| **$^{93}$Nb, $^{115}$In:** The total radiative thermal neutron-capture (n,γ) cross sections reported using the k0 method are in conflict with adopted values in the Atlas of Neutron Resonances. The IRDFF database values for $^{115}$In(n,γ) are also significantly discrepant with the Atlas value. | Verification of the total radiative thermal neutron-capture cross sections is required. Standalone methodologies that do not require decay-scheme normalizations will be required for independent verification. |
| **$^{23}$Na:** New data are needed to resolve ambiguities and support evaluations for material damage studies. | Total radiative neutron-capture (n,γ) data are discrepant in the fast neutron-energy region > 100 keV and new cross-section measurements are required. |
| **$^{55}$Mn, $^{58}$Fe:** New data are needed to resolve ambiguities and support evaluations for material damage studies. New data is needed to resolve ambiguities in the 10 keV to 1 MeV region for fast reactor neutrons to support evaluations for material damage studies. | Total radiative neutron-capture (n,γ) cross section measurements are needed from 10 keV to 1 MeV. |
| **$^{117}$Sn:** Enhanced understanding of materials damage. | Inelastic (n,n'γ) cross sections (total and partial) are needed to cover the energy response function from 0.3-3.0 MeV. |
| **A≈143** Isotopes: New data for high-yield fission fragments are needed for accurate prediction of inventory in used nuclear fuel assemblies, development of better physics models for calculation-based nuclear forensic tools, and neutron resonance transmission analysis. | Improved (n,f), (n,γ) thermal to fast, and (n,n'γ) fast, cross-section measurements are required. |
| **Am** and **Cm** Isotopes, **$^{237-239}$Np:** Improved data are needed for determination of spent-fuel isotopics, such as the production of $^{238}$Pu and $^{244}$Cm. | Improved thermal neutron-capture radiative (n,γ) cross-section measurements will facilitate this goal. |
| **$^{237-239}$Np**: New data for determining spent-fuel isotopics as well assessing weapons-usable material production. | Improved decay data, (n,f), and (n,n'γ) cross sections in the fast region of the neutron spectrum are required |
| **U, Pu:** Explosives detection diagnostics. | Delayed neutron-emission spectroscopy measurements. |

| Nuclides and Topic | Nuclear Data Need |
|---|---|
| **U, Pu** Fission Fragments**:** Better data is needed for enabling active interrogation technologies. | Fission-product yields needed from photofission measurements. |
| **$^{235,238}$U, $^{239,240}$Pu:** New data is required to assess potential methods for photon production from NRF. | Measurements of the (e,$\gamma$) electronic excitation cross section required from 0.5-4 MeV. |
| **$^{130,135}$I:** Accurate nuclear forensics relies on measuring actinide ratios in the debris. Blocked cesium and iodine products retain fission information and can be used to determine whether the fuel was uranium and plutonium, and if 14-MeV neutron-induced fission was involved. | The $^{130}$I and $^{135}$I thermal, fast, and 14-MeV fission yields need to be measured. |

# Matrix A.2: National Security + Counter-Proliferation + Isotope Production

| Nuclides and Topic | Nuclear Data Need |
|---|---|
| **$^{192}$Ir:** used in high dose-rate brachytherapy. | Alternative production mechanisms: High-energy protons on Platinum |
| **$^{131}$I:** used in the treatment of thyroid cancer. | Production rates may be investigated from $^{130}$Te(n,$\gamma$)$^{131}$Te->$^{131}$I; $^{130}$Te(d,p)$^{131}$Te->$^{131}$I; $^{130}$Te(d,n)$^{131}$I. |
| **$^{188}$Re:** used for palliative care of metastatic bone disease. | Alternative production: $^{187}$Re(d,p)$^{188}$Re. |
| **$^{153}$Sm:** used for palliative bone therapy. | Alternative production mechanisms require accurate data for the reaction cross section $^{150}$Nd($\alpha$,n)$^{153}$Sm. |
| **$^{186}$Re**: used for palliative bone therapy. | Alternative production mechanisms require accurate data for the reaction cross section $^{186}$W(p,n)$^{186}$Re. |
| **$^{90}$Y:** used in the treatment of liver cancer. | Alternative production mechanisms require accurate data for the reaction cross section $^{89}$Y(d,p)$^{90}$Y. |
| **$^{125}$I:** used in the treatment of prostate cancer. | Alternative production mechanisms require accurate data for the reaction cross section $^{125}$Te(p,n)$^{125}$I. |

## Matrix A.3: National Security + Nuclear Energy + Isotope Production

| Nuclides and Topic | Nuclear Data Need |
|---|---|
| **$^{206-209}$Bi:** Better data is needed for high-threshold reactions such as $^{209}$Bi(n,4n), and $^{209}$Bi(n,xn), where x>4, to help resolve discrepancies between the TENDL-2012 and ENDF/B-VII.1 evaluations. The level of discrepancy increases with x in (n,xn). A high-threshold $^{209}$Bi(n,xn) reaction may also find application as a NIF diagnostic. | Several high-accuracy cross-section measuremen reactions are desirable. |
| **$^{48}$Ti, $^{64}$Zn, $^{113}$In, $^{63}$Cu**: New dosimetry measurements will help guide future IRDFF evaluations; the current dosimetry evaluations are discrepant for several reactions in the 14 MeV neutron-energy region. | Improved dosimetry cross section data is needed near 14 MeV for the following reactions: $^{48}$Ti(n,x)$^{47}$Sc; $^{64}$Zn(n,p)$^{64}$Cu; $^{113}$In(n,n')$^{113m}$In; $^{63}$Cu(n,2n)$^{62}$Cu. |

## Matrix A.4: National Security + Counter-Proliferation

| Nuclides and Topic | Nuclear Data Need |
|---|---|
| **$^{56}$Fe:** CIELO high-priority nucleus. Improved evaluated nuclear data needed to create accurate ENDF-formatted files for general purpose transport applications, *e.g.*, criticality, shielding, and activation. | Innovative reactor systems require improved inelastic scattering cross section data to meet target a New measurements and evaluations are needed in the range 0.5-20 MeV to reduce uncertainty dow (depending on region). Substantial differences currently exist in the data libraries, *e.g.*, below 2 MeV between JEFF-3.1 and ENDF/B-VII.1 reaches 28%. |
| as above | Improved capture-gamma data from radiative neutron-capture required for nonproliferation applications (*e.g.* NDA screening): thermal – 20 MeV. High-energy primary gamma rays are particularly important. |
| as above | Double-differential neutron- and proton-emission cross section, *i.e.* (n,xn) and (n,xp), data needed in 20-200 MeV range to develop pre-equilibrium models. |
| **$^{235}$U:** as above | Radiative-capture data is poorly known in many regions and new (n,$\gamma$) measurements are needed for verification. The recent JENDL-4.0 evaluation lowered the cross section by over 25% in the 0.5-2 keV region. ENDF and JEFF libraries are also at odds with JENDL in the 3-5 keV region, and for 100-1000 keV. All evaluations need improving in the 10-70 keV region and recent findings suggest a lower capture cross section in the 100 eV to 2 keV region. |
| as above | Discrepant data evaluations for prompt and total neutron-multiplicities at 10-15 MeV indicate that new measurements and covariance analyses are needed. There are slight differences for total-thermal neutron multiplicities. |
| as above | Improved capture-gamma data from radiative neutron-capture required for nonproliferation applications (*e.g.* NDA screening): thermal – 20 MeV. Experimental high-energy primary gamma rays are particularly important; currently this information is nonexistent in the ENDF libraries. Assess HEU to LEU conversion analysis. |
| as above | Improved capture-gamma data from radiative neutron-capture required for nonproliferation applications (*e.g.* NDA screening): thermal – 20 MeV. Experimental high-energy primary gamma rays are particularly important; currently this information is nonexistent in the ENDF libraries. Assess HEU to LEU conversion analysis. |

| Nuclides and Topic | Nuclear Data Need |
|---|---|
| $^{239}$**Pu**: as above | Improved capture-gamma data from radiative neutron-capture required for nonproliferation applications (*e.g.* NDA screening): thermal – 20 MeV. Experimental high-energy primary gamma rays are particularly important; currently this information is nonexistent in the ENDF libraries. Currently there exists only one measurement in the 200 keV – 1 MeV region. Reaction theory will also be needed for cross-section evaluations owing to the paucity of data above 100 keV. |
| as above | Criticality deviations between prediction and measurement could point to possible deficiencies in the PFNS as well as (n,2n) cross sections. Deviations are particularly pronounced for outgoing neutron energies above 10 MeV and would benefit from further studies in this region. |

## Matrix A.5: Counter-Proliferation + Nuclear Energy

| Nuclides and Topic | Nuclear Data Need |
|---|---|
| $^{239}$**Pu:** CIELO high-priority nucleus. Improved evaluated nuclear data needed to create accurate ENDF-formatted files for general purpose transport applications, *e.g.*, criticality, shielding, and activation. | Evaluated total thermal neutron multiplicity values in data libraries are more than one standard deviation lower than the evaluated constant. This discrepancy needs to be addressed. |
| **Reactor Neutrinos:** Nuclear reactors provide an intense source of neutrinos up to 10 MeV and permit the study of neutrino oscillations. Two major problems are facing reactor-neutrino physics: (i) The short baseline reactor-neutrino anomaly which reveals a 6% deficit in the antineutrino flux at all short-baseline experiments; (ii) A shoulder (bump) observed at E=4.5-6.5 MeV in all current reactor neutrino experiments. The evaluations in ENDF and JEFF give different predictions because yields for the important fission products are different. | Improved experimental measurements of fission products that dominate the high-energy spectrum need to be measured to address these issues. |

## Matrix A.6: Nuclear Energy + Isotope Production

| Nuclides and Topic | Nuclear Data Need |
|---|---|
| **Te, Ru, $^{103}$Rh, $^{154,155}$Eu, $^{140,141}$Ce, Sn:** There is no radiative-capture (n,$\gamma$) gamma-ray production data in ENDF/BVII.1 for these target nuclides. This information is important for assessing heating limits from capture gammas for isotope production irradiations at MURR. | Total and partial radiative neutron-capture (n,$\gamma$) cross-section measurements from thermal to 20 MeV incident neutron energies. |

## Matrix A.7: National Security + Isotope Production

| Nuclides and Topic | Nuclear Data Need |
|---|---|
| **$^{1}$H:** CIELO high-priority nucleus. Improved evaluated nuclear data needed to create accurate ENDF-formatted files for general purpose transport applications, *e.g.*, criticality, shielding, and activation. | Precision determination (1-2%) of both total and double-differential elastic scattering cross sections at high-incident neutron energies (10-20 MeV), with emphasis on data at small center of mass scattering angles. |
| **$^{235}$U:** as above | Integral validation of the prompt-fission neutron spectrum (PFNS) for fast criticality assemblies using the (n,2n) reaction. Future work is needed to understand the differences between results from LANL and CEA in France. Better understanding of shape of PFNS is desirable. |

| Nuclides and Topic | Nuclear Data Need |
|---|---|
| as above | Discrepancies between ENDF/B-VII evaluations and IAEA-WPEC-CSEWG standards need to be resolved for radiative-capture (RC) in the 20-100 keV region. Uncertainties in RC cross section need to be reduced to 1-3% from 20 eV to 25 keV for innovative reactor design. New measurements are needed to improve understanding of RC cross section from a 1-$10^5$ eV. |
| **$^{239}$Pu:** as above | Resonance-parameter analyses to improve modeling of plutonium-solution critical assemblies and angular distribution measurements from resonance fission neutrons for high-fidelity criticality simulations. |
| **$^{252}$Cf:** Validation data are needed for several reactions for the $^{252}$Cf SF Standard Neutron Benchmark Field. | Cross section data which are currently lacking for this standard include: $^{238}$U(n,$\gamma$), $^{58}$Fe(n,$\gamma$), $^{45}$Sc(n,$\gamma$), $^{64}$Zn(n,p), $^{31}$P(n,p), $^{10}$B(n,x)$\alpha$, $^{54}$Fe(n,$\alpha$), $^{23}$Na(n,2n), $^{186}$W(n,$\gamma$), $^{115}$In(n,n'), $^{54}$Fe(n,2n), $^{75}$As(n,2n). Cross section data with large discrepancies: $^{232}$Th(n,f), $^{238}$U(n,2n). |
| **$^{252}$Cf:** There are also some known issues for the following reaction cross sections pertaining to the $^{252}$Cf SF standard: $^{197}$Au(n,$\gamma$), due to room-return neutrons; $^{90}$Zr(n,2n), due to Th contamination; $^{96}$Zr(n,2n) due to contributions from $^{94}$Zr(n,$\gamma$). | Greater accuracy is needed for cross-section measurements for the following reactions to help resolve ambiguities: $^{197}$Au(n,$\gamma$), $^{90}$Zr(n,2n), and $^{96}$Zr(n,2n). |
| **$^{235}$U:** Validation data are needed for several reactions for the $^{235}$U thermal fission Reference Neutron Benchmark Field. | Cross section data which are currently lacking for this standard include: $^{45}$Sc(n,$\gamma$), $^{93}$Nb(n,$\gamma$), $^{58}$Fe(n,$\gamma$), $^{109}$Ag(n,$\gamma$), $^{115}$In(n,2n), $^{65}$Cu(n,2n), $^{52}$Cr(n,2n), $^{23}$Na(n,2n), $^{46}$Ti(n,2n), $^{54}$Fe(n,2n), $^{59}$Co(n,3n), $^{186}$W(n,$\gamma$). Cross section data with large discrepancies: $^{103}$Rh(n,n'), $^{63}$Cu(n,$\gamma$), $^{58}$Ni(n,2n), $^{238}$U(n,$\gamma$), $^{169}$Tm(n,2n), $^{55}$Mn(n,2n). |

| Nuclides and Topic | Nuclear Data Need |
|---|---|
| **$^{235}$U:** It is important to compare calculated (C) and experimentally measured (E) $k_{eff}$ to advance evaluated library accuracy. Deviations from C/E=1 could point to (n,2n) cross section and/or PFNS deficiencies. Cross-section data for several high-threshold reactions are needed to establish consistency between ENDF/B-VI and measured values for the $^{235}$U ($n_{th}$,f) Reference Neutron Benchmark Field. | Calculated and experimentally-measured cross section ratios (C/E) that deviate significantly from unity, or have very high associated uncertainties, and require improved $k_{eff}$ data: $^{63}$Cu(n,2n), $^{58}$Ni(n,2n), $^{90}$Zr(n,2n), $^{127}$I(n,2n), $^{93}$Nb(n,2n), $^{48}$Ti(n,p), $^{56}$Fe(n,p), $^{32}$S(n,p), $^{64}$Zn(n,p), $^{27}$Al(n,p), $^{24}$Mg(n,p), $^{27}$Al(n,$\gamma$), $^{19}$F(n,2n), $^{59}$Co(n,2n), $^{55}$Mn(n,2n), $^{63}$Cu(n,$\alpha$), $^{51}$V(n,p), $^{46}$Ti(n,p), $^{47}$Ti(n,p). |
| **$^{109}$Ag, $^{232}$Th, $^{235,238}$U:** Validation data in the 30-keV Maxwellian Averaged Cross Section (MACS) neutron field are lacking. | Radiative-capture (n,$\gamma$) cross-section measurements needed at 30-keV incident neutron energy. |
| **$^{103m}$Rh, $^{140}$La, $^{186,187}$W:** For improved IRDFF evaluations, improved electromagnetic emission-probability data are needed. | X-ray emission probability measurements around 20 keV in $^{103m}$Rh. LEPS measurements required. |
| as above | Gamma-ray intensities for all lines below 1596 keV in $^{140}$La. Radiative-capture $^{139}$La(n,$\gamma$) studies will greatly facilitate this need. |
| as above | Gamma-ray intensities for transitions at around 473.5 and 685.8 keV in $^{187}$W. Radiative $^{186}$W(n,$\gamma$) studies will greatly facilitate this need. |
| **$^{23}$Na:** New data are needed to resolve ambiguities and support evaluations for material damage studies. | New (n,2n) cross-section measurements are required. |

## Matrix A.8: Single Area Nuclear Data Needs

| Nuclides and Topic | Nuclear Data Need | Area |
|---|---|---|
| **$^{16}$O**: CIELO high-priority nucleus. Improved evaluated nuclear data needed to create accurate ENDF-formatted files for general purpose transport applications, e.g., criticality, shielding, and activation. | Improved capture-gamma data from radiative neutron-capture required for nonproliferation applications (e.g. NDA screening): thermal – 20 MeV. High-energy primary gamma rays are particularly important | National Security |
| **$^{16}$O**: as above | Better radiative-capture data at 30 keV will also benefit nuclear astrophysics applications. | Astro-physics |
| **$^{16}$O**: as above | Thermal elastic scattering cross sections in libraries are discrepant with experimental data; possibly due to Doppler broadening (often neglected) but may have a 3% effect at room temperature. High-accuracy measurement and R-Matrix analyses needed. | National Security |
| **$^{56}$Fe**: CIELO high-priority nucleus. Improved evaluated nuclear data needed to create accurate ENDF-formatted files for general purpose transport applications, e.g., criticality, shielding, and activation. | Alpha-particle pre-equilibrium cluster emission via $(n,\alpha)$ need to be measured. Further high-energy $(n,\alpha)$ cross-section measurements may be useful for verification and validation of the data libraries because gas production in structural materials causes serious embrittlement problems in reactors. | Nuclear Energy |
| **Li, N, B, C, $^{16}$O, $^{17}$O, $^{18}$O, $^{19}$F, $^{23}$Na**: Inelastic cross-section data for low-Z elements are needed for explosives detection. | Inelastic $(n,n'\gamma)$ cross section measurements at energies >1.5 MeV, 2.5 MeV, and 14.1 MeV are required. | National Security |
| **$^{13}$C**: The National Ignition Facility (NIF) & LLNL requires diagnostic information to infer the carbon ablator mix. | Inelastic $(n,n'\gamma)$ and radiative-capture $(n,\gamma)$ cross sections are needed. | Fusion |
| **$^{124,136}$Xe**: NIF diagnostics measurements at LLNL:. Xe isotope tracers measure of ablator penetration into the fuel and identify its location in the ablator. | Accurate $^{124}$Xe(n,2n)$^{123}$Xe and $^{136}$Xe(n,2n)$^{135}$Xe cross sections are needed from fluence of 14-MeV neutrons. | Fusion |
| **$^{169}$Tm**: High-threshold NIF activation diagnostic | Greater precision $^{169}$Tm(n,3n)$^{167}$Tm reaction cross section is required. | Fusion |
| **$^{241}$Am**: Used in well logging for hydrocarbon exploration. There is a concern regarding certain $^{241}$Am production mechanisms due to the production of the parent nuclide $^{241}$Pu. | Possible large-scale production investigated using the reaction $^{238}$U$(\alpha,n)^{241}$Pu->$^{241}$Am requires accurate knowledge of $^{238}$U$(\alpha,n)$ cross section. | Industry |

# Appendix B: Nuclear Reaction Data Needs

Many of the workshop presentations on nuclear data needs listed specific nuclear reaction cross sections (including partial $\gamma$-ray production cross sections) that are of importance for applications. Although many are discussed in more detail elsewhere in this whitepaper, it may be useful for future reference to collect these reactions in a single appendix. They are listed below, by application area.

## Isotope Production Needs

1. Charged-particle reactions for the production of medical isotopes at low energies (E < 30 MeV):
   - $^{45}$Sc(p,n)$^{45}$Ti; $^{52}$Cr(p,n)$^{52}$Mn; $^{54}$Fe(d,n)$^{55}$Co; $^{67}$Zn(p,α)$^{64}$Cu; $^{72}$Ge(p,n)$^{72}$As, $^{74}$Se(d,n)$^{75}$Br; $^{86}$Sr(p,n)$^{86}$Y; $^{120}$Te(p,n)$^{120}$I

2. Charged-particle reactions for the production of medical isotopes at intermediate energies (30-100 MeV) organized by reaction:
   - $^{45}$Sc(p,2n)$^{44}$Ti, $^{69}$Ga(p,2n)$^{68}$Ge, $^{125}$Te(p,2n)$^{124}$I
   - $^{59}$Co(p,3n)$^{57}$Ni, $^{75}$As(p,3n)$^{73}$Se, $^{85}$Rb(p,3n)$^{83}$Sr, $^{122}$Te(p,3n)$^{120}$I, $^{88}$Sr(p,3n)$^{86}$Y, $^{121}$Sb(p,3n)$^{119}$Te/$^{119}$Sb, $^{133}$Cs(p,3n)$^{131}$Ba
   - $^{55}$Mn(p,4n)$^{52}$Fe, $^{71}$Ga(p,4n)$^{68}$Ge, $^{75}$As(p,4n)$^{72}$Se
   - $^{133}$Cs(p,5n)$^{128}$Ba
   - $^{127}$I(p,6n)$^{122}$Xe
   - $^{nat}$Br(p,x)$^{72}$Se, $^{nat}$In(p,x)$^{110}$Sn, $^{122}$Te(p,x)$^{118}$Sb, $^{232}$Th(p,x)$^{225,227}$Ac,$^{225}$Ra
   - $^{nat}$Sb(p,xn)$^{119}$Te/$^{119}$Sb, $^{nat}$La(p,xn)$^{134}$Ce/$^{134}$La
   - $^{68}$Zn(p,$\alpha$n)$^{64}$Cu
   - $^{68}$Zn(p,2p)$^{67}$Cu, $^{124}$Xe(p,2p)$^{123}$I
   - $^{124}$Xe(p,pn)$^{123}$Xe
   - (p,x) reaction on $^{94\text{-}98}$Mo for impurities in $^{99m}$Tc production
   - $^{107}$Ag(p,$\alpha$n)$^{103}$Pd
   - $^{116}$Cd(α,3n)$^{117m}$Sn; $^{192}$Os(α,3n)$^{193m}$Pt

3. Nuclear data needed for radionuclides produced using spallation, deuteron break-up and/or fission neutrons:
   - $^{36}$S(n,x)$^{32}$Si
   - $^{nat}$Cl(n,x)$^{32}$Si, $^{37}$Cl(n,x)$^{32}$Si
   - $^{nat}$Zn(n,x)$^{67}$Cu, $^{68}$Zn(n,x)$^{67}$Cu, $^{70}$Zn(n,x)$^{67}$Cu
   - $^{226}$Ra(n,2n)$^{225}$Ra
   - $^{232}$Th(n,x)$^{225}$Ac, $^{232}$Th(n,x)$^{227}$Ac
   - $^{32}$S(n,p)$^{32}$P; $^{47}$Ti(n,p)$^{47}$Ca, $^{64}$Zn(n,p)$^{64}$Cu; $^{67}$Zn(n,p)$^{67}$Cu; $^{89}$Y(n,p)$^{89}$Sr, $^{105}$Pd(n,p)$^{105}$Rh; $^{149}$Sm(n,p)$^{149}$Pm, $^{153}$Eu(n,p)$^{153}$Sm, $^{159}$Tb(n,p)$^{159}$Gd; $^{161}$Dy(n,p)$^{161}$Tb; $^{166}$Er(n,p)$^{166}$Ho; $^{169}$Tm(n,p)$^{169}$Er; $^{175}$Lu(n,p)$^{175}$Yb; $^{177}$Hf(n,p)$^{177}$Lu
4. High-energy photon-induced reactions
   - $^{68}$Zn(γ,p)$^{67}$Cu; $^{100}$Mo(γ,n)$^{99}$Mo; $^{104}$Pd(γ,n)$^{103}$Pd; $^{124}$Xe(γ,n)$^{123}$Xe; $^{232}$Th(γ,f)$^{99}$Mo; $^{238}$U(γ,f)$^{99}$Mo

5. Nuclear data needed for alternative $^{99m}$Tc production
   - $^{100}$Mo(d,3n)
   - $^{232}$Th(p,f)
   - $^{100}$Mo(d,p2n)
   - $^{100}$Mo(n,2n)
   - $^{100}$Mo(p,pn) - data on long-lived impurities

- o $^{100}Mo(p,2n)$ - data on long-lived impurities
- o (p,x) reaction on $^{94\text{-}98}Mo$ for impurities in $^{99m}Tc$ production

6. Nuclear data needed for optimizing $^{252}Cf$ production
   - o $^{245}Cm(n,\gamma)$, $^{247}Cm(n,\gamma)$, $^{248}Cm(n,\gamma)$
   - o $^{249}Bk(n,\gamma)$
   - o $^{250}Cf(n,\gamma)$ and $^{250}Cf$ (n,f), $^{251}Cf(n,\gamma)$ and $^{251}Cf$ (n,f)
   - o $^{252}Cf(n,x)$ - resonance near 1 eV in particular!

7. Nuclear data needed for Super Heavy Element (SHE) target isotopes production
   - o $^{248}Cm(n,\gamma)$ low energy resonances
   - o $^{249}Bk(n,\gamma)$
   - o $^{250}Cf$ (n,γ) and $^{250}Cf$ (n,f)
   - o $^{251}Cf(n,\gamma)$ and $^{251}Cf$ (n,f); first resonance varies greatly by library

8. Nuclear data needed in the production of $^{238}Pu$
   - o n+$^{238}Np$ fission products and related uncertainties (priority)
   - o $^{238}Np(n,f)$
   - o $^{237}Np(n,\gamma)$

9. Nuclear data needed in the production of medical isotopes at MURR;
   Gamma yield spectrum for incident neutron energies in the thermal to 10 MeV range for target heat generation rates:
   - o *Te(n,γ) (production of $^{131}I$)
   - o *Mo(n,γ)
   - o *Ru(n,γ)
   - o $^{103}Rh(n,\gamma)$
   - o $^{154}Eu(n,\gamma)$
   - o $^{155}Eu(n,\gamma)$
   - o $^{141}Ce(n,\gamma)$
   - o $^{140}Ce(n,\gamma)$
   - o *Sn(n,γ)

   Note: * indicates stable isotopes

   Excitation function for incident neutron energies from thermal to 10 MeV
   - o $^{177}Lu(n, \gamma)$, $^{154}Eu(n, \gamma)$, $^{155}Eu(n, \gamma)$, $^{77}As(n, \gamma)$, $^{72}As(n, \gamma)$, $^{141}Ce(n, \gamma)$, $^{192}Ir(n, \gamma)$, $^{198}Au(n, \gamma)$, $^{199}Au(n, \gamma)$
   - o

# Dosimetry Needs

1. $^{117}Sn(n,n')$, covering energy response 0.3 – 3.0 MeV

2. Data to support new evaluations
   - o $^{23}Na(n,\gamma)$, discrepant in fast neutron region, > 100 keV
   - o $^{23}Na(n,2n)$
   - o $^{27}Al(n,2n)$

3. Address discrepancies:
   - o $^{55}Mn(n,\gamma)$ cross section from 10 keV to 1 MeV
   - o $^{58}Fe(n,\gamma)$ reaction in the 10 keV to 1 MeV energy region for fast reactor
   - o $^{237}Np(n,f)$ and $^{241}Am(n,f)$ measurements between LANL and n-TOF (CERN) on the plateau
   - o Some 14-MeV dosimetry reactions ($^{48}Ti(n,x)^{47}Sc$, $^{64}Zn(n,p)^{64}Cu$, $^{113}In(n,n')$, $^{63}Cu(n,2n)^{62}Cu$)
   - o Thermal capture for $^{93}Nb$, $^{115}In$
   - o $^{209}Bi(n,4n)$; all $^{209}Bi(n,xn)$ for x=4,5,6,7

4. Need small uncertainties on all dosimetry reactions

5. Validation data in $^{252}$Cf spontaneous fission standard benchmark neutron field
   - Data lacking on $^{238}$U(n,γ), $^{58}$Fe(n,γ), $^{31}$P(n,p), $^{10}$B(n,X)α, $^{54}$Fe(n,α), $^{23}$Na(n,2n), $^{186}$W(n,γ), $^{115}$In(n,n'), $^{54}$Fe(n,2n), $^{75}$As(n,2n), $^{45}$Sc(n,γ), and $^{64}$Zn(n,p)
   - 14 other reactions from IRDFF library
   - Data with large discrepancy $^{232}$Th(n,f) and $^{238}$U(n,2n)
   - Data with outliers (4 reactions)

6. IRMM Exploratory Study of Validation Data in $^{252}$Cf Standard Neutron Benchmark Field
   - Issues with existing $^{197}$Au(n,γ) due to room return
   - Issues with existing $^{90}$Zr(n,2n) due to Th contamination
   - Issue with existing $^{96}$Zr(n,2n) due to $^{94}$Zr(n,γ) contribution

7. Validation data in $^{235}$U thermal fission reference benchmark neutron field
   - Data lacking $^{45}$Sc(n,γ), $^{115}$In(n,2n), $^{65}$Cu(n,2n), $^{52}$Cr(n,2n), $^{23}$Na(n,2n), $^{46}$Ti(n,2n), $^{54}$Fe(n,2n), $^{59}$Co(n,3n), $^{186}$W(n,γ), $^{93}$Nb(n,γ), $^{58}$Fe(n,γ), $^{109}$Ag(n,γ), and 6 other reactions from IRDFF library
   - Data with large discrepancies; $^{103}$Rh(n,n'), $^{63}$Cu(n,γ), $^{58}$Ni(n,2n), $^{238}$U(n,γ), $^{169}$Tm(n,2n), and $^{55}$Mn(n,2n)
   - Data with outliers (5 reactions)

8. Validation data in 30 keV MACS neutron field, data is lacking; $^{109}$Ag(n,γ), $^{232}$Th(n,γ), $^{235}$U(n,γ), and $^{238}$U(n,γ)

9. Test and improve decay characteristics for radionuclides in new IRDFF reactions:
   - $^{55}$Co
   - $^{56}$Co
   - $^{94}$Nb
   - $^{114m}$In
   - $^{117m}$Sn
   - $^{195}$Au

10. Gamma Emission Probabilities
    - $^{103m}$Rh -- X-ray emission probability around 20 keV
    - $^{140}$La -- gamma intensities for lines below 1596 keV
    - $^{187}$W -- gamma intensities of 2 lines (473.53 keV and 685.81 keV)

11. Important Isotopes
    - $^{69}$Ga, $^{71}$Ga, $^{75}$As (ASTM E722)
    - $^{56}$Fe, $^{54}$Fe (ASTM E693)

12. Uncertainty in recoil spectrum
    - Recoil spectrum characterization in cross sections (MF=6)
    - $^{69}$Ga, $^{71}$Ga, $^{75}$As
    - Fe isotopes
    - Validate/test use of calculated cross section libraries, *e.g.* TENDL, to characterize this uncertainty component and Scope "model defect"

13. Other dosimetry reactions identified by text-mining the EXFOR database
    - Reactions that produce $^{7}$Be: $^{7}$Li(p, n), $^{12}$C(p, X)$^{7}$Be and $^{9}$Be(p, X)$^{7}$Be
    - Reaction that produces $^{11}$C: $^{12}$C(p, X)$^{11}$C
    - Reaction that produce $^{24}$Na: $^{27}$Al($^{12}$C, X)$^{24}$Na
    - Reaction that produces $^{56}$Co: $^{nat}$Fe(p, X)$^{56}$Co

- Reaction that produces $^{61}Cu$: $^{nat}Cu(p, X)^{61}Cu$
- Reaction that produces $^{51}Cr$: $^{51}V(p, n)$
- Reactions that use $^{nat}Mo$ as a target material: $^{nat}Mo(p, X)^{96}Tc$ and $^{nat}Mo(\alpha, X)^{97}Ru$

# Inertial Confinement Fusion Data Needs

1. Accurate, temperature-dependent fusion reactivity for light ions is of primary importance to describe thermonuclear burn.
   - d(t,α)n, t(t, α)2n, d(d,t)p, d(d,$^{3}$He)n, d($^{3}$He, α)p
   - d-t, t-t, d-d, d-t-$^{3}$He and d-$^{3}$He gas fills are all used.

2. Energy loss of fusion-generated alpha particles in hot dense plasmas must be accurately assessed (engine of ignition). Radiochemical neutron activation and neutron time-of-flight diagnostics validate stopping power models.
3. Diagnostics for degraded implosion performance.
   - Xe dopants to probe ablation front instabilities.
   - Br(d,2n)Kr to probe ablator/cold fuel and ablator/hot core mix.
   - Alpha particle induced reactions to probe hot core mix: $^{6}$Li, $^{9}$Be, $^{10}$B (best one), $^{12}$C, $^{14}$N, $^{16}$O, $^{19}$F, $^{20}$Ne, $^{23}$Na, $^{24}$Mg, $^{27}$Al.

4. Gamma-ray diagnostics for performance and ablator/fuel instabilities.
   - Total yield from d-t fusion γ branching ratio at 17.6 MeV.
   - $^{12}$C(n,n'γ) 4.4 MeV time-integrated emission provides hydrocarbon areal densities (remaining mass). Cross section at 14 MeV must be accurate.
   - Does $^{13}$C(n,n'γ) have strong emission near 4 MeV? If not, then a useful mix diagnostic is possible.

5. Solid Radiochemistry Diagnostic (SRC) is currently an NIF diagnostic complementary to $^{12}$C-γ GRH detection (CH ρr).
   - Ratio of $^{198}$Au/$^{196}$Au from the activated hohlraum.
   - (n,γ)/(n,2n):low energy neutrons/14 MeV neutrons.

6. RIF that make n's in addition to d-d, d-t and t-t: $^{2}$H(n,2n), $^{12}$C(n,n2α), $^{3}$H(n,2n), $^{13}$C(n,2n), $^{3}$H(p,n), $^{16}$O(n,nα), $^{28}$Si(n,np), $^{29}$Si(n,2n), $^{2}$H(p,pn), $^{30}$Si(n,2n), $^{28}$Si(n,nα), $^{3}$He(t,np)

# Appendix C: Historical Perspective

The purpose of this appendix is to provide a short review of the evolution of research in Nuclear Data, specifically for applications to Nuclear Energy. We will also sketch the present situation and its vulnerabilities, consider the impact and implications for any future developments of nuclear energy, and suggest perspectives and a few initiatives to insure continuity in this strategic research domain.

## Background: The Early Years

In 1940s/50s, nuclear data needed for calculations of engineering parameters of emerging nuclear power plants were assessed, with large efforts, both theoretically and experimentally while in the 1960s/70s numerous computer codes for design and safety assessments were developed that needed large amounts of computer readable nuclear data.

Initially each country developed its own data and formats overseen by local technical committees. The ENDF project became the leading project and international collaboration resulted in extension of data and adoption of a single (evolving) format, the one associated with ENDF.

The success of these activities, focused on the needs of industry and utilities, has been a strong factor behind the reliable operation already of the first reactor generation all over the world.

The increasing needs related to the fuel cycle assessment and to the more and more stringent safety requirements, together with a growing demand from industry to optimize operation and to reduce margins, and new needs triggered by the active investigation in the early 1970s of new reactor concepts and in particular of fast reactors, did have several important consequences:

| Year | ENDF/B Release | Years between releases |
|---|---|---|
| 1968 | ENDF/B-I | |
| 1970 | ENDF/B-II | 2 |
| 1972 | ENDF/B-III | 2 |
| 1974 | ENDF/B-IV | 2 |
| 1978 | ENDF/B-V | 4 |
| 1990 | ENDF/B-VI | 12 |
| 2006 | ENDF/B-VII | 16 |

**Table 1.** ENDF release history.

1. The establishment of comprehensive nuclear databases, both experimental (EXFOR, or the “exchange format” data base, designed to allow transmission of nuclear data measurements between the Nuclear Reaction Data Centers) and evaluated, as the ENDF (Evaluated Nuclear Data Files) project in the US. As for the ENDF project, a rapid sequence of improved versions was released in the 1960s and early 1970s: The table 1 lists the ENDF release history. Initially each country developed its own data and formats overseen by local technical committees. International collaboration resulted in extension of data and adoption, as indicated above, of a single (evolving) format.

2. The deployment of outstanding experimental facilities for the nuclear data measurement (*e.g.* the ORELA facility in the USA and the GELINA facility in Europe, among many others in particular at LANL in the US, but also in the former Soviet Union)

3. ***A better and long-lasting connection was established between reactor designers and both reactor and nuclear physicists, in order to optimize the production of complete, reliable, user-oriented and validated nuclear data.*** The use of reactor physics critical facilities was expanded, sometimes with as primary objective the performance of integral experiments designed specifically to improve nuclear data, in particular in support of fast reactors (ZPR, ZPPR, MASURCA, BFS, FCA, SNEAK, ZEBRA *etc*.), but also in support of commercial reactors (*e.g.* EOLE, VENUS, PROTEUS *etc*. in Europe)

4. A new branch of reactor physics was developed, to formalize in a rigorous way the relation between reactor physics oriented integral data and nuclear data for applications (sensitivity theory based on generalized perturbation theory, sensitivity/uncertainty analysis, target accuracies assessment *etc*.)

## Decline and Renaissance

In the 1980s there was a significant reduction in nuclear data related activities in the US (somewhat related to *e.g.* the FR program cancellation) and, for different reasons, also in Europe (*e.g.* the shut-down of SUPERPHENIX) while in parallel there was some consolidation or emergence of new nuclear data projects: JEFF (Europe, Korea…), JENDL (Japan), CENDL (China), BROND/ROSFOND (Russia + former Soviet Union states). At the same time there was a progressive shutdown of critical facilities, reducing drastically the number of new data-oriented integral experiments. These reductions did also imply that over more than a decade, the training and hiring of nuclear data and experimental reactor physicists was also drastically reduced practically everywhere in the OECD countries.

During the 1990s, new issues became of high importance, mostly related to waste management that emerged as a key issue for any future development of nuclear energy (see *e.g.* the OMEGA program in Japan). Advanced fuel cycle challenges, as well as a renewal of interest for subcritical systems (*e.g.* ADS), the investigation of new fuels and materials and of new fuel cycle concepts and strategies, resulted in new research programs. ***It was quickly realized that most of the new challenges required a very significant enlargement of the traditional nuclear databases. Due to the reduction in effort, funding and manpower over the previous decade or more, collaborative projects were considered by most countries as the only feasible approach to addressing this need, with OECD-NEA and IAEA playing a key role.***

For example, the NEA's nuclear data evaluation co-operation activities involve the following evaluation projects: ENDF (United States), JENDL (Japan), ROSFOND/BROND (Russia), JEFF (other Data Bank member countries) and CENDL (China) in close co-operation with the Nuclear Data Section of the International Atomic Energy Agency (IAEA). The NEA Working Party on International Nuclear Data Evaluation Cooperation (WPEC) was established in 1989 to promote the exchange of information on nuclear data evaluations, measurements, nuclear model calculations, validation, and related topics, and to provide a framework for co-operative activities between the participating projects. The working party assesses nuclear data improvement needs via the Nuclear Data High Priority Request List (HPRL), which is an internationally agreed compilation of the most important nuclear data requirements and addresses these needs by initiating joint evaluation and/or measurement efforts.

**However, it should be kept in mind that these are volunteer projects, and partners need to obtain their own funding for work. In this sense, neither long-term commitment of manpower nor continuity of research directions is *a priori* guaranteed, unless there is strong support and a clear long-term vision for each of the participating national groups.**

## New Paradigms: Uncertainty Assessment and Science-Based Validation

Starting in the 2000s, partly at the request of industry, and due to a new awareness of nuclear data end-users, nuclear data uncertainty impact studies were performed, using the tools of sensitivity-uncertainty analysis mentioned above. These studies concluded that current uncertainties in nuclear data should be significantly reduced, to receive the full benefit of the advanced modeling and simulation initiatives that had been launched worldwide. At the same time, efforts towards the development of advanced simulations had been initiated with significant funding, in particular in the US.

**However, it was quickly realized that only a parallel effort in advanced simulation and in nuclear data improvement could provide designers with the more general, well-validated calculational tools needed to meet the strict new high-accuracy targets for design and safety. It was also realized that no simulation tool, whatever the degree of sophistication (*e.g.* new Monte Carlo methods and approaches), could replace well-designed, science-oriented validation experiments.**

The interest expressed by industry, regulators and by the scientific community for the sensitivity-uncertainty impact analysis (see *e.g.* the US DOE Office of Science "NUCLEAR PHYSICS AND RELATED COMPUTATIONAL SCIENCE R&D FOR ADVANCED FUEL CYCLES WORKSHOP" Washington DC, August 2006), did encourage and accelerate the development of data covariance assessments, using new science-based tools. This new effort has spread worldwide, with the US initiatives, led by BNL, in the forefront, producing spectacular results.

# Appendix D: Capabilities

The need to generate nuclear data for applications can arise from either a lack of key information, or from a situation where discrepant experiments limit confidence in evaluation. In some cases, only modest precision is required for improvement, while in others increasingly precise data provides greater benefit for the application. In some situations, modest improvements in the quality of available nuclear data can be gained using straightforward and simple experimental approaches; while in others improvements can only be obtained by significant rethinking of experimental techniques. One concept that became clear in the workshop was that no one facility was capable of addressing the entire spectrum of nuclear applications.

Fortunately, the capabilities and facilities available in the United States for applied nuclear science are robust diverse. In some cases, such as the Gaerttner LINAC Center at RPI, the detector and beam characteristics are focused on the production of data relevant for nuclear energy. Others, such as the Weapons Neutron Research (WNR) facility at LANL and the National Ignition Facility (NIF) at LLNL, emphasize national security needs such as stockpile stewardship and counter-proliferation. In contrast, facilities like ANL and NSCL have broad reaching capabilities that can potentially contribute to either curiosity- or application-driven projects.

That being said, while the primary focus of curiosity-driven low-energy nuclear science involves studying nuclei far from the valley of stability, the needs of the applications communities presented in this workshop tended to focus more on neutron-induced reactions on stable nuclei, with the notable exceptions being charged particle reactions for medical isotope production. Since neutron beam are amongst the first radioactive beams, most of the neutron facilities discussed in the workshop utilized "secondary beams" formed from either charged-particle induced nuclear reaction products (LANL, RPI, TUNL, Ohio, Kentucky RPI, LBNL *etc*.) or from fission at reactors, such as MURR and HFIR at ORNL. The US is fortunate to host such a wide range of neutron beam facilities.

One of the challenges facing a researcher interested in performing neutron reaction studies is to choose which facility provides the optimal blend of neutron beam characteristics (pulse structure, flux, energy range) and detector capabilities to obtain the required data. One of the speakers at the workshop (Darren Bleuel, LLNL) attempted to help in this decision making process by producing a comparison of neutron capabilities at different pulsed beam facilities. Figure 6 below shows the flux and energy spectrum of a number of neutron sources available to the applications community. These include the thick-target deuteron breakup neutron source at LBNL, the Weapons Nuclear Research (WNR) facility at LANL (green curve), and the Gelina neutron source in Brussels. A "typical" monoenergetic CW neutron source, the UC Berkeley quasi-monoenergetic High-Flux Neutron Generator (HFNG) is presented for comparison purposes. It should be noted that Dr. Bleuel's comparison was by no means comprehensive, in that it excluded a number of other important neutron sources, such as the ($\gamma$,n) neutron source at RPI. Fortunately, these facilities are well described in their own sections of this appendix.

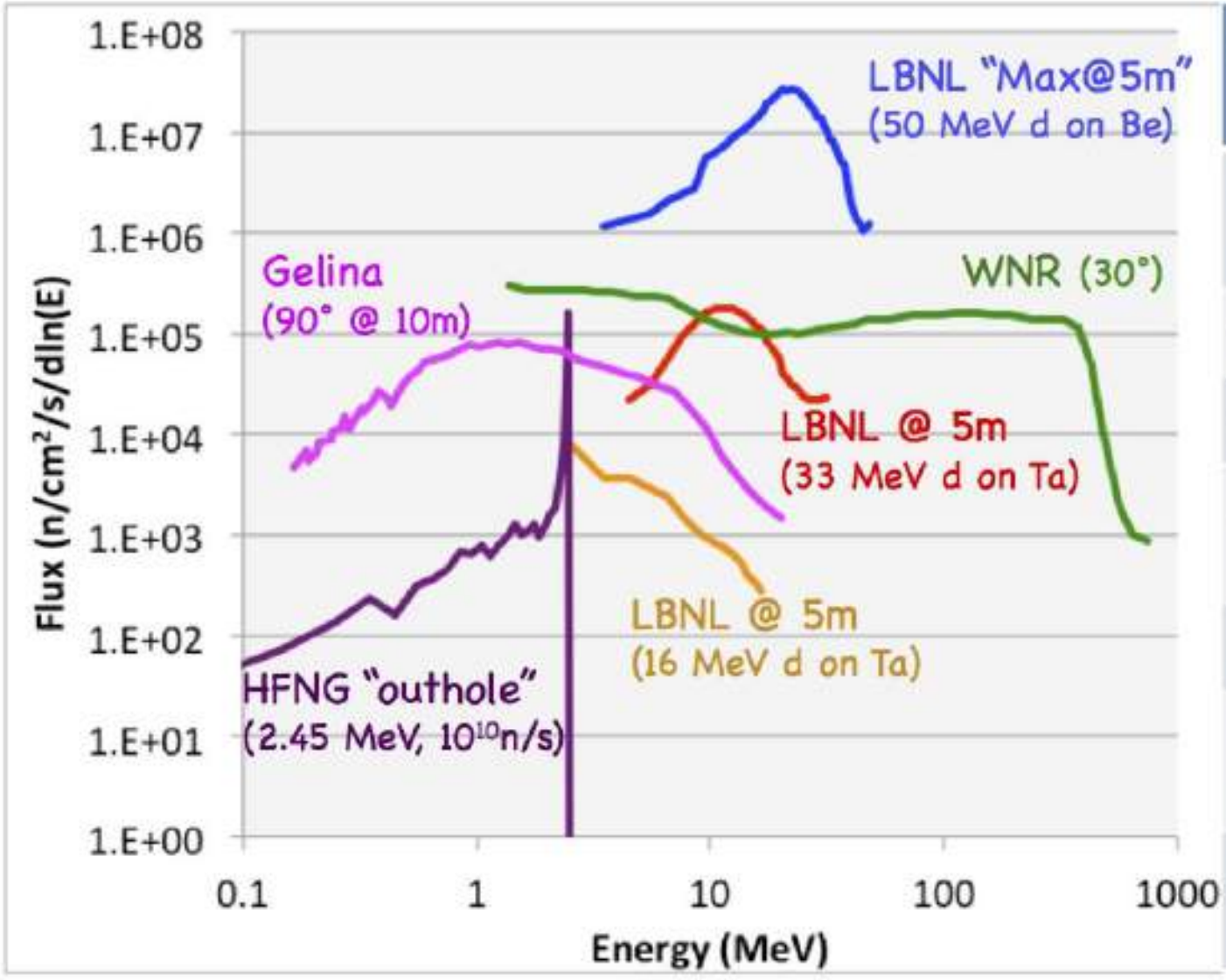

| Source | Tot. Flux (n/cm2/s) |
|---|---|
| 50 MeV d on Be (10μA, 5m) | $2x10^7$ |
| WNR (30°) | $10^6$ |
| Gelina (90° @ 10m) | $2x10^5$ |
| 33 MeV d on Ta (2μA @ 5m) | $2x10^5$ |
| HFNG (2.27m) | $10^4$ |
| 16 MeV d on Ta (2μA @ 5m) | $5x10^3$ |
| Kentucky T(p,n) (@ 2m) | $2x10^3$ |

**Figure 6.** Comparison of the neutron flux available at several neutron facilities (from the talk by Bleuel).

Many of the neutron facilities described in this work utilize light charged particles (p, d, t, $^3$He, or $\alpha$). This is a "happy coincidence" in that the much of the nuclear data needs relevant to medical isotope production center on light-ion production cross section measurements. This potentially allows a number of facilities described in this whitepaper to serve the needs of all three major applications topics (Nuclear Energy, National Security, and Isotope Production). Examples of facilities in this category include the 88-Inch cyclotron at LBNL, the tandem accelerator at TUNL and the Edwards Accelerator Lab at Ohio University.

A "third class" of facility discussed here is the High Intensity Gamma Source (HIGS), which produces monoenergetic photon beams through the use of a free electron laser: This provides a unique capability for measuring ($\gamma$,$\gamma$') and ($\gamma$,n) cross sections. These cross sections are needed for a number of national security applications, and were specifically called out as requiring additional measurement in the talks by *Quiter* and *Cerjan*.

Along with issues such as beam and detection capabilities and sensitivities, the issues of beam-time allocation and detector/spectrometer availability are non-negligible. While some facilities operate as user facilities with rather straightforward opportunities for collaboration in connection with beam availability, others operate utilizing highly competitive Program Advisory Committees that review the scientific merit of any proposed experimental work, and others may use a cost-center model, in which beam-time charges of tens to hundreds of thousands of dollars per week are typical.

The goal of this Appendix is to provide a review of the capabilities at many of the facilities available for applied nuclear science research in the US that can be used by experimentalists who are planning to carry out applications-relevant nuclear data measurements. The editors of this whitepaper attempted to keep this list as broad as possible, including a number of facilities that were not presented in great detail at the workshop due to time constraints. Although the list is undoubtedly incomplete, every effort was made to have it be representative of the broad spectrum of facilities at hand. Lastly, it should be noted that most of the text in the individual facility descriptions was provided by the points-of-contact (POC) at each institution, and that the editors performed only minor revision of the content. Users of this Appendix are encouraged to contact the listed facility POC for additional information.

# Appendix D.1: Argonne National Laboratory, Atlas/CARIBU Facility

| |
|---|
| **General Description:** US DOE low-energy nuclear physics national user facility. Provides stable and radioactive beams at low and Coulomb barrier energy. |
| **Accelerator:** ATLAS heavy-ion superconducting linac |
| **Beams:**<br>• All stable beams from proton to uranium at high intensity and energies up to 20 MeV/u for the lightest beams and 10 MeV/u for the heaviest<br>• Over 500 mass separated beams of neutron-rich isotopes produced by $^{252}$Cf fission, available at low energy or reaccelerated to 2-15 MeV/u<br>• In-flight produced light radioactive beams one or two neutrons away from stability at energies of 5-20 MeV/u<br>Beam time is allocated by PAC. |
| **Research Focus (relevant to applications):** measurement of properties (mass, beta-delayed neutrons/gammas) of fission fragments, accelerator mass spectrometry of heavy elements, single particle structure, surrogate reactions |
| **Present detector array capabilities (relevant to applications):** Canadian Penning trap mass spectrometer, beta-delayed neutron trap, X-array and tape station, Gammasphere, HELIOS, MANTRA AMS system |
| **Contact person:** Guy Savard |

*Prepared by Guy Savard*

The ATLAS superconducting linac provides stable beams with intensities up to 10 puA at energies up to 10-20 MeV/u, well suited for studies of nuclear structure relevant to fundamental nuclear physics, astrophysics and nuclear physics applications. The CARIBU facility is a source of neutron-rich isotopes for ATLAS, making available mass separated beams of fission fragments from $^{252}$Cf fission. The unique gas catcher technology used at CARIBU allows even the most refractory short-lived fission fragments to be extracted, mass separated, and made available as clean beams for experiments. These beams are available to all experimental stations and equipment at ATLAS. Unique instrumentation such as Gammasphere, the CPT mass spectrometer, the beta-delayed neutron trap, and HELIOS, can be used to provide key information on the ground state, excited states, and decay properties of neutron-rich nuclei. The facility has available target stations to host during experimental campaign other instruments built by the community. In addition, ATLAS itself can be used as an AMS system and is particularly well suited to study the heaviest elements (*e.g.* the MANTRA project). A layout of the ATLAS facility is shown below.

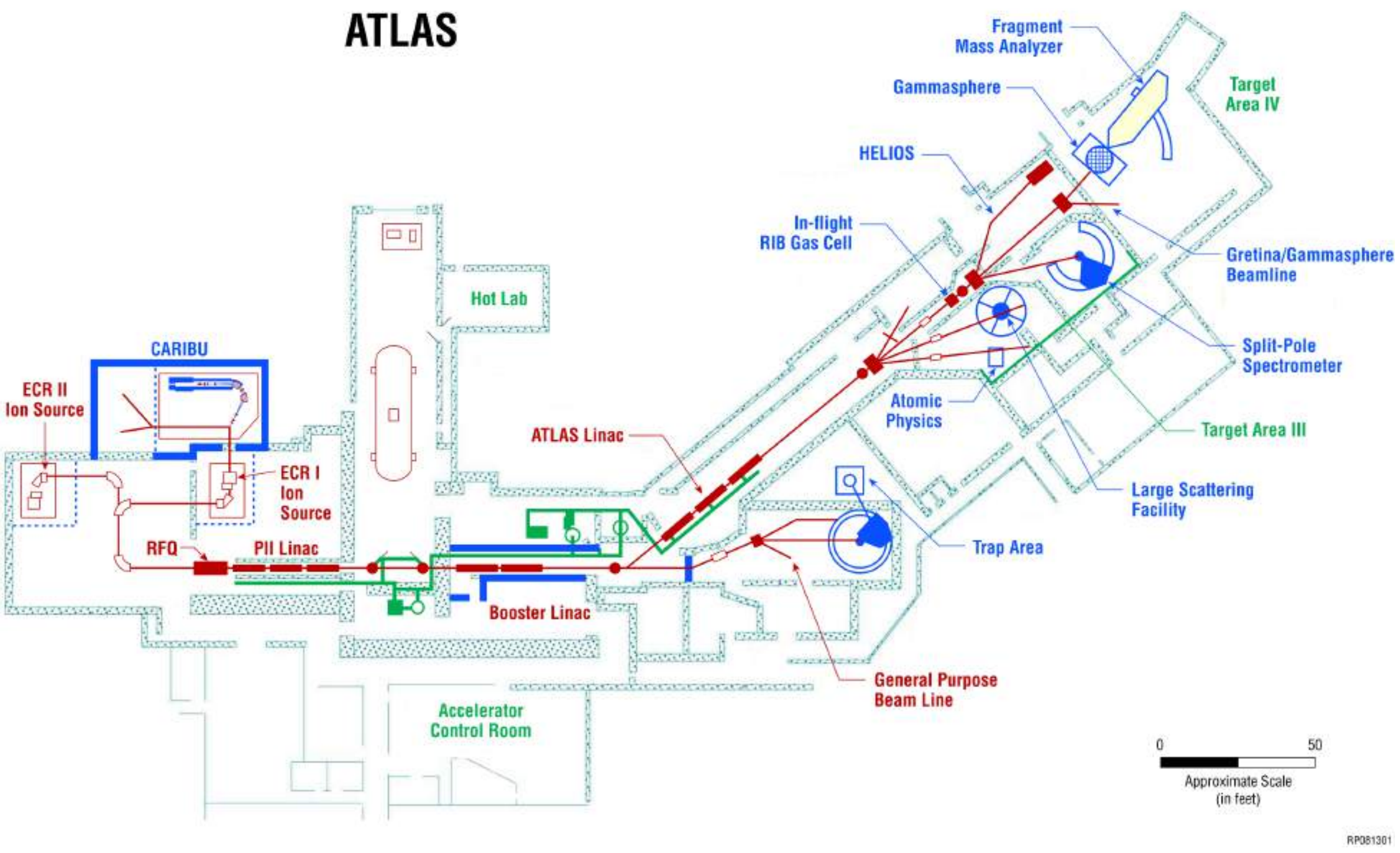

**Figure 7.** The ATLAS floor plan.

The facility operates typically 5000-6000 hrs/yr of stable and CARIBU reaccelerated beams, in addition to another 2000 hrs/yr of low-energy CARIBU beams. Beam time is allocated based on PAC recommendation with the PAC meeting typically twice a year. More information is available at www.phy.anl.gov/atlas/.

# Appendix D.2: Brookhaven National Laboratory, Brookhaven Linac Isotope Producer (BLIP)

**General Description:** Radionuclide Production for DOE Isotope Program is part of the Collider-Accelerator Department at Brookhaven National Laboratory; not a user facility but maintaining limited funding and staff for collaborative research

**Beams:** 40-200 MeV, 0.1 – 140 μA proton beams; Raster beam under development and due to be completed in FY 2016.

**Additional Capabilities:** Hot cell facilities for remote manipulation of intense sources, radiochemical characterization and separations, expertise in gamma-ray spectroscopy and thermal analysis of targets and machining of target material and cans.

**Research Focus:** Isotope production and R&D for radiochemical separations.

**Contact person:** Cathy Cutler: email: ccutler@bnl.gov Phone: +1 (631) 344-3873

*Prepared by Suzanne V. Smith*

This program uses the Brookhaven Linac Isotope Producer (BLIP), and the associated radiochemistry laboratory and hot cell complex in Building 801 to develop, prepare, and distribute to the nuclear medicine community and industry some radioisotopes that are difficult to produce or are not available elsewhere. The BLIP, built in 1972, was the world's first facility to utilize high-energy protons for radioisotope production by diverting the excess beam of the 200 MeV proton LINAC that injects protons into the Booster synchrotron for injection into the AGS then RHIC for the high energy nuclear physics program. After several upgrades BLIP continues to serve as an international resource for the production of selected isotopes that are generally unavailable elsewhere. The Linac is capable of accelerating H-ions to produce 66, 90, 118, 140, 162, 184 or 202 MeV protons at 37-48 mA current for 425 μs duration with a 6.67 Hz repetition rate. In 2015 FY, with the initial phase of the Linac Intensity Upgrade project complete, the Linac has reached currents of 142 μA. A hot-cell in Bld 931, situated over target area, is used to transfer the two target assembly boxes to and from the irradiation area. The target boxes can house up to four targets in each, however degraders can also be used to tune the beam to the desired energy on the target. AIP funded project to raster the proton beam will be completed in 2016. This upgrade will allow more heat sensitive targets to be irradiated at higher currents. BLIP operates usually concurrently with the RHIC polarized proton program and BLIP receives about 90% of the available beam pulses.

The irradiated targets are transported to Building 801, which contains chemical processing capabilities, which include Target Processing Facilities with 7 hot-cells with manipulators, one cold chemistry, 3 radiochemistry and an instrumentation laboratory. The latter laboratory has three gamma spectrometers and an ICP-OES and ICP-MS, set-up for the characterization of radioactive samples. Additional available research capabilities include four radiochemistry laboratories and 2-4 Hot-Cells. Other available instrumentation include a gamma counter, 5 fumehoods, HPLC, balances, centrifuges, glove boxes, machining capabilities and thermal analysis to target materials.

# Appendix D.3: Brookhaven National Laboratory, Tandem Van De Graaff

| |
|---|
| **General Description:** A flexible and user-friendly facility for providing high quality ion beams for a variety of uses. |
| **Beams:** Two large 15 MeV electrostatic accelerators which deliver ion beams covering most of the periodic table |
| **Please visit our website for additional information:** https://www.bnl.gov/tandem/ |

*Prepared by Chuck Carlson*

## Ions Beams for Science and Technology

A wide range of ion species and energies are delivered to the users on a full cost-recovery basis, mainly for industrial and space related applications. Rapid energy and ion changes, well-controlled intensities, high quality beams and extraordinary reliability make this a very versatile and user-friendly facility. At the same time these accelerators have been used for many years as the heavy ion pre-injectors for two larger BNL user facilities (RHIC and NSRL).

## Testing of Electronics for Space Applications

Cosmic rays striking microelectronics on a spacecraft can cause errors in the operation of critical devices. Spacecraft reliability therefore requires the testing of all such devices used in onboard electronics to establish their sensitivity to cosmic radiation. Such testing is done in the laboratory by placing components in an accelerator particle beam. At the Brookhaven Tandem, the well-characterized beams and the large variety of ions allow these tests to be performed with high precision and great detail, leading to improved designs and further testing. The Single Event Upset Test Facility developed in collaboration with NASA, NRL, NSL and USASDC has made, and continues making significant contributions to the Space Program.

## Fabricating Filter Materials

Plastic films used in the fabrication of nano- and micro-pore filters for ultra-pure water filtration and for specialized medical and biological applications are bombarded with heavy ions in a chamber owned by GE HealthCare. These materials are used in a large variety of medical tests, biology investigations, microchip tissue growth, fabrication, and find important applications.

## Radiobiology Research Facility

Complementing the NASA Radiations Effects Facility (NSRL) at BNL, we have recently developed a lower ion energy radiobiology research facility at the Tandem. Low energies may be of particular interest since the high energy ions lose energy when traversing spacecraft materials and produce the maximum damage just before coming to rest in the astronauts' bodies. Thus, energies lower than most present in the primary cosmic ray spectrum are appropriate to cover the range of maximum LET (the Bragg peak) but, due to their short ranges, they are only useful to perform studies with thin samples such as cell cultures. Figure 8 illustrates the very large range of LET values and respective penetration depths in water (or tissue) for iron beams from 10 KeV per nucleon to 1 GeV per nucleon.

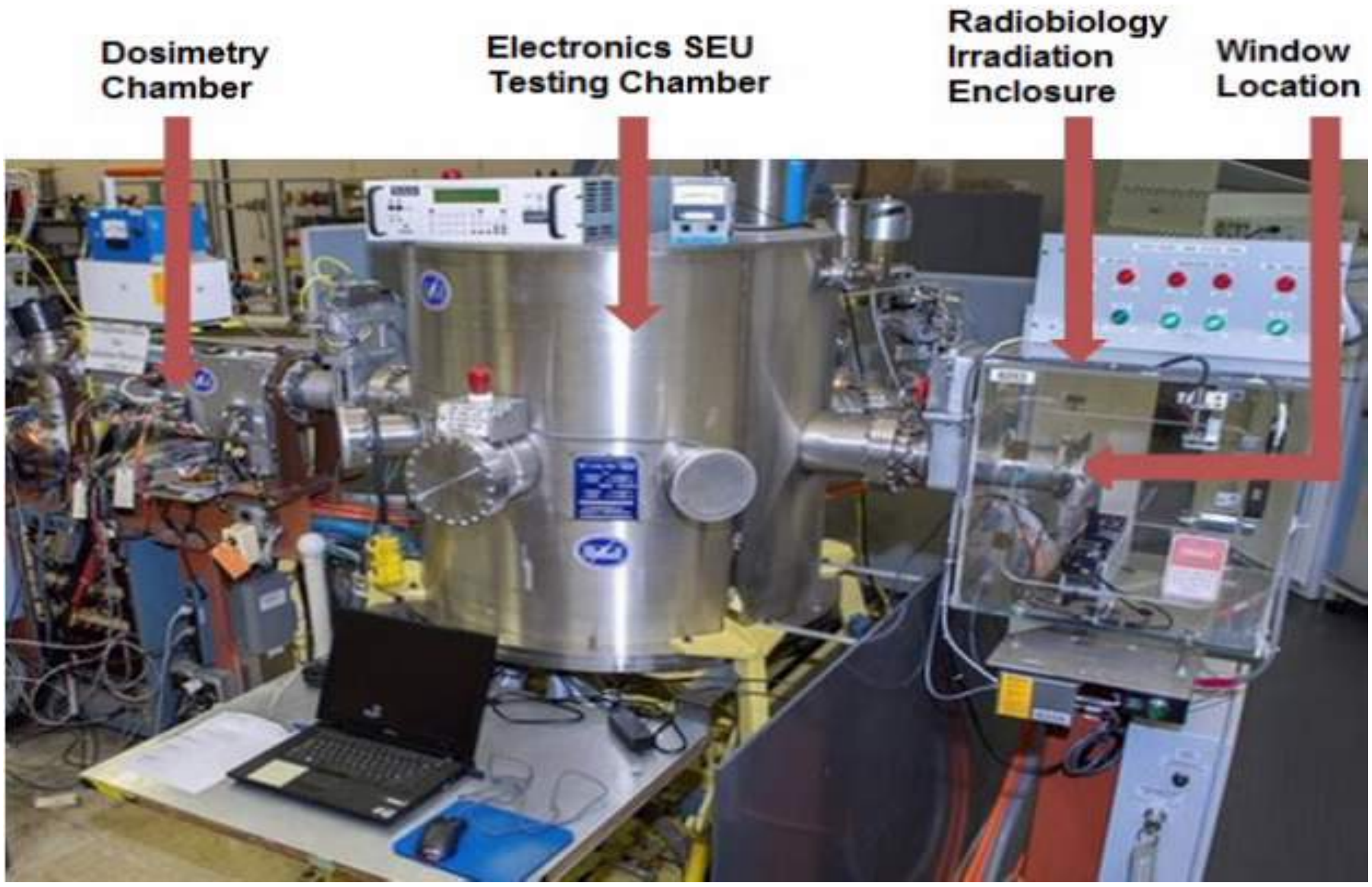

**Figure 8.** The radiobiology irradiation enclosure is shown at the right of the picture, the Single Event Upset Test Facility (SEUTF) chamber in the middle and the dosimetry chamber used for both at the left.

# Appendix D.4: Florida State University, John D. Fox Accelerator Laboratory

| |
|---|
| **General Description:** University Accelerator Laboratory; Research program driven by local faculty in collaboration with various university and laboratory groups |
| **Accelerators:** 9 MV Tandem, 8 MV Superconducting Linac |
| **Beams:** Stable beams of Masses 1-50, up to 4-8 MeV/u energy; Radioactive beams produced in-flight at RESOLUT facility, masses 6-30 |
| **Additional Capabilities:** Compton-suppressed γ-detector array; ANASEN active target detector system; RESONEUT neutron detector setup<br>Soon: High-resolution high-acceptance magnetic spectrograph |
| **Research Focus:** Nuclear Structure studies using high-resolution γ-spectroscopy; Nuclear Astrophysics studies with radioactive and stable beams; Development of advanced detector systems for exotic beam experiments |
| **Contact person:** I. Wiedenhöver, (850)-644-1429 iwiedenhover@physics.fsu.edu |

*Prepared by I. Wiedenhöver*

The John D. Fox laboratory operates a two-stage accelerator comprised of a 9 MV FN tandem accelerator and an 8 MV superconducting linear accelerator (Linac). The FN tandem is injected by either a NEC SNICS-II cesium sputter ion source, for most beams created from solid chemicals, or an NEC RF-discharge source for beams generated from gaseous materials, most importantly $^3$He and $^4$He. Among the beams available from the sputter source is the radioactive isotope $^{14}$C.

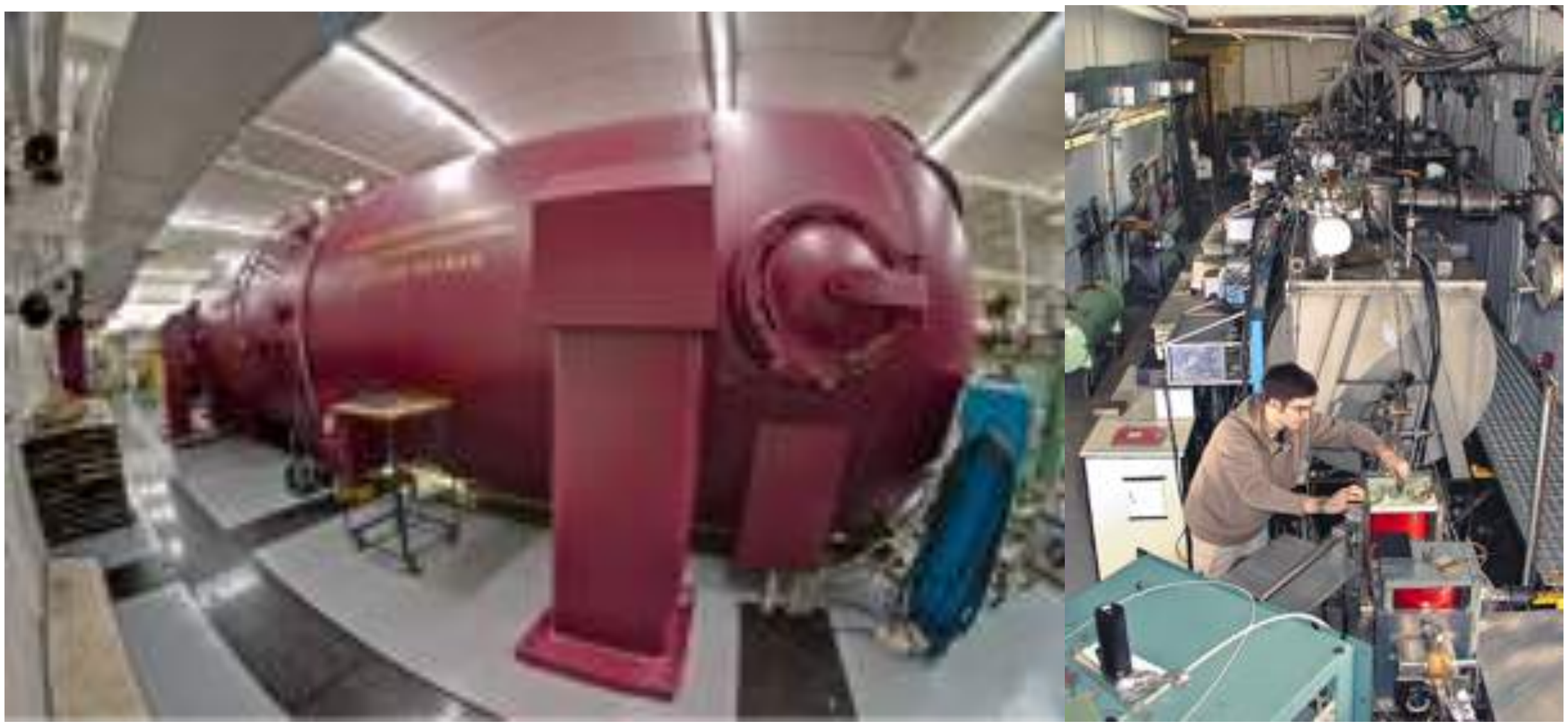

**Figure 9.** The FSU Tandem accelerator

The beams from the Tandem are injected into the Linac, which more than doubles their energy. The superconducting linear accelerator consists of twelve accelerating resonators installed in three cryostats, plus buncher and re-buncher. The resonators are niobium-on-copper "split-ring" resonators produced by Argonne National Laboratory. The cryostats were designed and built at FSU.

The laboratory has developed an upgrade plan to increase the energy and mass-range of beams available for experiments. The upgrade entails the increase of cryogenic capacity by the addition of a second liquid Helium refrigerator (completed 2013), and the addition of two cryostats to the Linac.

A recent focus of the laboratory operations is on experiments with radioactive beams created in RESOLUT, an in-flight radioactive beam facility, which uses beams from the Tandem-Linac to create beams of exotic, radioactive isotopes. The isotopes, which are created through a nuclear reaction in the production target, are separated in mass by the combined effect of the electrical fields in a superconducting RF-resonator and the magnetic fields of the spectrograph.

The laboratory has developed advanced detector systems for research with radioactive beams. One example is the ANASEN device, which was developed in collaboration with a group from Louisiana State University. ANASEN is an active-target detector for the efficient study of resonances in exotic nuclei, either for nuclear structure or nuclear astrophysics. ANASEN will be used both at FSU and the re-accelerated beam facility of the NSCL.

The FSU laboratory is in the process of installing a high-resolution magnetic Split-Pole spectrograph, which had previously been used at the Yale Nuclear Structure Laboratory. The device is projected to be commissioned in the summer of 2016. The research with this device will focus on the spectroscopy of resonances for nuclear astrophysics.

Our group is one of the seven founding members of ARUNA, the Association for Research with University Nuclear Accelerators. ARUNA's goal is to support and enhance the research and education programs enabled by University laboratories.

For up to date information on the laboratory and its science program, visit http://fsunuc.physics.fsu.edu

# Appendix D.5: Idaho National Laboratory

**General Description:**

- **ATR:** Fuels and materials test reactor
- **NRAD:** TRIGA® Mark II tank-type research reactor.
- **MANTRA program:** integral reactor physics experimental program to infer the neutron capture cross sections of actinides and fission products in fast and epithermal spectra.

**Contact persons:** Giuseppe Palmiotti
208 526-9615, Giuseppe.Palmiotti@inl.gov
Gilles Youinou, 208 526-1049,
Gilles.Youinou@inl.gov

## The Advanced Test Reactor (ATR)

The ATR is located at the ATR Complex on the INL site and has been operating continuously since 1967. The primary mission of this versatile facility was initially to serve the U.S. Navy in the development and refinement of nuclear propulsion systems. However, in recent years, the ATR has been used for a wider variety of government- and privately-sponsored research.

The designation of the ATR as a National Scientific User Facility (NSUF) provides nuclear energy researchers access to world-class facilities to support the advancement of nuclear science and technology. The ATR NSUF accomplishes this mission by offering state-of-the-art experimental irradiation testing and PIE facilities and technical assistance in design and safety analysis of reactor experiments. ATR general characteristics and some approximate irradiation performance data are summarized in Tables I and II, respectively.

| | |
|---|---|
| **Reactor:** | |
| Thermal power | 250 $MW_{th}$ [a] |
| Power density | 1.0 MW/L |
| Maximum thermal neutron flux | $1.0 \times 10^{15}$ $n/cm^2$-sec [b] |
| Maximum fast flux | $5.0 \times 10^{14}$ $n/cm^2$-sec [b] |
| Number of flux traps | 9 |
| Number of experiment positions | 68 [c] |
| **Core:** | |
| Number of fuel assemblies | 40 |
| Active length of assemblies | 4 feet |
| Number of fuel plates per assembly | 19 |
| Uranium-235 content of an assembl | 1,075 g |
| Total core load | 43 kg [d] |
| **Coolant:** | |
| Design pressure | 2.7 Mpa (390 psig) |
| Design temperature | 115°C (240°F) |
| **Reactor Coolant:** | |
| Light water maximum coolant flow ra | 3.09 $m^3/s$ (49,000 gpm) |
| Coolant temperature (operating) | <52°C (125°F) inlet, 71°C (160°F) outlet |

a. Maximum design power. ATR is seldom operated above 110 MWth
b. Parameters are based on the full 250 MWth power level and will be proportionally reduced for lower reactor power levels.
c. Only 66 of these are available for irradiations.
d. Total U-235 always less due to burn-up.

**Table 2**. ATR general characteristics.

The ATR has large test volumes in high-flux areas. Designed to permit simulation of long neutron radiation exposures in a short period of time, the maximum thermal power rating is 250 MWth with a maximum unperturbed thermal neutron flux of 1.0 x 1015 $n/cm^2$–s. Since most recent experimental objectives generally do not require the limits of its operational capability, the ATR typically operates at much lower power levels. Occasionally, some lobes of the reactor are operated at higher powers that generate higher neutron flux.

The ATR is cooled by pressurized (2.5 MPa [360 psig]) water that enters the reactor vessel bottom at an average temperature of 52°C (125°F), flows up outside cylindrical tanks that support and contain the core, passes through concentric thermal shields into the open upper part of the vessel, then flows down through the core to a flow distribution tank below the core. When the reactor is operating at full power, the primary coolant exits the vessel at a temperature of 71°C (160°F).

The unique design of ATR (Figure 11) control devices permits large power variations among its nine flux traps using a combination of control cylinders (drums) and neck shim rods. The beryllium control cylinders contain hafnium plates that can be rotated toward and away from the core, and hafnium shim rods, which withdraw vertically, can be individually inserted or withdrawn for minor power adjustments. Within bounds, the power level in each corner lobe of the reactor can be controlled independently to allow for different power and flux levels in the four corner lobes during the same operating cycle.

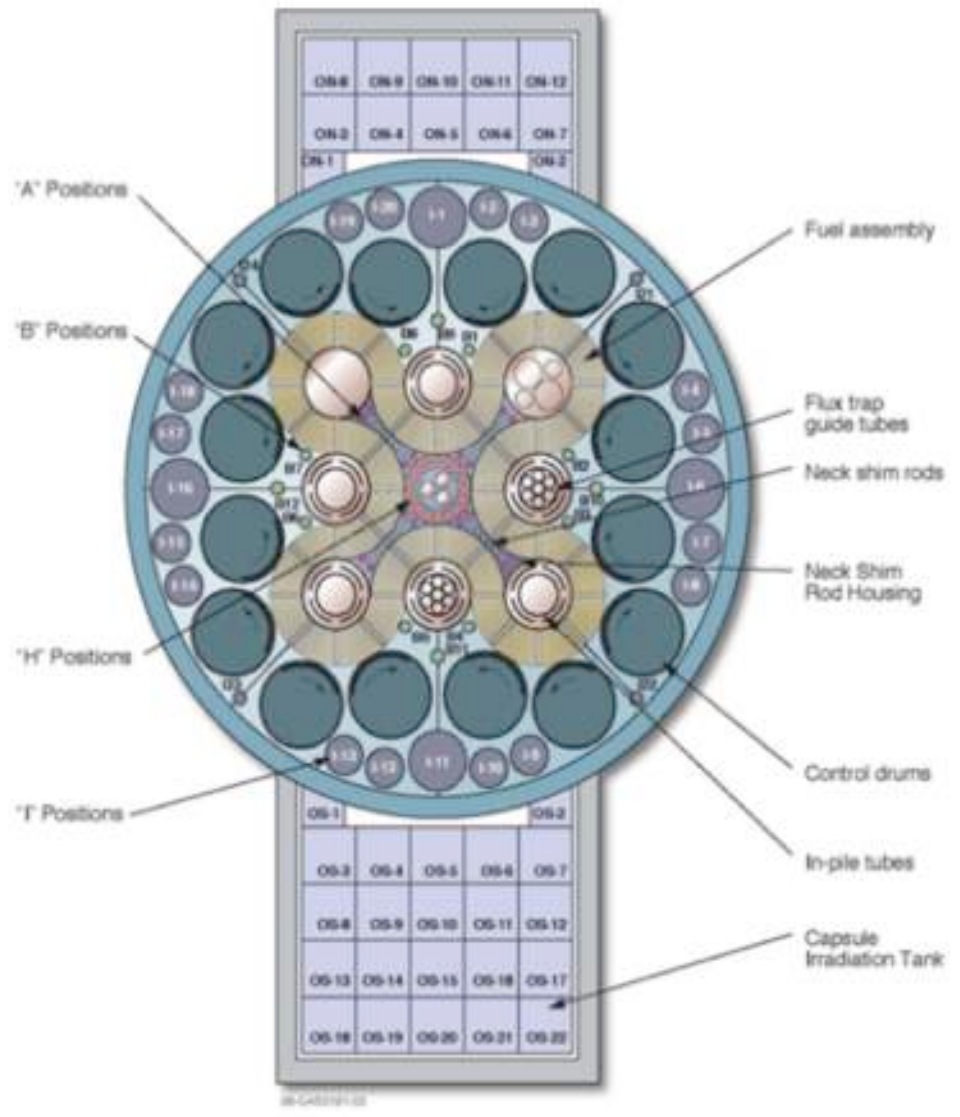

**Figure 11**. ATR Core cross section.

Neutron flux in the ATR varies from position to position and along the vertical length of the test position. It also varies with the power level in the lobe(s) closest to the irradiation position. Thermal and fast flux intensity values listed in Table 3 are at the core mid-plane for a reactor power of 110 MWth and assume a uniform reactor power of 22 MWth in each lobe.

| Positions | Diameter (in.)[a] | Thermal Flux ($n/cm^2$-s)[b] | Fast Flux (E>1 MeV) ($n/cm^2$-s) | Typical Gamma Heating W/g (SS)[c] |
|---|---|---|---|---|
| **Northwest and Northeast Flux Traps** | 5.250 | $4.4x10^{14}$ | $2.2x10^{14}$ | |
| **Other Flux Traps** | 3.000[d] | $4.4x10^{14}$ | $9.7x10^{13}$ | |
| **A-Positions** | | | | |
| (A-1 - A-8) | 1.590 | $1.9x10^{14}$ | $1.7x10^{14}$ | 8.8 |
| (A-9 - A-12) | 0.659 | $2.0x10^{14}$ | $2.3x10^{14}$ | 8.8 |
| * **(A-13 - A-16)** | **0.500** | $\mathbf{2.0x10^{14}}$ | $\mathbf{2.3x10^{14}}$ | 8.8 |
| **B-Positions** | | | | |
| * **(B-1 - B-8)**[f] | **0.875** | $\mathbf{2.5x10^{14}}$ | $\mathbf{8.1x10^{13}}$ | 6.4 |
| * **(B-9 - B-12)** | **1.500** | $\mathbf{1.1x10^{14}}$ | $\mathbf{1.6x10^{13}}$ | 5.5 |
| **H-Positions** | | | | |
| (H-1 - H-16) | 0.625 | $1.9x10^{14}$ | $1.7x10^{14}$ | 8.4 |
| **I-Positions** | | | | |
| * **Large (4)** | **5.000** | $\mathbf{1.7x10^{13}}$ | $\mathbf{1.3x10^{12}}$ | 0.66 |
| * **Medium (16)** | **3.500** | $\mathbf{3.4x10^{13}}$ | $\mathbf{1.3x10^{12}}$ | |
| * **Small (4)** | **1.500** | $\mathbf{8.4x10^{13}}$ | $\mathbf{3.2x10^{12}}$ | |
| **Outer Tank Position** | | | | |
| ON-4 | Var[e] | $4.3x10^{12}$ | $1.2x10^{11}$ | 0.15 |
| ON-5 | Var[e] | $3.8x10^{12}$ | $1.1x10^{11}$ | 0.18 |
| ON-9 | Var[e] | $1.7x10^{12}$ | $3.9x10^{10}$ | 0.07 |
| OS-5 | Var[e] | $3.5x10^{12}$ | $1.0x10^{11}$ | 0.14 |
| OS-7 | Var[e] | $3.2x10^{12}$ | $1.1x10^{11}$ | 0.11 |
| OS-10 | Var[e] | $1.3x10^{12}$ | $3.4x10^{10}$ | 0.05 |
| OS-15 | Var[e] | $5.5x10^{11}$ | $1.2x10^{10}$ | 0.20 |
| OS-20 | Var[e] | $2.5x10^{11}$ | $3.5x10^{9}$ | 0.01 |

a. Position diameter. Capsule diameter must be smaller

b. Average speed 2,200 m/s.

c. Depends on configuration

d. Current east, center, and south flux trap configurations contain seven guide tubes with inside diameters of 0.694 in.

e. Variable; can be either 0.875, 1.312, or 3.000 in.

f. B-7 is the location of the Hydraulic Shuttle Irraidation System

* Positions available for experiment irradiation in FY-2009

**Table 3**. Approximate peak flux values for ATR capsule positions at 110 MWth (22 MWth in each lobe).

## Neutron Radiography Reactor (NRAD)

The neutron radiography (NRAD) reactor is a TRIGA® (Training, Research, Isotopes, General Atomics) Mark II tank-type research reactor located in the basement, below the main hot cell, of the Hot Fuel Examination Facility (HFEF) at the Idaho National Laboratory (INL). It is equipped with two beam tubes with separate radiography stations for the performance of neutron radiography irradiation on small test components.

The NRAD reactor is currently under the direction of the Battelle Energy Alliance (BEA) and is operated and maintained by the INL and Hot Cell Services Division. It is primarily used for neutron radiography analysis of both irradiated and un-irradiated fuels and materials. Typical applications for examining the internal features of fuel elements and assemblies include fuel pellet separations, fuel central-void formation, pellet cracking, evidence of fuel melting, and material integrity under normal and extreme conditions.

The NRAD core is designed for steady-state operation with or without in-core and/or in-tank experiments. The combined reactivity worth of all removable experiments within the reactor tank is limited to less than $0.50.

The NRAD reactor is a TRIGA-conversion-type reactor originally located at the Puerto Rico Nuclear Center (PRNC). It was converted to a TRIGA-FLIP-(Fuel Life Improvement Program)-fueled system (70% 235U) in 1971. The 2-MW research reactor was closed in 1976 and then a portion of the TRIGA reactor fuel elements and other components (with a single radiography beam line) were moved in 1977 by the US Department of Energy (DOE) to Argonne National Laboratory (West) in Idaho Falls, Idaho. The NRAD reactor was first brought to critical in October 1977, and then became operational in 1978. A second beam line was added in 1982.

The NRAD reactor (Figure 12) is a 250 kW TRIGA LEU conversion reactor that is a water-moderated, heterogeneous, solid-fuel, tank-type research reactor. The reactor is composed of fuel in three- and four element clusters that can be arranged in a variety of lattice patterns, depending on reactivity requirements. The grid plate consists of 36 holes, on a 6-by-6 rectangular pattern, that mate with the end fittings of the fuel cluster assemblies.

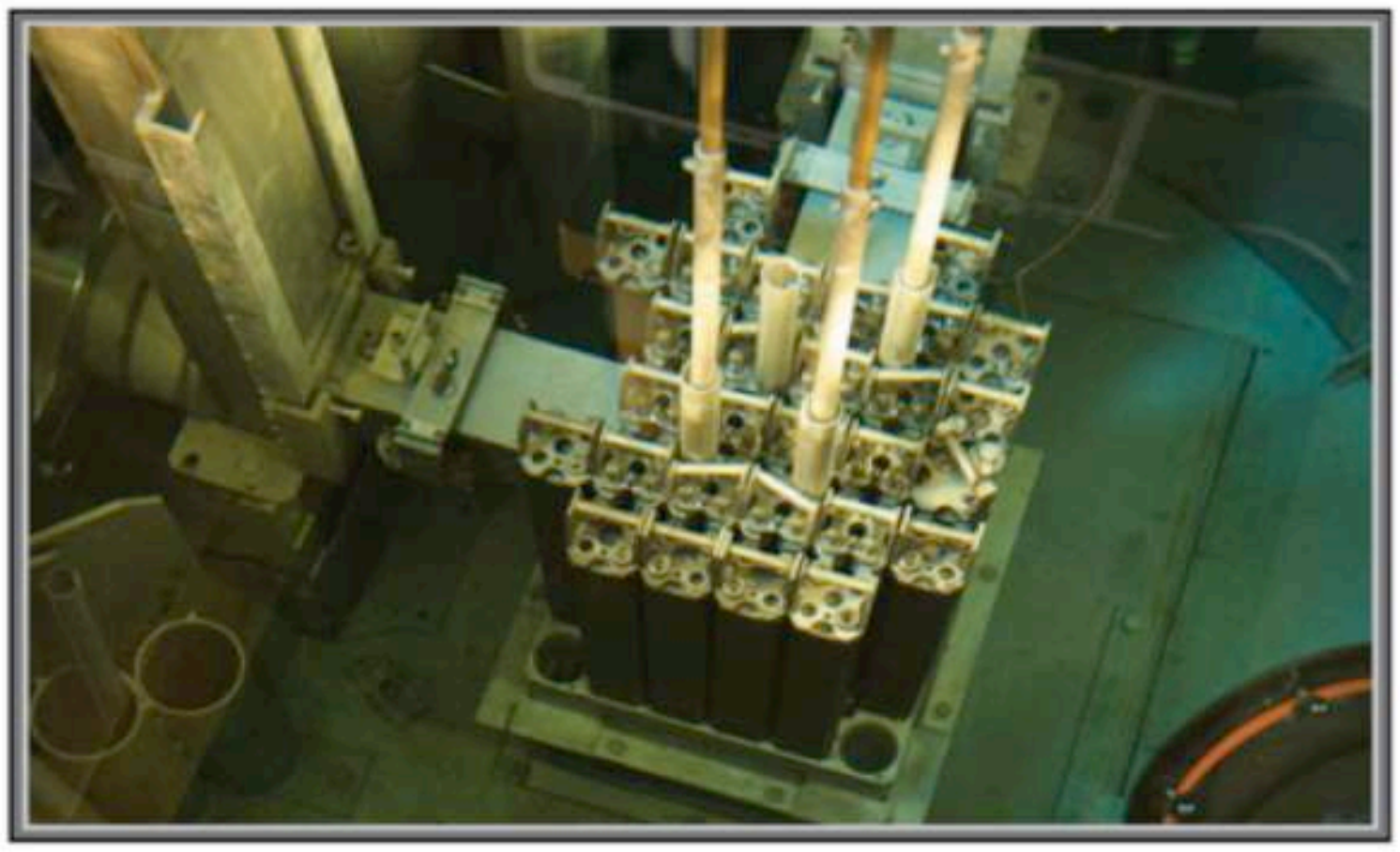

**Figure 12**. In-tank view of the NRAD reactor core.

The NRAD LEU core configuration contains 60 fuel elements, two water-followed shim control rods, and one water-followed regulating rod (Figure 3). A water hole is provided as an experimental irradiation position. The NRAD reactor uses graphite neutron reflector assemblies located along the periphery grid plate locations. The number and position of fuel-element and reflector assemblies can be varied to adjust core reactivity.

## MANTRA Program

The MANTRA (Measurements of Actinide Neutron Transmutation Rates with Accelerator mass spectrometry) experimental program is the first reactor physics integral experiment performed in the USA in more than 20 years. It aims at obtaining integral information about neutron cross sections for actinides that are important for advanced nuclear fuel cycles. Its principle is to irradiate very pure actinide samples in the Advanced Test Reactor (ATR) at INL and, after a given time, determine the amount of the different transmutation products. The determination of the nuclide densities before and after neutron irradiation allows inference of the effective neutron capture cross-sections. The following actinides have been irradiated: $^{232}$Th, $^{233}$U, $^{235}$U, $^{236}$U, $^{238}$U, $^{237}$Np, $^{239}$Pu, $^{240}$Pu, $^{242}$Pu, $^{244}$Pu, $^{241}$Am, $^{243}$Am, $^{244}$Cm and $^{248}$Cm. The irradiated fission products are: $^{149}$Sm, $^{153}$Eu, $^{133}$Cs, $^{103}$Rh, $^{101}$Ru, $^{143}$Nd, $^{145}$Nd and $^{105}$Pd. In order to obtain effective neutron capture cross sections corresponding to different neutron spectra, three sets of actinide samples were irradiated: the first one is filtered with cadmium and the other two are filtered with enriched boron of different thicknesses (5 mm and 10 mm). The neutron capture reactions on $^{10}B$ and $^{113}Cd$ have large cross-sections and strongly impact the neutron spectrum (see Figure 13) allowing the samples to be irradiated in epithermal and fast neutron spectra whereas the unfiltered neutron spectrum is largely thermal. The total flux levels in the samples are, respectively, about $2\times10^{14}$ n/cm$^2$s and $10^{14}$ n/cm$^2$s with the cadmium filter and the boron filters. The cadmium-filtered and the 5 mm boron-filtered irradiations were completed in January 2013 after, respectively, 55 days and 110 days in the reactor. The last irradiation with the 10 mm boron-filtered was completed in January 2014 after 110 days in the reactor.

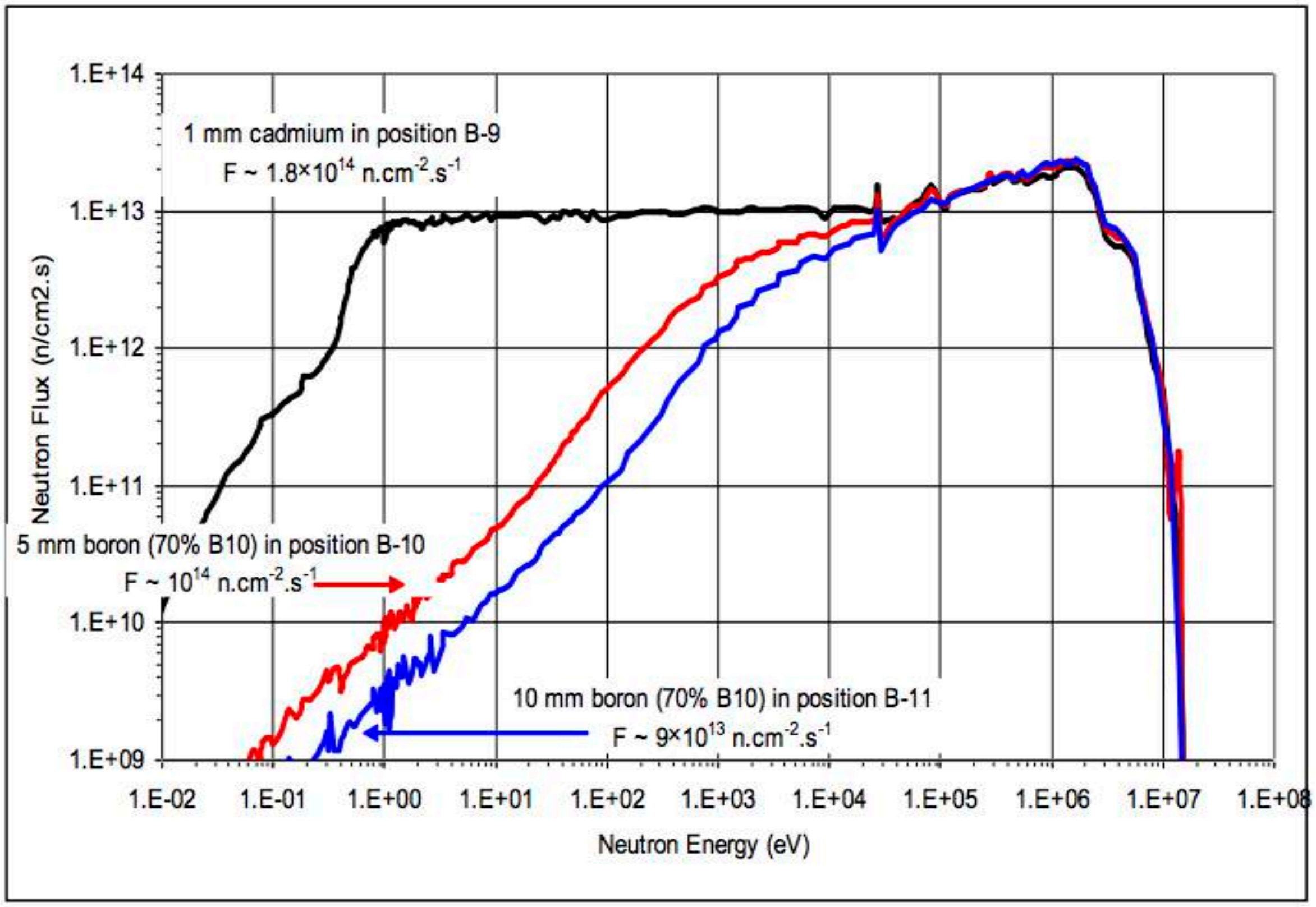

**Figure 13.** Neutron flux in the samples with boron and cadmium filters as calculated with MCNP.

The Post Irradiation Examination (PIE) was carried out both at INL and ANL using, respectively, the newly acquired Multi-Collector ICPMS and the Accelerator Mass Spectrometry at the ATLAS facility. The use of these two independent measurement techniques is benefiting both the reactor physicists interested in the neutron cross sections, by providing them with two sets of independent measurements, and also the experimentalists in charge of both facilities, by providing them with a consistent benchmark of their respective techniques. The results of detailed MCNP calculations are currently being compared with the measured isotopic ratio present in the irradiated samples.

Even though we expect the MANTRA experimental program to be a success, there is already a need for a second phase (MANTRA-2) of such a type of experiments. There are good reasons justifying this statement. First there are several actinide samples that, for different reasons, have not been irradiated, specifically $^{238}Pu$, $^{241}Pu$, and $^{241}Cm$ (irradiated only with thin filters). Moreover, at the time of an anticipated MANTRA-2 campaign efficient mass separators should be available at INL. This would allow purifying samples of isotopes already irradiated in MANTRA and avoiding one of the program's main concerns: contamination from other isotopes during post irradiation analysis.

Finally, due to the limited space available, in most cases only one sample per isotope (and in a couple of cases two) was irradiated in MANTRA. For the sake of comparison: in the French irradiation experiments PROFIL at least three, in PROFIL-2 even six samples of the same isotope were irradiated. This approach is justified by the fact that in certain cases during the post irradiation analysis, and due to bad manipulation, some samples may become contaminated. While for MANTRA, a low failure rate is expected, a MANTRA-2 campaign would provide the opportunity for repeating the compromised irradiation of the respective isotopes.

In complementing the MANTRA campaign, a separate experimental program performed at the NRAD facility would provide a wealth of integral experimental data in support of nuclear data validation and uncertainty quantification efforts. The INL NRAD is a TRIGA reactor that has enough space to allow the introduction of thick neutron filters (including $^{238}U$ blocks) allowing simulating the full gamut of neutron spectra from thermal, epithermal, soft fast, to hard fast. The systematic measurement of fission rate spectral indices using fission micro-chamber would enhance the knowledge on a vast range of actinides (both major and minor). Moreover, in this facility reactivity sample oscillation measurements could be performed with the help of an Idaho State University (ISU) apparatus (open and closed loop) that could be easily installed at NRAD. These measurements of actinides samples in different spectra would be invaluable for the validation and uncertainty quantification of cross sections needed for advanced fuel cycles analyses.

# Appendix D.6: University of Kentucky Accelerator Laboratory

| |
|---|
| **General Description:** University facility with research programs in nuclear structure, neutron-induced reactions, and neutron cross section measurements |
| **Accelerator:** 7-MV Van de Graaff Accelerator |
| **Beams:** pulsed beams with high currents of light ions (protons, deuterons, $^{3}$He, and $^{4}$He ions); secondary neutrons |
| **Experimental focus:** neutron scattering reactions with neutron time-of-flight and gamma-ray detection |
| **Present detector array capabilities:** HPGe gamma-ray detectors and various neutron detectors |
| **Contact person:** Steven W. Yates, yates@uky.edu, 859-257-4005 |

*Prepared by Steven W. Yates and Erin E. Peters*

The University of Kentucky Accelerator Laboratory (UKAL) is one of the premier facilities for studies with fast (MeV) neutrons. The laboratory opened in 1964 and the accelerator underwent a major upgrade in the 1990's. Over the last 5 decades, the facilities have been used for research in nuclear physics, as well as for homeland security and corporate applications.

The UK 7-MV single-stage model CN Van de Graaff accelerator is capable of producing pulsed beams of protons, deuterons, $^{3}$He, and $^{4}$He at energies up to 7 MeV. The beam is pulsed at a frequency of 1.875 MHz and can also be bunched in time such that each pulse has a FWHM of ≈1 ns. Secondary neutron fluence may also be produced by reaction of protons or deuterons with tritium or deuterium gas. Nearly monoenergetic neutrons with energies between ≈ 0.1 – 23 MeV may be produced with fluxes up to $10^{9}$ neutrons/s depending on the reaction employed. The pulsed beam allows for use of time-of-flight methods. Both neutron and gamma-ray detection are available. Figure 1 shows the typical setup for neutron detection. For more detailed information, see Refs. [1] and [2].

The research performed at the UKAL has been funded continuously by the U. S. National Science Foundation for more than 50 years and includes fundamental science studies of nuclear structure and reactions. In recent years, the laboratory has also received funding from the U. S. Department of Energy in support of a more application-based project for neutron cross section measurements.

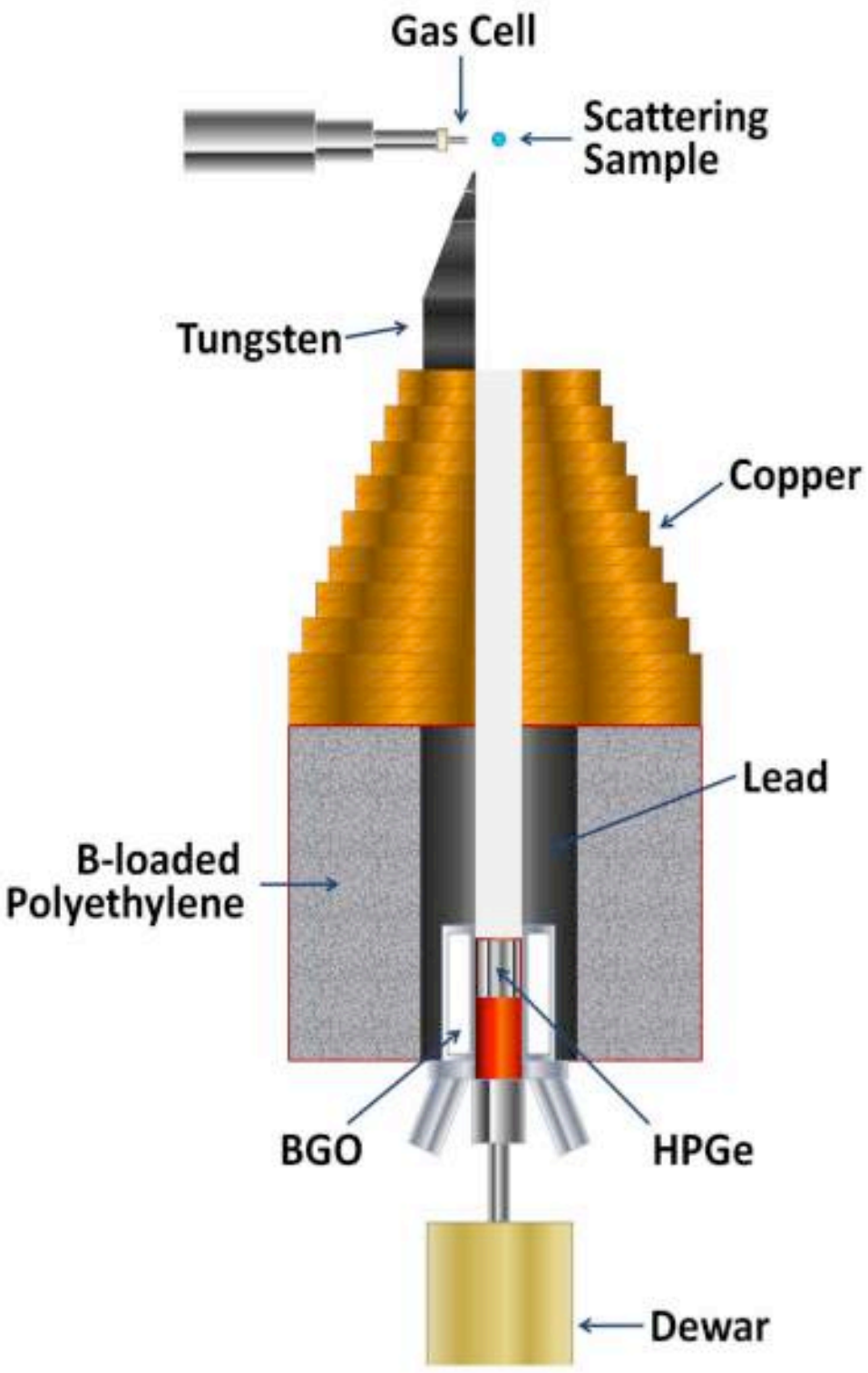

**Figure 14.** Typical experimental setup for neutron time-of-flight measurements.

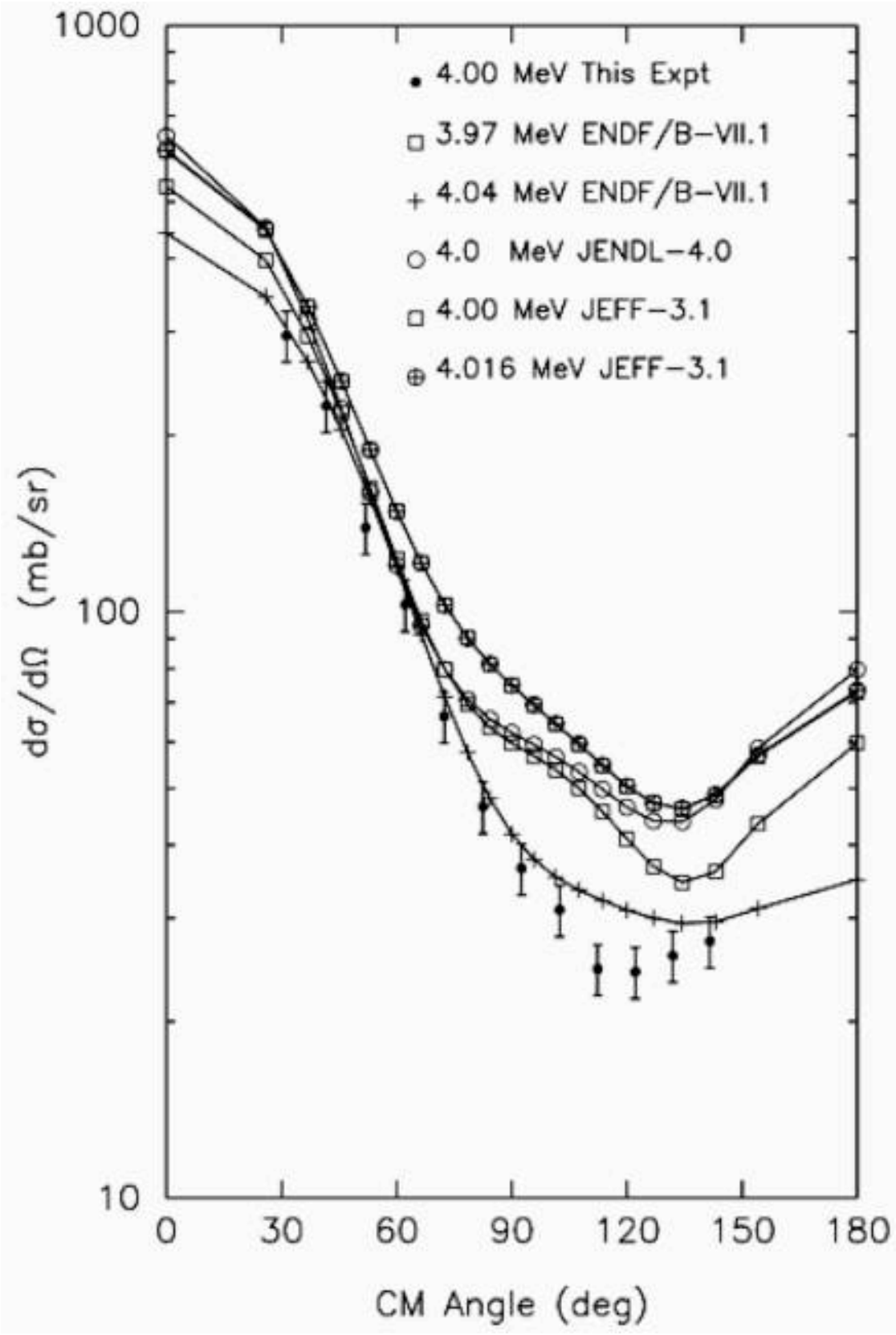

**Figure 15.** Comparison of 4.00-MeV elastic scattering cross sections for $^{23}Na$ with those from various nuclear libraries [2].

The Advanced Fuels Program of the Department of Energy sponsors research and development of innovative next generation light water reactor (LWR) and future fast systems. Input needed for both design and safety considerations for these systems includes neutron elastic and inelastic scattering cross sections that impact the fuel performance during irradiations, as well as coolants and structural materials. The goal of this project is to measure highly precise and accurate nuclear data for elastic/inelastic scattered neutrons. The high-precision requirements identified in the campaign supported by nuclear data sensitivity analyses have established a high priority need for precision elastic/inelastic nuclear data on the coolant $^{23}Na$ and the structural materials $^{54}Fe$ and $^{56}Fe$. Measurements of cross sections over an energy region from 1 to 9 MeV are desired. The measurements for $^{23}Na$ were recently published [2] and example data are shown in Figure 15; measurements for the stable iron isotopes are in progress.

The major theme of this applied science program is affirming the accuracy of the recommended cross sections found in the nuclear libraries, such as ENDF, JENDL, and JEFF and generating additional data where none exists. Often, the discrepancy between library values is greater than the covariance implies for the individual libraries. In other situations, the measured data on which the libraries are based is simply non-existent.

Gamma-ray production cross sections are also of interest for neutrinoless double-beta decay (0νββ). The experimental signature of 0νββ is a discrete peak at the energy of the Q value of the decay. It is possible that neutrons may inelastically scatter from surrounding materials or those composing the detector and produce background gamma rays in the region of the Q value, which would obscure the observation of this speculated but as yet unobserved process. Experiments have been performed to identify and measure cross sections for such background gamma rays for the 0νββ candidates $^{76}$Ge [4] and $^{136}$Xe [5].

Other applications-based programs have been established with collaborators from multiple institutions who are interested in detector development and/or characterization. Groups from the University of Guelph, the University of Nevada Las Vegas, and the University of Massachusetts at Lowell have all performed experiments which utilize the monoenergetic neutron capabilities in order to perform detector tests and characterizations. The Guelph group characterized deuterated benzene liquid scintillators, which will now be employed in the DESCANT array at TRIUMF [3].

Scientists with commercial interests, for example, Radiation Monitoring Devices in Watertown, MA, also visit the laboratory to make use of the monoenergetic neutrons. Projects range from development of radiation detecting materials to imaging systems. In addition to the typical nuclear physics markets, their detection systems are deployed in medical diagnostic, homeland security, and industrial non-destructive testing applications.

See the laboratory web page at http://www.pa.uky.edu/accelerator/ for an expanded description of the facilities, the research programs, and recent results from UKAL.

## References

1. P. E. Garrett, N. Warr, and S. W. Yates, J. Res. Natl. Inst. Stand. Technol. 105, 141 (2000).
2. J. R. Vanhoy, S. F. Hicks, A. Chakraborty, B. R. Champine, B. M. Combs, B. P. Crider, L. J. Kersting, A. Kumar, C. J. Lueck, S. H. Liu, P. J. McDonough, M. T. McEllistrem, E. E. Peters, F. M. Prados-Estévez, L. C. Sidwell, A. J. Sigillito, D. W. Watts, S. W. Yates, Nucl. Phys. A, 939, 121 (2015).
3. V. Bildstein, P. E. Garrett, J. Wong, D. Bandyopadhyay, J. Bangay, L. Bianco, B. Hadinia, K. G. Leach, C. Sumithrarachchi, S. F. Ashley, B. P. Crider, M. T. McEllistrem, E. E. Peters, F. M. Prados-Estévez, S. W. Yates, J. R. Vanhoy, Nucl. Instrum. Meth. A 729, 188 (2013).
4. B. P. Crider, E. E. Peters, T. J. Ross, M. T. McEllistrem, F. M. Prados-Estévez, J. M. Allmond, J. R. Vanhoy, and S. W. Yates, EPJ Web of Conferences 93, 05001 (2015).
5. E. E. Peters, T. J. Ross, B. P. Crider, S. F. Ashley, A. Chakraborty, M. D. Hennek, A. Kumar, S. H. Liu, M. T. McEllistrem, F. M. Prados-Estévez, J. S. Thrasher, and S. W. Yates, EPJ Web of Conferences 93, 01027 (2015).

# Appendix D.7: Lawrence Berkeley National Laboratory and the University of California Berkeley; 88-Inch Cyclotron and the High Flux Neutron Generator (HFNG)

**General Description:**

- 88-Inch Cyclotron: Sector-focused K-150 cyclotron coupled to 3 ECR ion source
- HFNG: Dual ion-source based self-loading DD neutron source

**Beams:** Protons to uranium @ $E/A \leq 20$ MeV/amu; neutrons @ $E_n \leq 60$ MeV

**Additional Capabilities:** BGS recoil separator, FIONA

**Research Focus:** Heavy element nuclear structure; Applied Nuclear Science

**Contact person:** Cyclotron Director: Larry Phair (LWPhair@lbl.gov); HFNG Contact: Lee Bernstein (LABernstein@berkeley.edu)

*Prepared by L. Bernstein and L. Phair*

## Executive Summary

The 88-Inch Cyclotron (the "88") at Lawrence Berkeley National Laboratory (LBNL) [1] is a variable energy, high-current, multi-particle cyclotron capable of accelerating ions ranging from protons to uranium at energies approaching and exceeding the Coulomb barrier. Maximum currents on the order of 10 particle•μamperes, with a maximum beam power of 2 kW, can be extracted from the machine for use in experiments in 7 experimental "caves". Beam currents up to the mA level could also be developed through the use of internal ion sources and targets. In addition to single-isotope beams the cyclotron can produce mixed-ion "cocktail" beams for use in electronic upset and damage studies. The cyclotron can also produce high-intensity pulsed, neutron beams whose energy can be determined via time-of-flight with flux $\leq 10^7$ n/s/cm$^2$ (DE/E≈5% at $E_n$=10 MeV), or broad spectrum (DE/E≈50%) with flux up to $\leq 10^{13}$ n/s/cm$^2$ via thick target deuteron breakup. Neutrons can also be provided in Berkeley using the DD-based High Flux Neutron Generator (HFNG) located at the adjacent UC-Berkeley department of nuclear engineering.

The cyclotron also has an array of research equipment developed for heavy-element research including the Berkeley Gas-filled Separator (BGS) and the FIONA ion trap. Lastly, a wide variety of mobile neutron, particle and gamma-ray detectors together with a mobile data acquisition system are present at the cyclotron for use in user experiments.

## General Considerations of 88-Inch Cyclotron

The 88 was originally envisioned as a high-current, variable energy, light-ion accelerator for nuclear physics and nuclear chemistry studies, as well as for the production of isotopes used in scientific research. It started operation in 1961 and has maintained its position as a premier stable-beam facility through periodic upgrades, especially to its ion sources [2]. These ion sources have enabled acceleration of an ever-increasing variety of heavy-ion beams up to, and beyond, the Coulomb barrier. Protons, deuterons, and alpha particle beams are available up to maximum energies of 55, 65, and 130 MeV, respectively. For extracted beams the operational upper limits of current intensities are not known since we restrict running to a maximum power of 1.5 kW. These administrative limitations are self-imposed. There is no reason that we cannot exceed these restrictions with proper planning and preparation. One can readily envision extracted beams of several tens of particle-microamperes. Development of a negative ion acceleration scheme combined with "stripping" would allow a clean extraction of intense proton beams (as recently demonstrated with the same cyclotron at Texas A&M University).

One consideration for even more intense beams of light ions is the use of internal targets. Indeed, this technique was used at the 88 in its early years to produce isotopes for research and there is no reason that the capability cannot be re-established. This would enable use of beams with intensities exceeding a milliampere (1000μA). This would open up great possibilities for production of isotopes. But then radioactive target handling and radiochemistry would need additional attention. The resulting power levels (tens of kW) make it the only charged particle accelerator facility currently in the DOE complex capable of large-scale isotope production using light-ion beams other than protons.

Beam-time at the 88-Inch cyclotron can be obtained either via purchase (≈$1500/hour), or by merit-based review provided by a local advisory committee.  Approximately 60% of the beam-time is reserved for nuclear science research.  Individuals interested in performing experiments at the 88-Inch should contact the user liaison, Mike Johnson (MBJohnson@lbl.gov), the cyclotron Larry Phair (LWPhair@lbl.gov) or the scientific director Paul Fallon (PFallon@lbl.gov).

## Instrumentation and facility layout

The 88-Inch Cyclotron is host a number of unique instruments and capabilities.  These include three electron cyclotron resonance (ECR) ion sources, featuring VENUS, the most powerful superconducting ECR ion source in the world.  These ECRs provide a range of highly-charged ions up to and including fully-stripped $U^{92+}$.  The cyclotron also plays host to the Berkeley Gas-filled Separator (BGS).  The BGS provides rejection of beam-like and fission fragment nuclides formed in heavy-ion reactions in excess $1{:}10^{12}$ for use in heavy-element research.  The back end of the BGS can accommodate an array of pixelated Micron "W2" Si detectors three "Clover" HPGe detectors for use in alpha- and gamma-decay spectroscopy of evaporation product nuclides.  Alternatively, the back end of the BGS can be coupled to the FIONA ion trap that can isolate a single charge-to-mass ratio fragment.

The 88-Inch also has a mobile data acquisition system that can be used to with the three in-house "clover" HPGe detectors and an array of 6-10 modular neutron detectors. LBNL is also a member of the clovershare program, providing access to an additional 6-10 detectors on a by-arrangement basis. Lastly, LBNL has a pair of well-calibrated shielded HPGe detectors located outside of the experimental caves that can be used to measure activities off-line for cross section or decay spectroscopy measurements.

The figure below shows the layout of the experimental capabilities at the cyclotron.

**Figure 16.** Experimental Cave Beam Lines Layout

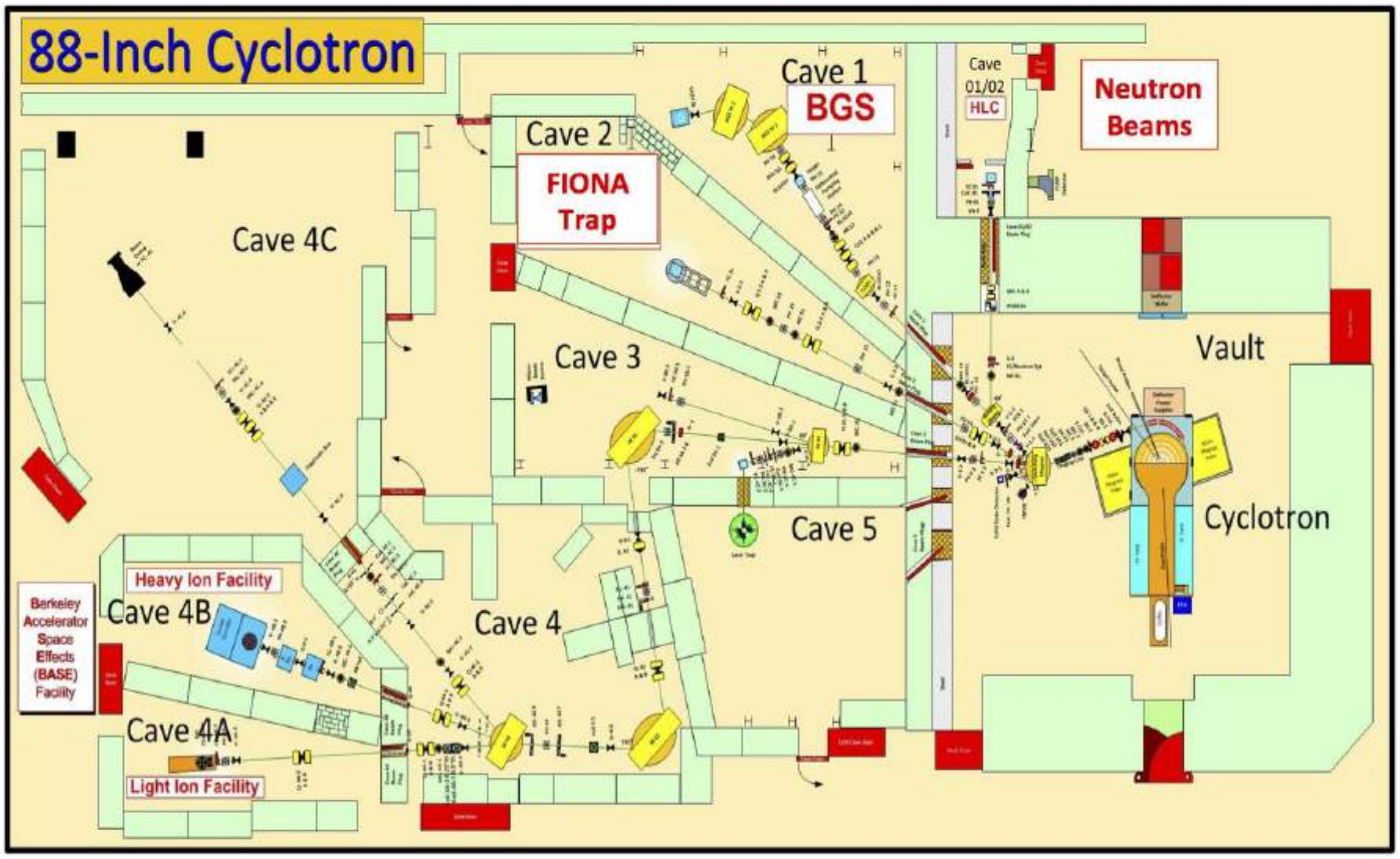

## The Berkeley Accelerator Space Effects (BASE) facility

In addition to the basic nuclear science research the 88-Inch cyclotron is host to the Berkeley Accelerator Space Effects (BASE) Facility. BASE provides well-characterized beams of protons, heavy ions, and other medium energy particles which simulate the space environment. The primary capability employed at BASE is a "cocktail" heavy-ion beam capable of performing damage and electronic upset studies over a range of thicknesses in materials. The table 4 lists the properties and constituents of these cocktail beams.

**Table 4.** BASE Facility Standard "Cocktail" Ion List. Standard "cocktails" (of species with similar charge-to-mass ratios) are listed along with their energy loss and range values.

| Ion | Cocktail (AMeV) | Energy (MeV) | Z | A | Chg. State | % Nat. Abund. | LET 0° | LET 60° | Range (Max) (μm) |
|---|---|---|---|---|---|---|---|---|---|
| | | | | | | | (MeV/mg/cm2) | | |
| **B** | **4.5** | 44.90 | 5 | 10 | +2 | 19.9 | **1.65** | 3.30 | 78.5 |
| **N** | **4.5** | 67.44 | 7 | 15 | +3 | 0.37 | **3.08** | 6.16 | 67.8 |
| **Ne** | **4.5** | 89.95 | 10 | 20 | +4 | 90.48 | **5.77** | 11.54 | 53.1 |
| **Si** | **4.5** | 139.61 | 14 | 29 | +6 | 4.67 | **9.28** | 18.56 | 52.4 |

| Ion | Cocktail (AMeV) | Energy (MeV) | Z | A | Chg. State | % Nat. Abund. | LET 0° | LET 60° | Range (Max) (μm) |
|---|---|---|---|---|---|---|---|---|---|
| **Ar** | **4.5** | 180.00 | 18 | 40 | +8 | 99.6 | **14.32** | 28.64 | 48.3 |
| **V** | **4.5** | 221.00 | 23 | 51 | +10 | 99.75 | **21.68** | 43.36 | 42.5 |
| **Cu** | **4.5** | 301.79 | 29 | 63 | +13 | 69.17 | **29.33** | 58.66 | 45.6 |
| **Kr** | **4.5** | 378.11 | 36 | 86 | +17 | 17.3 | **39.25** | 78.50 | 42.4 |
| **Y** | **4.5** | 409.58 | 39 | 89 | +18 | 100 | **45.58** | 91.16 | 45.8 |
| **Ag** | **4.5** | 499.50 | 47 | 109 | +22 | 48.161 | **58.18** | 116.36 | 46.3 |
| **Xe** | **4.5** | 602.90 | 54 | 136 | +27 | 8.9 | **68.84** | 137.68 | 48.3 |
| **Tb** | **4.5** | 724.17 | 65 | 159 | +32 | 100 | **77.52** | 155.04 | 52.4 |
| **Ta** | **4.5** | 805.02 | 73 | 181 | +36 | 99.988 | **87.15** | 174.30 | 53.0 |
| **Bi*** | **4.5** | 904.16 | 83 | 209 | +41 | 100 | **99.74** | 199.48 | 52.9 |
| **B** | **10** | 108.01 | 5 | 11 | +3 | 80.1 | **0.89** | 1.78 | 305.7 |
| **O** | **10** | 183.47 | 8 | 18 | +5 | 0.2 | **2.19** | 4.38 | 226.4 |
| **Ne** | **10** | 216.28 | 10 | 22 | +6 | 9.25 | **3.49** | 6.98 | 174.6 |
| **Si** | **10** | 291.77 | 14 | 29 | +8 | 4.67 | **6.09** | 12.18 | 141.7 |
| **Ar** | **10** | 400.00 | 18 | 40 | +11 | 99.6 | **9.74** | 19.48 | 130.1 |
| **V** | **10** | 508.27 | 23 | 51 | +14 | 99.75 | **14.59** | 29.18 | 113.4 |
| **Cu** | **10** | 659.19 | 29 | 65 | +18 | 30.83 | **21.17** | 42.34 | 108.0 |
| **Kr** | **10** | 885.59 | 36 | 86 | +24 | 17.3 | **30.86** | 61.72 | 109.9 |
| **Y** | **10** | 928.49 | 39 | 89 | +25 | 100 | **34.73** | 69.46 | 102.2 |
| **Ag** | **10** | 1039.42 | 47 | 107 | +29 | 51.839 | **48.15** | 96.30 | 90.0 |
| **Xe** | **10** | 1232.55 | 54 | 124 | +34 | 0.1 | **58.78** | 117.56 | 90.0 |
| **Au*** | **10** | 1955.87 | 79 | 197 | +54 | 100 | **85.76** | 171.52 | 105.9 |
| **He*** | **16** | 43.46 | 2 | 3 | +1 | 0.000137 | **0.11** | 0.22 | 1020.0 |
| **N** | **16** | 233.75 | 7 | 14 | +5 | 99.63 | **1.16** | 2.32 | 505.9 |
| **O** | **16** | 277.33 | 8 | 17 | +6 | 0.04 | **1.54** | 3.08 | 462.4 |
| **Ne** | **16** | 321.00 | 10 | 20 | +7 | 90.48 | **2.39** | 4.78 | 347.9 |
| **Si** | **16** | 452.10 | 14 | 29 | +10 | 4.67 | **4.56** | 9.12 | 274.3 |

| Ion | Cocktail (AMeV) | Energy (MeV) | Z | A | Chg. State | % Nat. Abund. | LET 0° | LET 60° | Range (Max) (µm) |
|---|---|---|---|---|---|---|---|---|---|
| **Cl** | **16** | 539.51 | 17 | 35 | +12 | 75.77 | **6.61** | 13.22 | 233.6 |
| **Ar** | **16** | 642.36 | 18 | 40 | +14 | 99.600 | **7.27** | 14.54 | 255.6 |
| **V** | **16** | 832.84 | 23 | 51 | +18 | 99.750 | **10.90** | 21.80 | 225.8 |
| **Cu** | **16** | 1007.34 | 29 | 63 | +22 | 69.17 | **16.53** | 33.06 | 190.3 |
| **Kr** | **16** | 1225.54 | 36 | 78 | +27 | 0.35 | **24.98** | 49.96 | 165.4 |
| **Xe*** | **16** | 1954.71 | 54 | 124 | +43 | 0.1 | **49.29** | 98.58 | 147.9 |
| **N** | **30** | 425.45 | 7 | 15 | +7 | 0.370 | **0.76** | 1.52 | 1370.0 |
| **O** | **30** | 490.22 | 8 | 17 | +8 | 0.04 | **0.98** | 1.96 | 1220.0 |
| **Ne** | **30** | 620.00 | 10 | 21 | +10 | 0.27 | **1.48** | 2.96 | 1040.0 |
| **Ar** | **30** | 1046.11 | 18 | 36 | +17 | 0.337 | **4.87** | 9.74 | 578.1 |

Additionally, BASE is unique in having beams parallel enough to support microbeams, used to probe increasingly miniaturized semiconductor parts with new modes of failure. The National Security Space (NSS) community and researchers from other government, university, commercial, and international institutions use these beams to understand the effect of radiation on microelectronics, optics, materials, and cells. Space missions utilizing the BASE Facility include Voyager, the Space Shuttle, Solar Dynamics Observatory, Mars Spirit and Opportunity rovers, Galileo (Jupiter), Cassini (Saturn), and the new James Webb Space Telescope, currently preparing for launch in 2018.

## References

1. http://cyclotron.lbl.gov/
2. D. Leitner, C. M. Lyneis, T. Loew, *et al.*, Rev. Sci. Instrum. **77**, 03A303 (2006).

## The High Flux Neutron Generator at UC-Berkeley

In addition to the 88-Inch cyclotron, nuclear researchers working in Berkeley can utilize the High Flux Neutron Generator (HFNG) on the UC-Berkeley campus. The HFNG is a dual-ion source-based DD neutron generator located in a 62”-thick concrete enclosure in Etcheverry Hall on the UC-Berkeley campus. Collaborative research, including radioactive material transport between LBNL and UC-Berkeley is facilitated by the designation of Etcheverry Hall as a location on the LBNL campus.

The HFNG uses a self-loading titanium-coated copper target to provide continuous operation. Voltages from 80-120 keV are used to accelerate beams from 1-50 mA onto the production target. The target is designed to allow the placement of samples in the center of the generator less than 5 mm from the DD reaction surfaces. Activation measurements can be performed using samples placed in the interior neutron production target with flux monitoring provided by external neutron moderators and the use of calibrated activation foils (indium, nickel etc.). In addition, the HFNG can be positioned to allow the extraction of an external beam of monochromatic 2.45 MeV neutron beam for use in prompt (n,n) and (n,n’γ) measurements. The HFNG currently runs at a total neutron output of $10^8$ n/s into 4π solid angle, but fluxes up to several $10^9$, to $10^{10}$ could be achieved if deemed necessary.

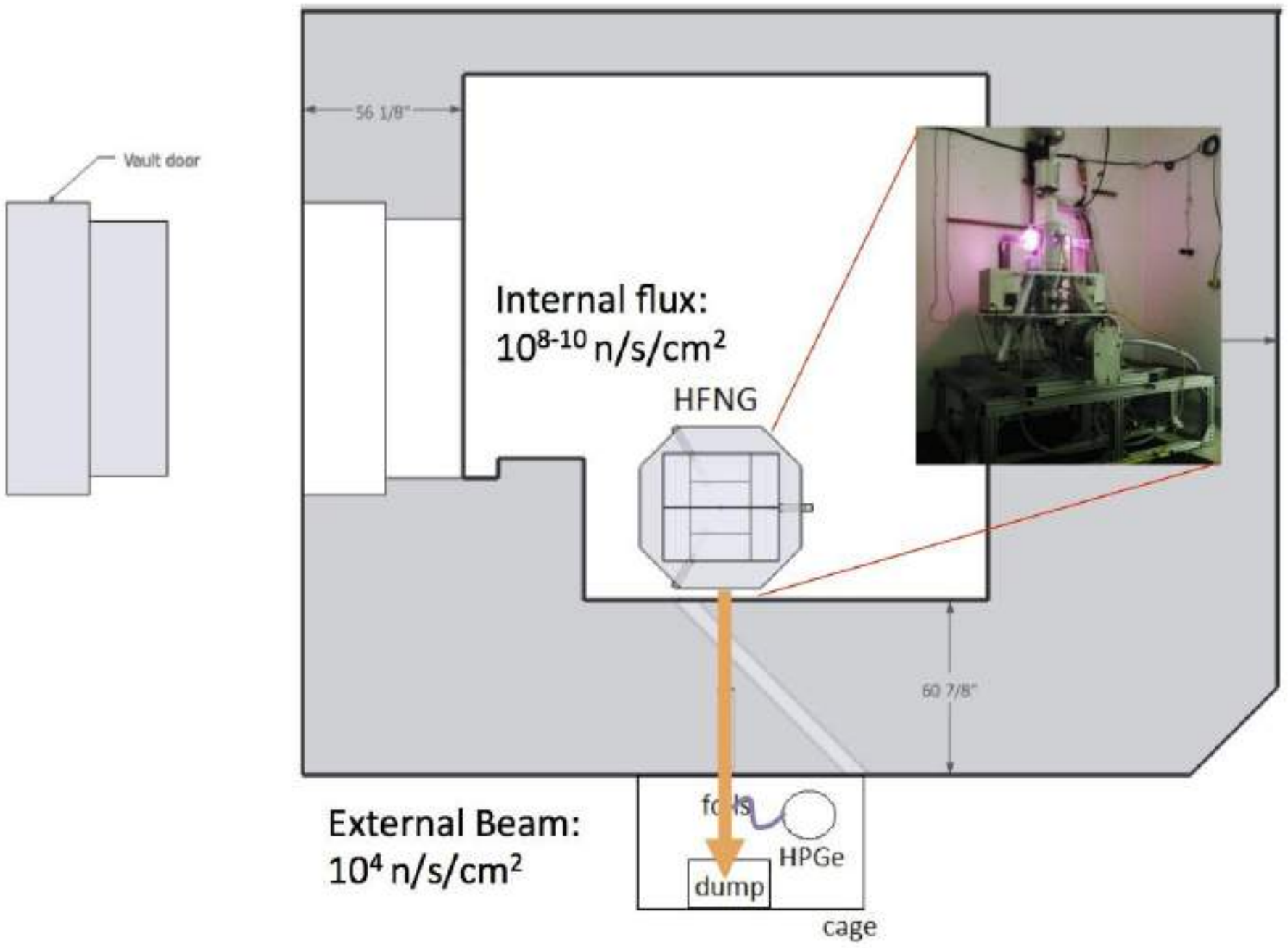

**Figure 17.** The HFNG facility layout, including the external beam-line. The inset shows a photo of the HFNG with its ion sources energized.

Equipment at the HFNG includes several HPGe, X-ray and proton-recoil detectors. Researchers can also utilize the adjacent teaching laboratories with on a by-arrangement basis The HFNG is run and maintained by students in the UC-Berkeley department of nuclear engineering. For information about running at the HFNG researchers should contact Lee Bernstein (labernstein@berkeley.edu).

# Appendix D.8: Los Alamos National Laboratory, Isotope Production Facility

**General Description:** Radionuclide Production for DOE Isotope Program housed in the LANSCE accelerator at Los Alamos National Laboratory; not a user facility but maintaining limited funding and staff for collaborative research

**Beams:** 40-100 MeV, 0.1 – 250 μA proton beams; Unmoderated $10^{13}$ $cm^{-1}$ $s^{-1}$ spallation neutron flux

**Additional Capabilities:** Hot cell facilities for remote manipulation of intense sources, radiochemical characterization and separations expertise, alpha/beta/gamma spectroscopy, 200-800 MeV protons at LANSCE-WNR

**Research Focus:** Isotope production, nuclear data for proton-induced reactions, radiochemical separations research.

**Contact person:** Eva Birnbaum; eva@lanl.gov; +1 505 665 7167

*Prepared by Jonathan W Engle*

The LANL Isotope Production Facility (IPF) is a dedicated target irradiation facility located at the Los Alamos Neutron Science Center (LANSCE), which accepts up to 100 MeV protons at beam currents up to 250 μA (and up to 450 μA in the future) to produce isotopes via LANL's 800-MeV accelerator. Three target slots allow target irradiation to be optimized by energy range for a particular isotope. Available beam time is estimated to be ~3000 hours / year.

The Los Alamos Hot Cell Radiological Facility is a cGMP compliant facility located at TA-48 consisting of 13 hot cells with a sample load shielding capacity of 1 kCi of 1 MeV gamma rays per cell for the remote handling of highly activated samples. The Hot Cells are equipped for separation, purification and wet chemistry activities with standard laboratory equipment, and the ability to perform radioassay of materials within the cells. The facility also contains fume hoods for radiological chemistry and reagent preparation. Available instrumentation includes counting capabilities described above, ICP-OES, HPLC, balances, centrifuges, and access to shared capabilities for materials diagnostics and characterization.

The LANL Count Room capability occupies more than 7000 square feet of LANL Building RC-1 at TA-48, and is dedicated to performing qualitative and quantitative assay of gamma, beta, and alpha-emitting radionuclides in a variety of matrices and over a wide range of activity levels. Founded in support of the US Testing Program, this facility is currently funded ~70% by a range of national security programs, and the balance in support of other internal and external customers. The Count-room's more than 65 systems include High Purity Germanium (HPGe) gamma- and X-ray spectrometers, alpha spectrometers and counters, and beta counters, operate 24x7x365, and perform more than 70,000 measurements annually.

# Appendix D.9: Los Alamos National Laboratory, Los Alamos Neutron Science Center (LANSCE)

| |
|---|
| **General Description:** US DOE NNSA National Laboratory, NNSA User Facilities, proton and neutron beams for basic and applied research in nuclear science, materials research, and fundamental science. Proposals submitted online are rated for scientific/applied merit by PAC. Proprietary proposals at Target 4 cost-recovery rates: \$11k/1st day, \$9k/day after 1st day. |
| **Accelerator:** Proton Linear Accelerator (100 MeV (IPF) and 211-800 MeV) dual $H^+$ and $H^-$ beams. |
| **Beams:**<br>• *Neutrons*: Target 4 - bare tungsten neutron production target, 6 flight paths 8 to 90 m, proton $\Delta t< 1$ ns<br>• *Neutrons:* Target 1 flux-trap water & $LH_2$ moderated – 3+ flight paths, 8 to 20 m<br>• *Neutrons*: Target 4 East Port – neutron irradiations – moderated or un-moderated, $10^{11}$ n/cm$^2$-s @ 0.7 m<br>• *Neutrons*: Target 4 60R pre-collimator neutron irradiations – $10^9$ n/cm$^2$-s @ 10 m<br>• *Protons*: Target 2 Blue Room – (low neutron return) 12 m dia. room, 211 – 800 MeV protons, 80 nA average, higher for LSDS or shielded target.<br>• *Protons*: Planned high current irradiations in Area A. |
| **Experimental focus:** neutron-induced nuclear reactions, fission studies, prompt reactions, activation and decay studies, isotope production cross sections, proton-induced nuclear reactions. Neutron imaging/CT Target 1 & Target 4, energy-selective imaging.<br><br>Proton flash radiography. Ultra-cold neutrons/fundamental physics. |
| **Detector arrays:** High-energy neutron PSD 54-detector array, Low-energy neutron 22-Li-glass array, fission time projection chamber, DANCE – 160 BaF2 array for (n,γ) |
| **Contact person:** LANSCE User Office; lansce-user-office@lanl.gov ; +1 505 665 1010 |

*Prepared by Ron Nelson & Steve Wender*

The Los Alamos Neutron Science Center (LANSCE) facilities for Nuclear Science consist of a high-energy "white" neutron source (Target 4) with 6 flight paths, three low-energy nuclear science flight paths at the Lujan Center (Target-1), and a proton reaction area (Target-2). The neutron beams produced at the WNR Target 4 complement those produced at the Lujan Center because they are of much higher energy and have shorter pulse widths. The 800 MeV proton beam of the LANSCE linear accelerator or linac drives the neutron sources. Proposals for beam time at the neutron production targets, Blue Room, and proton radiography facilities may be submitted for open research or proprietary work. See http://lansce.lanl.gov "Facilities" and "User Resources" tabs for details on the facilities and proposal submission.

Neutron beams with energies ranging from approximately 0.1 MeV to greater than 600 MeV are produced in Target-4. The Target-4 neutron production target is a bare unmoderated tungsten cylinder that is bombarded by the 800 MeV pulsed proton beam from the LANSCE linear accelerator and produces neutrons via spallation reactions. Because the proton beam is pulsed, the energy of the neutrons can be determined by time-of-flight (TOF) techniques. The time structure of the proton beam can be easily changed to optimize a particular experiment. Presently, Target-4 operates with a proton beam current of approximately 4 μA, 1.8 μs between pulses and approximately 35,000 pulses/sec. Target-4 is the most intense high-energy neutron source in the world and has 6 flight paths instrumented for a variety of measurements.

In the Target-2 area (Blue Room), samples can be exposed to the 800 MeV proton beam directly from the linac, or with more peak intensity with a beam that has been accumulated in the Proton Storage Ring (PSR). Although the total beam current is limited by the shielding in Target-2, the PSR beam provides significantly greater peak intensity than the direct beam from the accelerator. Target-2 is used for proton irradiations and hosts the Lead Slowing-Down Spectrometer (LSDS). Proton beams with energies as low as 211 MeV can be transported to Target-2.

At present there are three flight paths at the Lujan Center that are devoted to Nuclear Science research. Other flight paths are devoted to Materials Science research. These flight paths view a moderated target with both water and liquid hydrogen moderators, and have useful neutron fluxes that range from sub-thermal to approximately 500 keV.

With these facilities, LANSCE is able to deliver neutrons with energies ranging from a meV to several hundreds of MeV, as well as proton beams with a wide range of energy, time and intensity characteristics. The facilities, instruments and research programs are described briefly below.

## Overview of the Flight Paths

Each Flight Path name identifies the target and the direction of the flight path (FP) with respect to the proton beam. For example, 4FP15R is a FP (flight path) that starts at Target 4 and is 15 degrees to the right (15R) of the incoming proton beam. Figure 18 shows the layout of the flight paths.

The neutron fluxes available are shown in Figure 19.

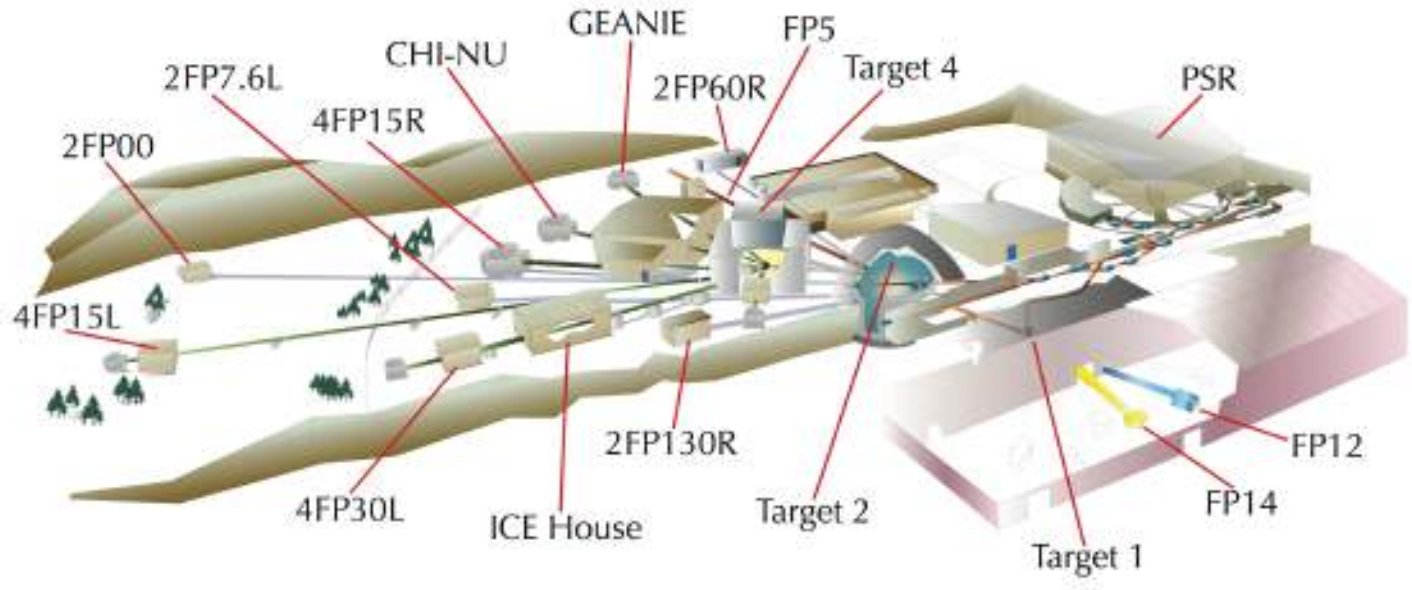

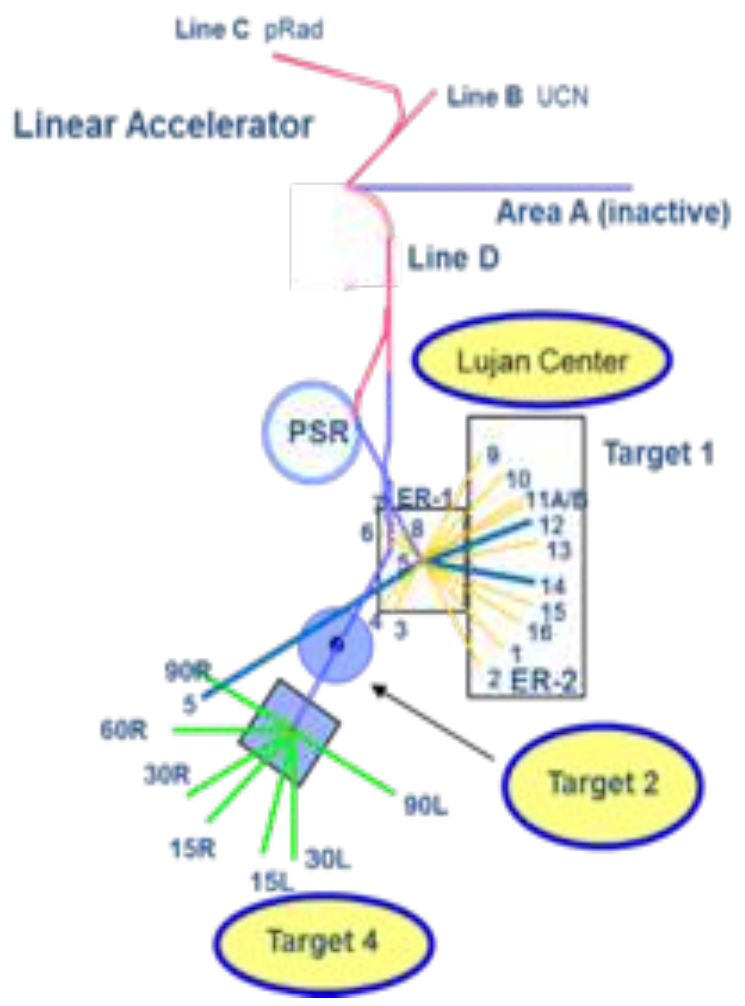

**Figure 18.** Two different views of the layout of the Target-1, 2, and 4 flight paths at the LANSCE neutron sources.

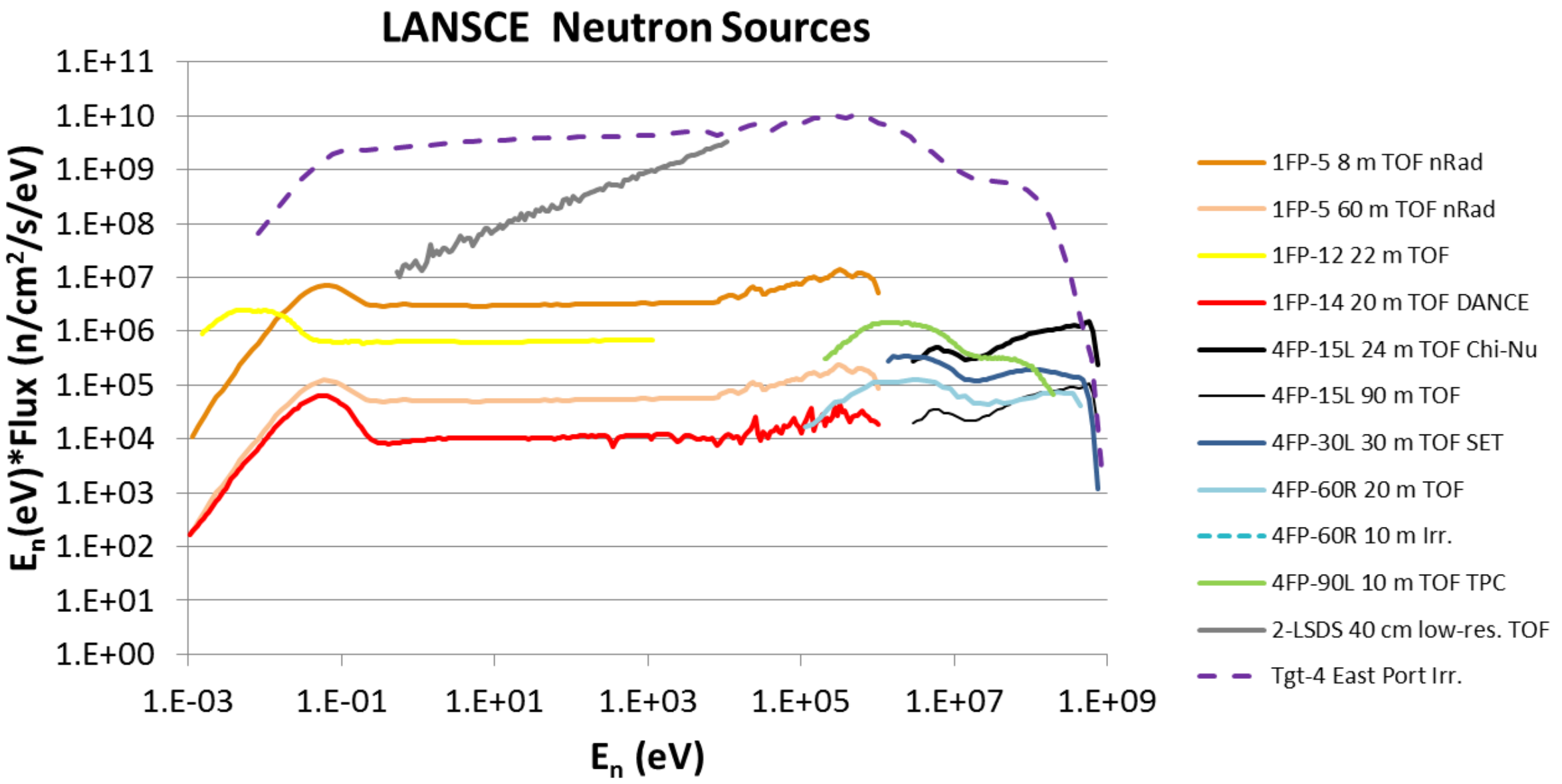

**Figure 19.** Graphs of the neutron flux times energy (also known as the flux/unit lethargy) for a representative sample of the neutron time-of-flight (TOF) and irradiation (Irr.) stations at LANSCE. The data are from measurements or calculations vetted against measurements

## Target 4 Flight Paths (FP)

For the Target-4 flight paths, the neutron spectrum depends on the angle of the flight path with respect to the proton beam with the higher-energy neutron flux greater at the more forward angles. Below we list the main activities that are presently being performed on each flight path.

- **4FP90L** is the location of the Time-Projection Chamber (TPC) that is used to measure fission cross sections to high precision.

- **4FP30L** The ICE House is ~20 m from the production target and is used by industry, universities, and national laboratories for semiconductor electronics testing (SET) to measure neutron-induced failures in devices.
- **4FP15L** has two experimental locations available at distances of 22 and 90 meters from the spallation target. Primarily used for the Chi-Nu experiments at 22 meters. Chi-Nu is measuring the fission neutron output spectrum. A low-neutron-return room is below the 22 m station. The 90 m flight path is used mostly for neutron detector development and calibration
- **4FP15R** is a general purpose flight path that is now being used for neutron radiography, the SPIDER detector (fission product yields) and the low-energy (n,z) (LENZ) experiment.
- **Industry, universities, and national laboratories primarily use 4FP30R or ICE II station at 15 m** for SET.
- **4FP60R** The 20 m station is used for gamma-ray spectroscopy and other experiments. An irradiation station using peripheral beam is available at 10 m.

## Target 2 (Blue Room)

- Target 2 is used for proton beam irradiation experiments. Beam is available directly from the linac or from the proton storage ring (PSR). Present and past experiments include:
- A lead slowing-down spectrometer (LSDS) provides very large effective neutron fluxes in the energy range from ~1 eV to ~10 keV with low neutron energy resolution for measuring cross sections with ultra-small samples.
- Pulsed beam experiments to simulate intense neutron environments for semiconductor certification.
- Proton irradiation of detectors and radiation-hardness testing of components for the Large Hadron Collider at CERN.
- Measurement of radioisotope production cross sections for the Isotope Production Facility (IPF) at LANSCE (see the IPF contribution to this report).

## Target 1 Lujan Center Flight Paths

- **FP5** is a water-moderated general purpose flight path that is currently being used for neutron radiography. It has two detector areas: one at approximately 10m in ER-1 and the second at a distance of 60 m that is reached from the Target-4 yard. The 60 m station has a large field of view.
- **FP14** is the location of the Detector for Advanced Neutron Capture Experiments (DANCE). It consists of a 4-π array of $BaF_2$ scintillators designed for neutron capture measurements on sub-milligram and radioactive samples. These measurements support radiochemical detector cross section measurements for Defense Programs, and experiments for nuclear astrophysics.
- **FP12** is a cold-moderator flight path currently used by the SPIDER spectrometer to measure fission fragment yields. FP12 has a neutron guide.

## Other Experimental Areas

**Target-4 East Port** provides a mechanism for irradiating samples in the intense broad spectrum neutron field at 0.7 m from the Target-4 neutron production target. Samples can be moved from the irradiation position to a storage position by remote control.

**Proton Radiography Facility** The pRad facility provides fast imaging of static and dynamic systems. See http://lansce.lanl.gov/pRad/index.shtml for more information.

**Ultra-Cold Neutron (UCN) Facility** State-of-the-art UCN Facility See http://lansce.lanl.gov/UCN/index.shtml

# Appendix D.10: Michigan State University, National Superconducting Cyclotron Laboratory

**General Description:** University-based, national user facility focused on basic research in low-energy nuclear science, accelerator science, fundamental symmetries and societal applications.

**Accelerators:** 2 coupled cyclotrons, one linear reaccelerator

**Beams:** Over 1000 rare isotopes produced both neutron-rich and neutron deficient.

- Primary beam rates are available from: http://www.nscl.msu.edu/users/beams.html
- Secondary beams rates can be calculated with LISE available at: https://groups.nscl.msu.edu/a1900/software/lise++/

Beam time is allocated by PAC.

**Experimental focus (relevant to applications):**

- Beams of most isotopes of data interest
- Decay spectroscopy
- Neutron capture rate inference on short-lived rare isotopes
- Isotope Harvesting

**Present detector array capabilities (relevant to applications):**

- Decay spectroscopy station
- Total absorption gamma-ray spectrometer
- Proof-of-principle isotope harvesting station

**Contact person:** Sean Liddick

*Prepared by Sean Liddick*

Facility provides unique access to rare isotopes over a broad energy range including thermal, few MeV/nucleon to ~100 MeV/nucleon.  It includes a large complement of state-of-the art experimental equipment for study of nuclear properties and reactions.

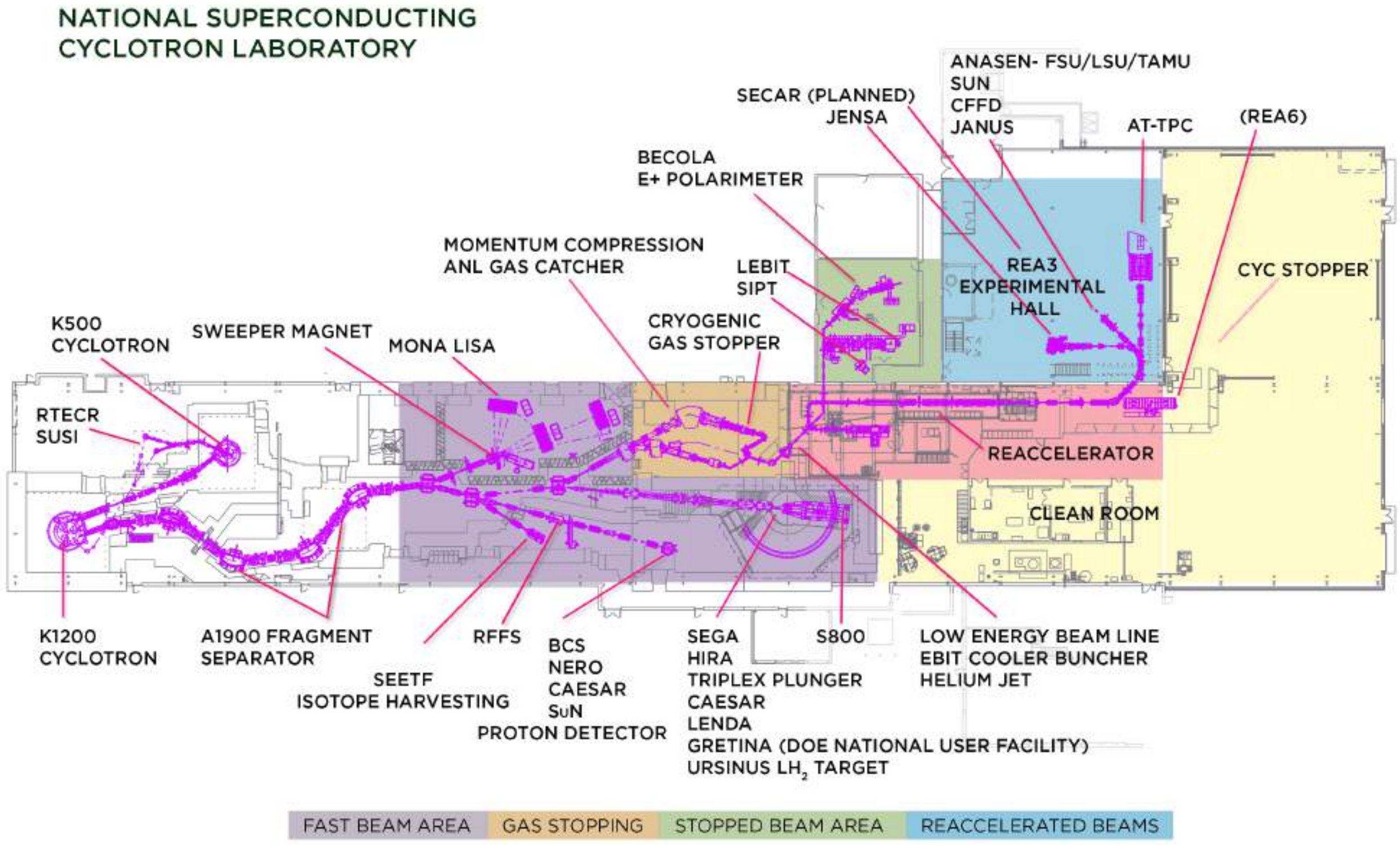

**Figure 20.** General layout of experimental equipment at NSCL for use with fast, stopped, and reaccelerated beams. See http://www.nscl.msu.edu/users/equipment.html for more detail.

# Decay Spectroscopy

*Motivation:* Decay spectroscopy provides a number of quantities of interest for the low-energy nuclear science community such has half-lives, delayed neutron-branching ratios, and delayed gamma-ray transitions. Absolute gamma-ray intensities can be obtained based on ion-by-ion counting of the radioactive ion beam and the beta-delayed gamma rays are used to elucidate the low-energy level scheme of the daughter nucleus. High- and low-resolution delayed gamma-ray studies can be used to infer average electron and gamma-ray energies emitted following beta decay.

*Detection System:* The detection system consists of either a central Si or Ge detector for ion and beta-decay electron detection [1,2]. Multiple ancillary arrays existed for delayed emissions including gamma-rays and neutrons [3,4,5,6].

*Recent Results:* Conversion electron emission from an isomer state was monitored in $^{68}$Ni to extract E0 monopole transition strengths [7]. Decays of various neutron-rich isotopes were studied to determine low-energy level schemes and identify gamma and beta-emitting isomeric states [8]. Total absorption spectroscopy addressed deficiencies in previously reported decay scheme of $^{76}$Ga into $^{76}$Ge.

## References

1. "*Beta counting system for fast fragment beams*", J. I. Prisciandaro *et al.*, Nucl. Instrum. Meth. Phys. Res. A **505**, 140 (2002).
2. "*High Efficiency Beta-decay Spectroscopy using a Planar Germanium Double-Sided Strip Detector*", N. Larson *et al.*, Nucl. Instrum. Methods in Phys. Res. A, **727**, 59 (2013).
3. "*Thirty-two-fold segmented germanium detectors to identify gamma rays from intermediate-energy exotic beams*", W.F. Mueller *et al.*, Nucl. Instrum. Meth. in Phys. Res. A, **466**, 492 (2001).

4. "*The neutron long counter NERO for studies of beta-delayed neutron emission in the r-process*", J. Pereira *et al.*, Nucl. Instrum. Meth. in Phys. Res. A, **618**, 275 (2010).
5. "*Half-lives and branchings for beta-delayed neutron emission for neutron-rich Co-Cu isotopes in the r-process*", P. Hosmer *et al.*, Phys. Rev. C, **82**, 025806 (2010).
6. "*SuN: Summing NaI gamma-ray detector for capture reaction measurements*", A. Simon *et al.*, Nucl. Instrum. Meth. in Phys. Rev. A, **703**, 16 (2013).
7. "*Shape coexistence in Ni-68*", S. Suchyta *et al.*, Phys. Rev. C **89**, 021301 (2014).
8. "*Low-energy level schemes of $^{66,68}$Fe and inferred proton and neutron excitations across Z = 28 and N = 40*", S. Suchyta *et al.*, Phys. Rev. C, **87**, 014325 (2013).

## Neutron Capture Rates of Short-Lived Rare Isotopes

*Motivation*: Neutron capture rates impact a wide variety of fields including nuclear astrophysics, national security, and nuclear power generation. The need for neutron capture rates on short-lived nuclei has motivated a number of indirect techniques. At NSCL, a new technique has been developed to infer neutron capture rates by determining the basic nuclear properties of radioactive ions.

*Technique:* The detection system consists of a small beta-decay-electron sensitive detector inserted into a large total absorption gamma-ray spectrometer called the Summing NaI detector (SuN) [1] at NSCL. Radioactive ions are produced and delivered to SuN and the resulting beta-delayed gamma rays are detected. Gamma-ray emission from highly excited states in the daughter nucleus is used to extract the functional form of the gamma-ray strength and nuclear level density. These quantities are inserted into Hauser-Feshbach calculations to infer neutron capture rates.

*Recent Results:* The technique has been applied to the neutron capture of $^{75}$Ge, which is unstable ($t_{1/2}$ = 83 min), see Figure 21 [2]. Further work is anticipated in neutron-rich Fe and Sr regions for nuclear astrophysics and national security applications.

### References

1. "*SuN: Summing NaI gamma-ray detector for capture reaction measurements*", A. Simon *et al.*, Nuclear Instrum. Methods in Phys. Rev. A, **703**, 16 (2013)
2. "*Novel Technique for constraining r-process (n,γ) reaction rates*", A. Spyrou *et. al.*, Phys. Rev. Lett. **113**, 232502 (2014).

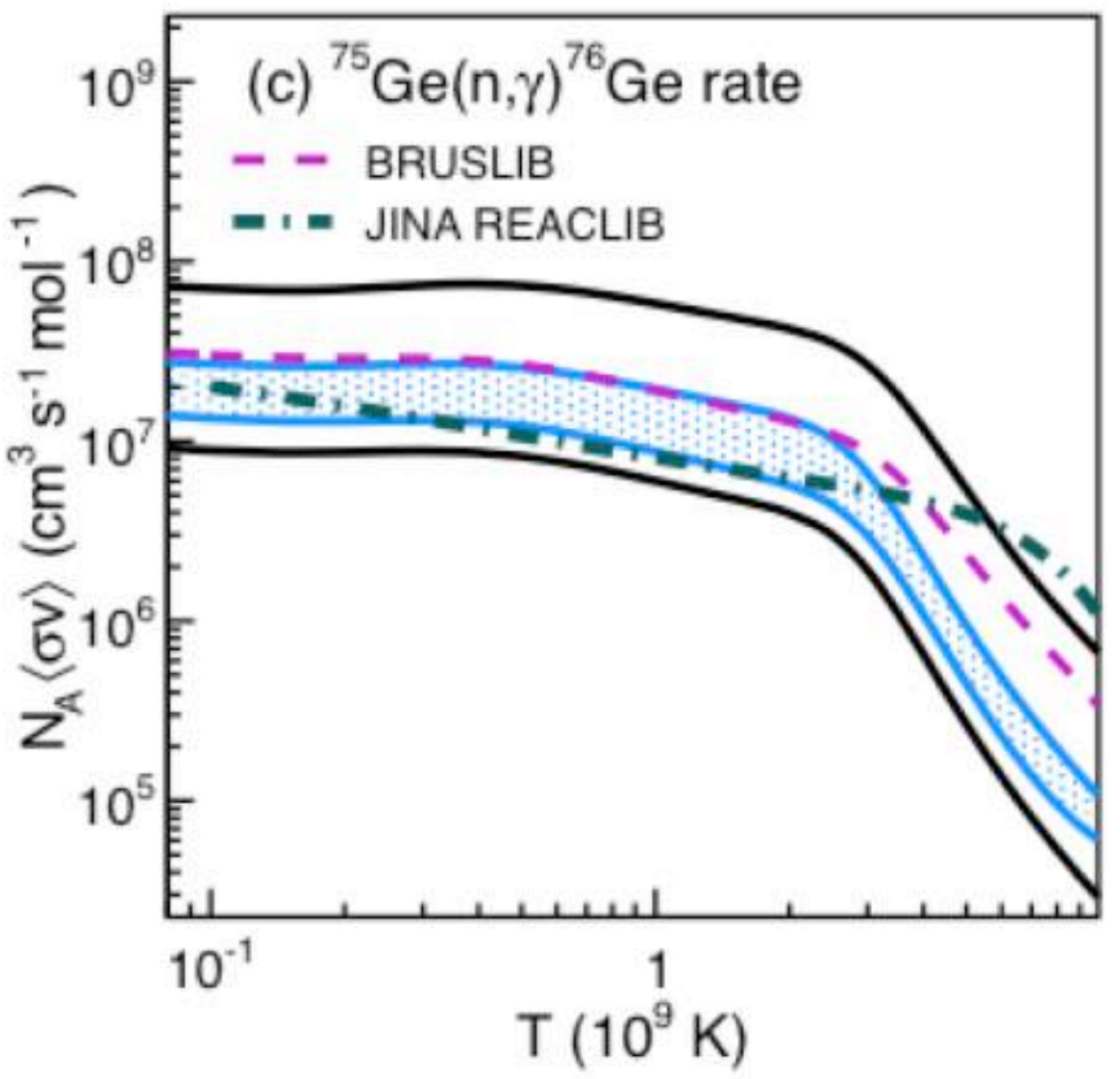

**Figure 21.** Maxwellian-averaged reaction rate as a function of stellar temperature compared to rates from BRUSLIB and JINA REACLIB and TALYS limits.

## Isotope Harvesting

*Motivation:* The vast majority of rare isotope beams used in experiments at the NSCL and that will be produced at FRIB only live for a few seconds or less. However, a very large number of longer-lived isotopes that have important uses in medical research (and other applications) are not collected during normal operations. The long-term possibilities for isotope harvesting have been assessed in an ongoing series of user workshops. A collaboration of researchers at Hope College and Washington University in St. Louis are working with NSCL researchers to develop systems and to solve problems associated with harvesting the unused isotopes at now at the NSCL, and eventually FRIB, for off-line experiments.

*Detection System:* The team from Hope College designed and built an end-station to fill, irradiate and collect samples of 100 milliliters of water. The collection system does not have any metal parts in contact with the water so that only metallic elements delivered by the beam will remain in the water. The group from Washington University in St. Louis developed chemical processing schemes to purify the various elements, removing all the unwanted activities that might be present, and to chemically attach the collected radioisotopes to biological molecules for testing. The next step in this work is the construction of a new system to collect long-lived isotopes from the cooling water in the NSCL A1900 beam blocker. The beam blocker is at the exit of the first large bending magnet of the fragment separator and is often used to intercept the unused primary beam.

*Recent Results:* The first experiments produced and extracted the relatively easy isotope $^{24}Na$. Subsequently $^{67}Cu$ was extracted from a relatively pure sample and then this isotope was extracted from a very contaminated sample similar to what would be present in the NSCL and FRIB beam dumps. The $^{67}Cu$ was used to create a radioactive antibody that was injected into mice and the distribution of the activity in different biological materials was determined.

## References

1. Design and construction of a water target system for harvesting radioisotopes at the National Superconducting Cyclotron Laboratory, A. Pen, *et al*., Nucl. Instrum. Meth. A 747**,** 62 (2014).
2. Feasibility of Isotope Harvesting at a Projectile Fragmentation Facility: $^{67}Cu$, T. Mastren, *et al*., Nature/Scientific Research **4,** 6706 (2014).

# Appendix D.11: University of Missouri, MURR Research Reactor

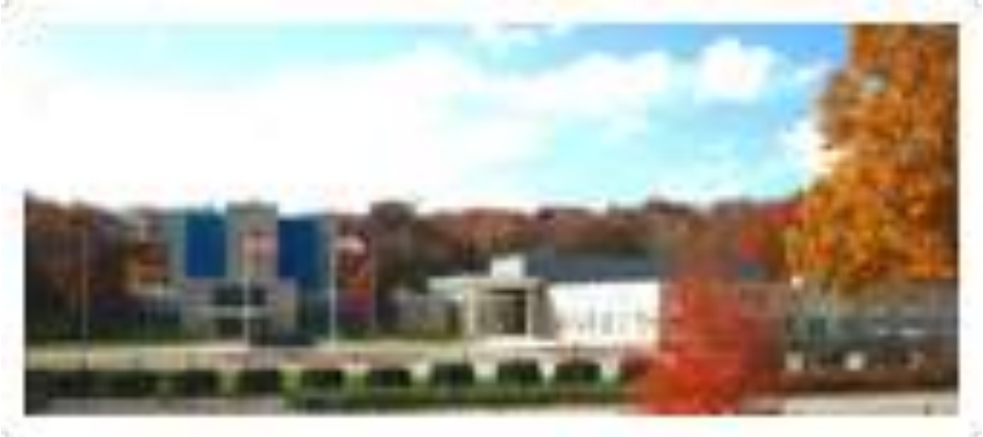

| |
|---|
| **General Description:** Multi-disciplinary, university-based research and educational reactor and cyclotron facility |
| **Beams:** Thermal-fast (fission) neutrons peak flux @6.5x$10^{14}$ n/cm$^2$s, 16MeV protons<br>Reactor power: 10MW(thermal) |
| **Additional Capabilities:** Neutron source-field irradiation, gamma-ray spectroscopy |
| **Research Focus:** Medical/industrial/research isotope production (DoE funded, Private industry funded), neutron scattering, metrology (elemental analysis, INAA) |
| **Contact person:** David Robertson, Les Foyto (see below) |

*Prepared by Nickie Peters*

The University of Missouri Research Reactor (MURR) is a multi-disciplinary research and educational facility providing a broad range of analytical, materials science and irradiation services to the research community and the commercial sector. Scientific programs include research in archaeometry, epidemiology, health physics, human and animal nutrition, nuclear medicine, radiation effects, radioisotope studies, radiotherapy, boron neutron capture therapy and nuclear engineering; and research techniques including neutron activation analysis, neutron and gamma-ray scattering and neutron interferometry. The MURR is situated on a 7.5-acre lot in the central portion of the University Research Commons, an 84-acre tract of land approximately one mile (1.6 km) southwest of the University of Missouri at Columbia's main campus (see Figure 22). The heart of this facility is a pressurized, reflected, open pool-type, light water moderated and cooled, heterogeneous reactor designed for operation at a maximum steady-state power level of 10 Megawatts thermal (see Figure 23) – the highest-powered university-operated research reactor in the United States.

## General Information

- 10 MW research reactor
- Operates 24 hours a day, seven days a week, 52 weeks a year
    - Uniquely operates on 52 weeks per year at full power
- Peak neutron flux: 6.5 x $10^{14}$ n/cm$^2$s
- 16 MeV cyclotron and laboratories

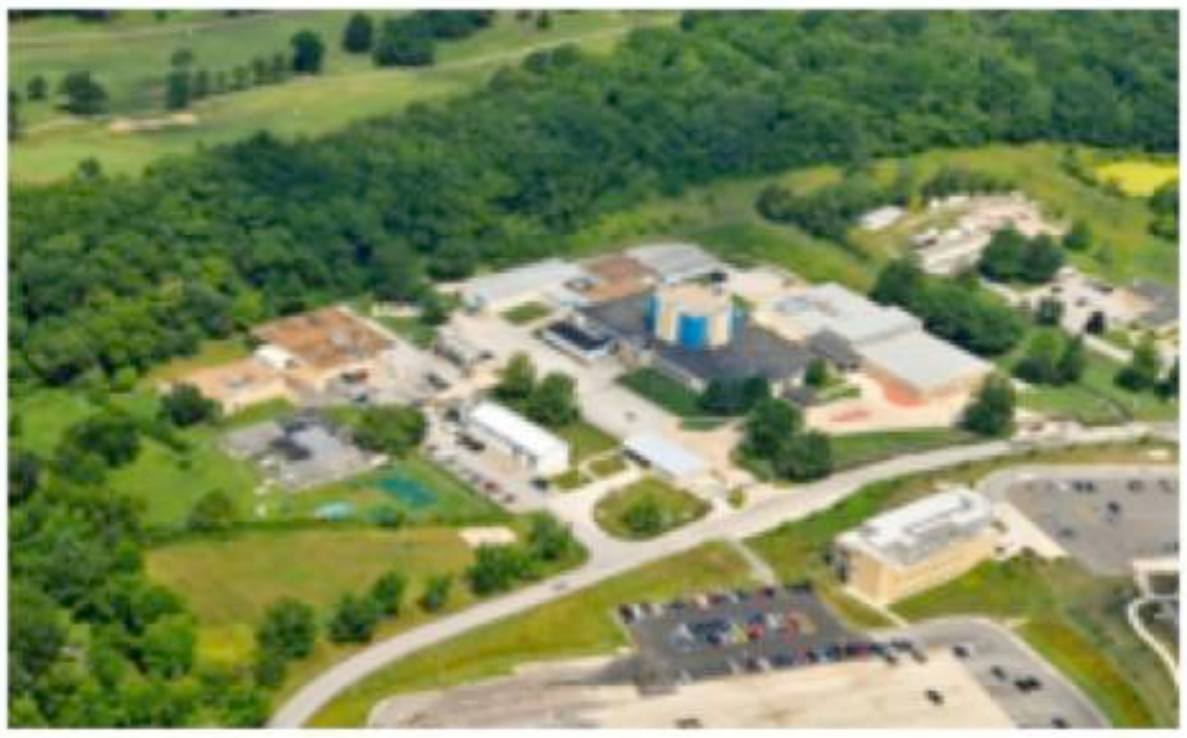

**Figure 22.** General layout of the MURR facilities

## MURR Experimental Layout and Description of Facility Utilization

The MURR has six types of experimental facilities designed to provide these services: the Center Test Hole (Flux Trap); the Pneumatic Tube System; the Graphite Reflector Region; the Bulk Pool Area; and the (six) Beamports. The first four types provide areas for the placement of sample holders or carriers in different regions of the reactor core assembly for the purpose of material irradiation. Some of the material irradiation services include transmutation doping of silicon, isotope production for the development of radiopharmaceuticals and other life-science research, and neutron activation analysis. The six beamports channel neutron radiation from the reactor core to experimental equipment, which is used primarily to determine the structure of solids and liquids through neutron scattering. The layout is depicted in Figure 24.

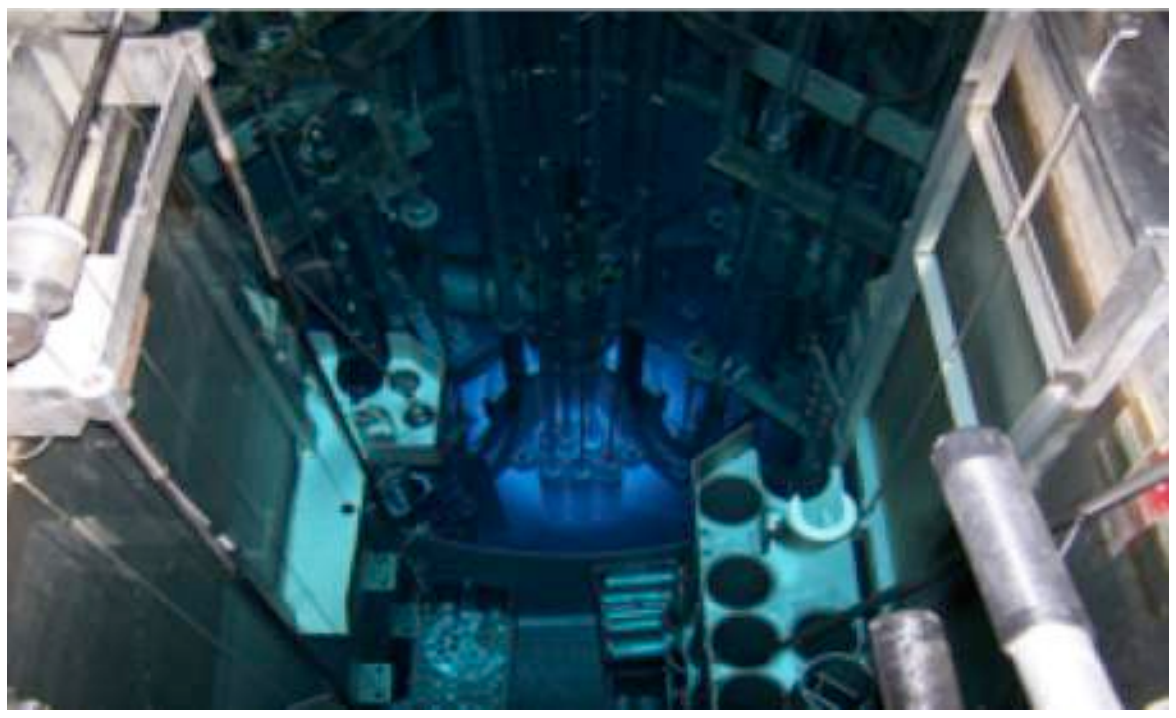

**Figure 23.** The MURR core

## MURR Research Activities and Concern for Improved Nuclear Data

Table 5 shows the list of nuclide with applications in medicine and material science industrial that were produced at MURR in 2014. Many of these important nuclides are lacking in their current nuclear data files, which hinder their production optimization. Each and every week MURR supplies the active ingredients for FDA-approved Quadramet® and TheraSpheres®; Ceretec™ (with Tc-99m), a diagnostic used to evaluate cerebral blood flow in patients and label white blood cells to identify infection; Quadramet® (with Sm-153), a therapeutic for easing pain associated with metastatic bone cancer; TheraSphere® (with Y-90), a glass microsphere used to treat patients with inoperable liver cancer.

**Table 5.** List of Nuclides Produced at MURR in 2014

| Isotopes Produced in 2014 | | |
|---|---|---|
| Au-198 | Ir-192 | Sb-122 |
| Au-199 | Kr-79 | Sb-124 |
| Ba-131 | Mo-99 | Sc-46 |
| Ca-45 | Na-24 | Se-75 |
| Cd-115 | P-32 | Sm-153 |
| Ce-141 | P-33 | Sn-117m |
| Co-60 | Pd-109 | Sr-89 |
| Cr-51 | Po-210 | W-181 |
| Cu-64 | Rb-86 | Y-90 |
| Fe-59 | Re-186 | Yb-169 |
| Lu-177 | Ru-103 | Zn-65 |
| Hg-203 | S-35 | Zr-95 |

Specifically, MURR isotope production research activities includes: Carrier free lanthanides indirect production (Lu-177, Pm-149, Ho-166) – a DOE Advanced Nuclear Medicine Initiative and Electromagnetic isotope separation (Sm-153); Mo-99 (n, gamma) production for novel generator technologies industry partnership with Northstar and fission production with uranium recycle - industry partnership with Northwest Medical Isotopes; Rh-105 carrier free from uranium fission using selective gas extraction - subcontract with General Atomics/DOE Isotope Program; Re-186 accelerator production and separations for high specific activity - DOE Isotope Program;As-72, As-77, Cu-67 production of high specific activity with target recycle- DOE Isotope

Program; Po-210 production and incorporation into nuclear batteries - private industry; Os-191 production and incorporation into device- industry partnership with CheckCap.

Complementary set of proton-rich isotopes for area medical facilities and researchers, such as F-18 (FDG) for PET scans and F-18 for clinical trials of new imaging agents, and Cu-64 for radiopharmaceutical research are produced in the 16 MeV cyclotron laboratories.

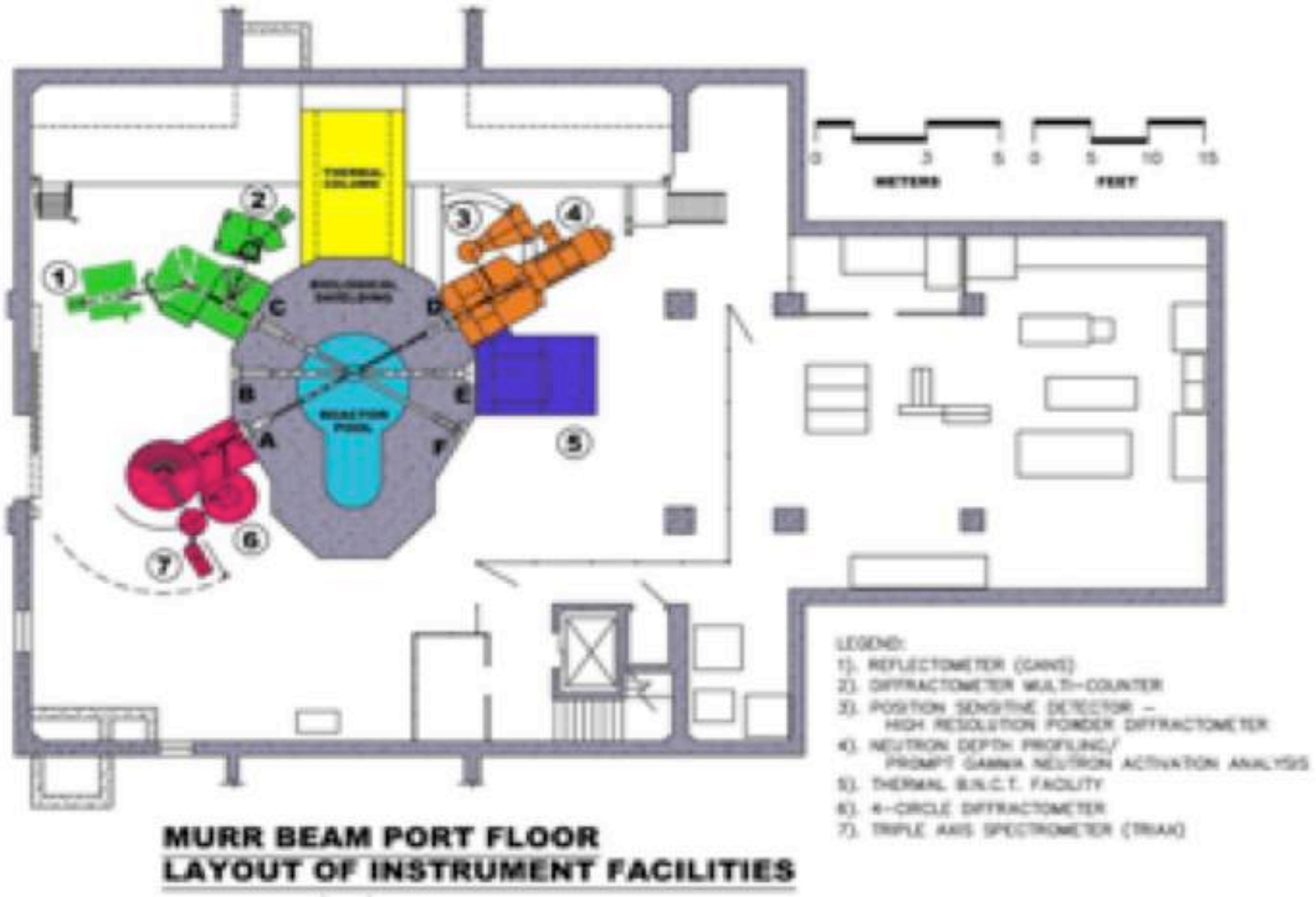

**Figure 24.** MURR Neutron Beam Port Floor layout

**Contacts: Les Foyto Tel: (573) 882-5276 e-mail:foytol@missouri.edu**

**David Robertson Tel: (573) 882-2240 e-mail:robertsonjo@missouri.edu**

**Nickie Peters Tel (573) 884-9561 e-mail:petersnj@missouri.edu**

# References

1. Perez, Pedro B. (2000). “University Research Reactors: Contributing to the National Scientific and Engineering Infracstructure from 1953 to 2000 and Beyond. National Organization of Test, Research and Training Reactors. http://www.trtr.org/links/trtr_february.html.
2. http://nsei.missouri.edu/ Nuclear Science and Engineering Institute
3. "http://murr.missouri.edu/operations.php". Retrieved 8 April 2012.
4. http://web.missouri.edu/~umcreactorweb/pages/rnr_milestones.pdf
5. Ralph Butler. "University of Missouri Research Reactor (MURR) License Renewal Experience" (PDF).
6. "http://archaeometry.missouri.edu/naa_applications.html" Retrieved 8 February 2014

# Appendix D.12: Notre Dame University, Nuclear Science Laboratory

**General Description:** University based accelerator laboratory

**Accelerators:**

- 10 MV Tandem Pelletron
- 5 MV 5U single ended Pelletron
- 3MV Tandem Pelletron (to be installed)
- TwinSol radioactive beam device

**Beams:** *Protons, alphas, and heavy ions. Light radioactive ions A<20 can be produced by the TwinSol facility*: Beams can be produced over a wide energy range at the FN tandem with terminal voltage up to 10MV. The typical beam intensities are in the µAmp range for protons and alpha particles, but lower for heavy ions. The 5U accelerator is equipped with a Nanogan ECR source capable of production of beams in higher ionization states. Typical beam intensities range in the ten to hundred µAmps.

**Experimental focus:** low energy nuclear reaction studies for nuclear astrophysics, nuclear structure physics, PIXE and PIGE material analysis, nuclear reaction studies for isotope production, activation and decay studies for nuclear astrophysics with application potential. AMS with long lived radioisotopes up to A=60, will be extended in near future.

**Present detector array capabilities (relevant to applications):** AMS capability, Ge-gamma and 3He neutron detector arrays, Silicon particle detector array, St. George recoil separator, helicital spectrometer under construction

**Contact person:** Michael Wiescher, Wiescher.1@nd.edu

*Prepared by Michael Wiescher*

The Nuclear Science Laboratory at Notre Dame is a university based accelerator lab whose main research focus is on nuclear astrophysics, radioactive beam physics and nuclear physics applications. The operation is funded through the National Science Foundation. The facility is not funded as a user facility, but welcomes users. There is no specific PAC process, but collaboration with the NSL faculty is recommended to facilitate user support. Presently 60% of the experiments are user based efforts.

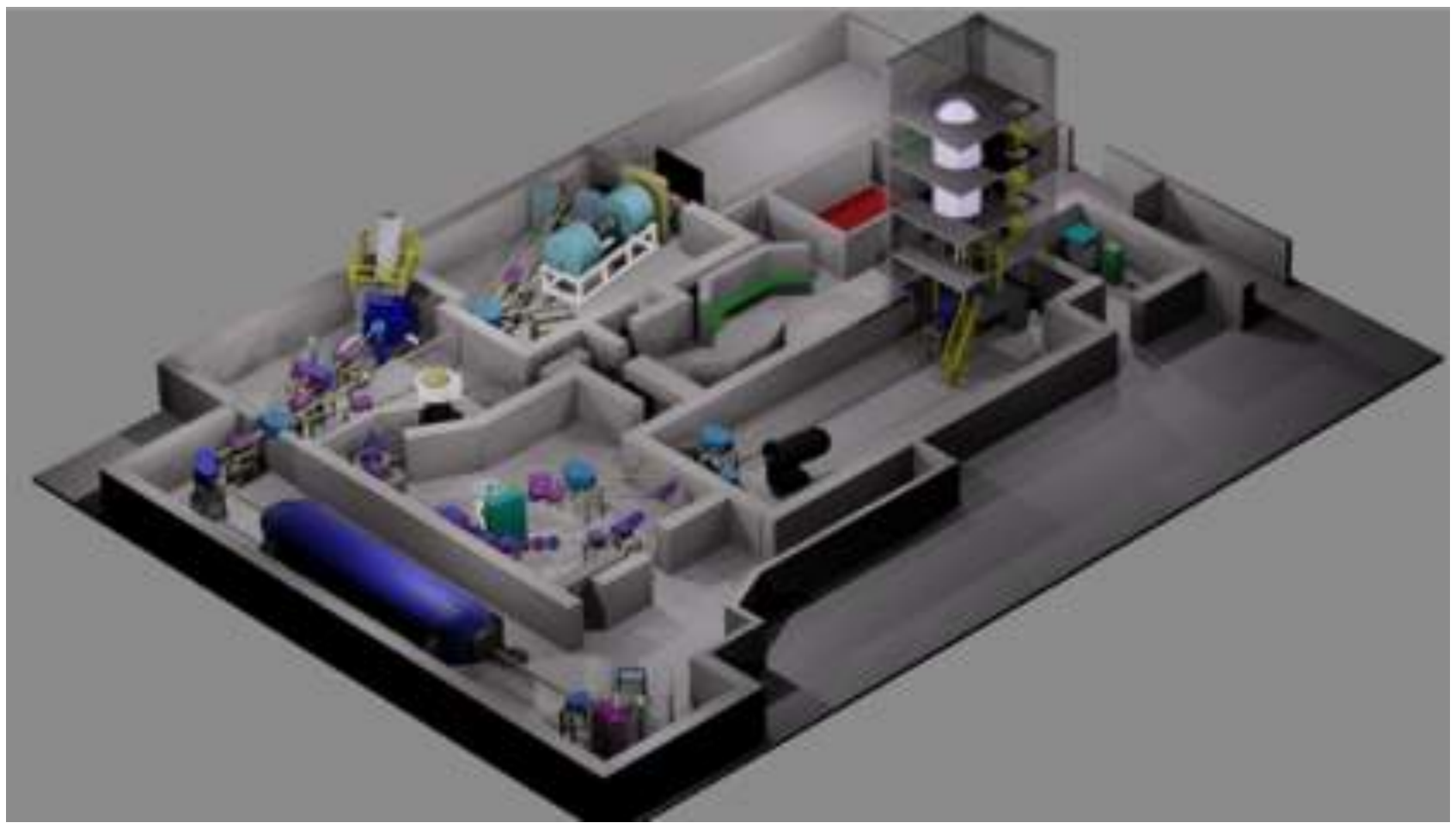

**Figure 25.** General layout of Notre Dame Nuclear Science Laboratory

The NSL operates a broad program in nuclear astrophysics, AMS physics and nuclear structure physics. The laboratory operates an FN Pelletron tandem accelerator and a high intensity 5MV single ended accelerator. Presently a 3 MV Pelletron tandem is being installed dedicated for nuclear application studies. Applications are presently focused on AMS techniques as well as on PIXE and XRF based material science applications. A new program on medical isotope studies has been formed and the purchase of a 25 MeV cyclotron is presently negotiated. The applied program will be substantially expanded in the near future with two new faculty positions. In terms of nuclear data the laboratory focuses primarily on nuclear astrophysics data such as low energy nuclear cross section measurements for stellar hydrogen, helium and carbon burning. This is complemented by nuclear reaction studies for determining nuclear reaction rates for explosive hydrogen burning environments.

# Appendix D.13: Oak Ridge National Laboratory, High Flux Isotope Reactor (HFIR)

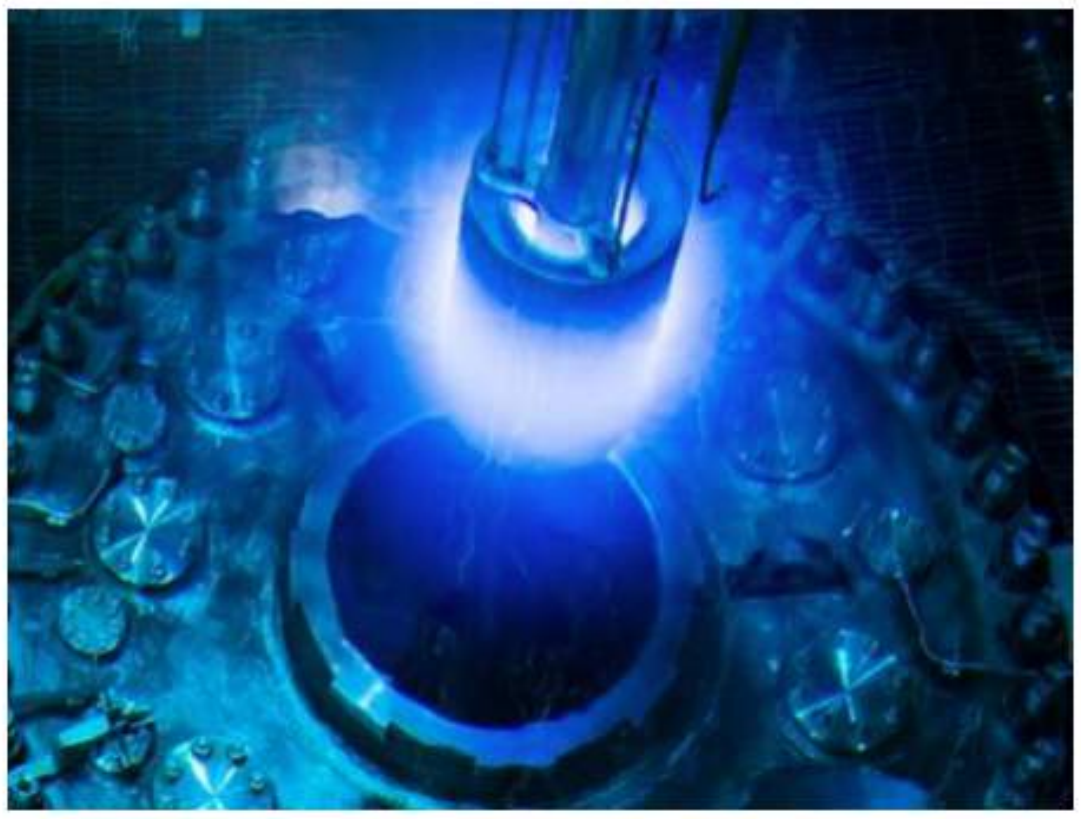

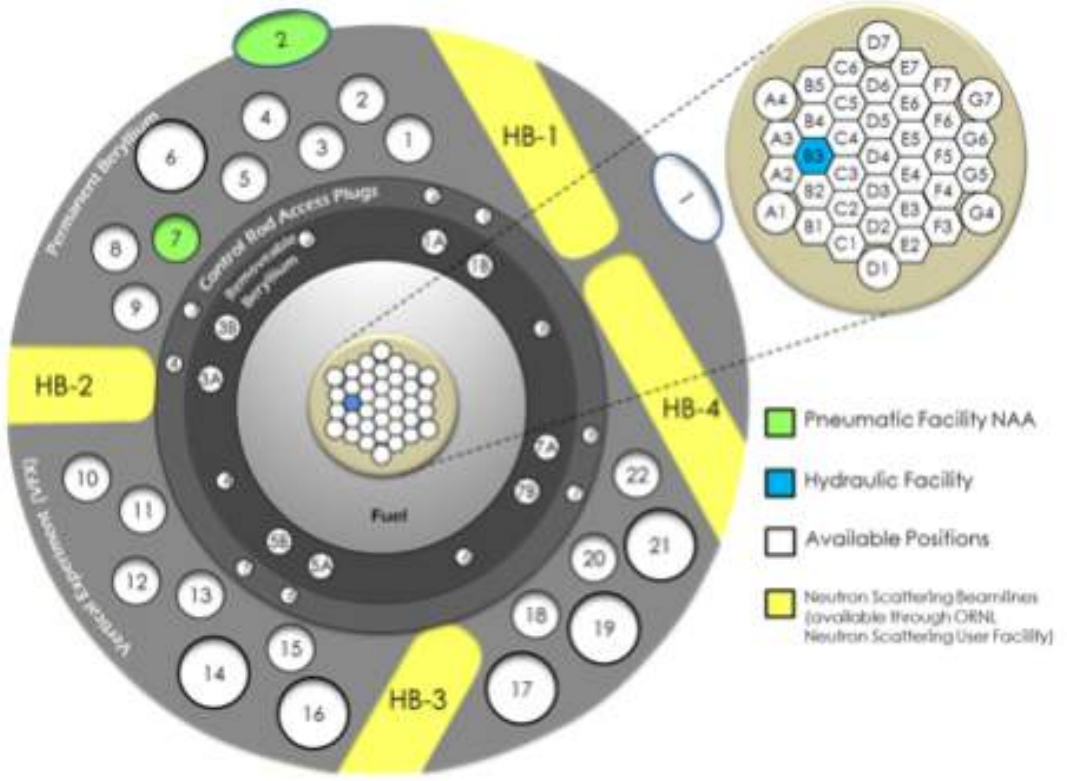

**General Description:**
85MW Research Reactor with very high neutron flux. Primary missions of

1. Neutron Scattering
2. Isotope Production
3. Materials Research
4. Nuclear Forensics

**Beams:**

- 4 Primary beamlines. (3 thermal and one cold).
- 12 Active instruments and 3 development instruments

**Additional Capabilities:**

- Isotope production/research
- Materials damage testing (neutron and gamma)
- Nuclear forensics via neutron activation analysis

**Contact persons:**
*In-vessel Irradiations:*
Chris Bryan (865.241.4336)
*Neutron Scattering User Program:*
Laura Edwards Morris (865.574.2966)
*Neutron Activation Analysis:*
David Glasgow (865.574.4918)
*Gamma Irradiations:*
Geoff Deichert (865.241.3946)

*Prepared by Chris Bryan*

The High Flux Isotope Reactor (HFIR) at Oak Ridge National Laboratory is one of the world's most powerful nuclear research reactor facilities. It is a versatile 85-MW isotope production and test reactor with the capability and facilities for performing a wide variety of irradiation experiments.

The neutron scattering research facilities at HFIR contain a world-class collection of instruments used for fundamental and applied research on the structure and dynamics of matter. HFIR is also used for medical, industrial, and research isotope production; research on neutron damage to materials; and neutron activation analysis to examine trace elements in the environment. Additionally, the building houses a gamma irradiation facility that uses spent fuel assemblies and is capable of providing high gamma doses for studies of the effects of radiation on materials.

## Neutron Scattering

Neutron scattering can provide information about the structure and properties of materials that cannot be obtained from other techniques such as X-rays or electron microscopes. There are many neutron scattering techniques, but they all involve the detection of particles after a beam of neutrons collides with a sample material. HFIR uses nuclear fission to release neutrons which are directed away from the reactor core and down four steady beams. Three of these beams use the neutrons as they are created (thermal neutrons), and one beam moderates (cools and slows) the neutrons with supercritical hydrogen, enabling the study of soft matter such as plastics and biological materials. The thermal and cold neutrons produced by HFIR are used for research in a wide array of fields of study, from fundamental physics to cancer research. The high neutron flux in HFIR produces the world's brightest neutron beams, which allow faster and higher resolution detection.

## Irradiation Materials Testing

HFIR provides a variety of in-core irradiation facilities, allowing for a wide range of materials experiments to study the effects of neutron-induced damage to materials. This research supports fusion energy and next-generation nuclear power programs, as well as extending the lifetime of the world's current nuclear power plants. HFIR has the unique ability to deliver the highest material damage in the

The HFIR Gamma Irradiation Facility is designed to expose material samples to gamma radiation using spent HFIR fuel elements. The facility offers high dose rates and custom sample environments for the most innovative research.

## Isotope Production

Isotopes play an extremely important role in the fields of nuclear medicine, homeland security, energy, defense, as well as in basic research. HFIR's high neutron flux enables the production of key isotopes that cannot be made elsewhere, such as $^{252}$Ca, $^{75}$Se, and $^{63}$Ni, among others. Additionally, HFIR will produce $^{238}$Pu, which is used to power NASA's deep space missions.

## Neutron Activation Analysis

Neutron Activation Analysis (NM) is an extremely sensitive technique used to determine the existence and quantities of major, minor and trace elements in a material sample for applications including forensic science, environmental monitoring, nonproliferation, homeland security, and fundamental research.

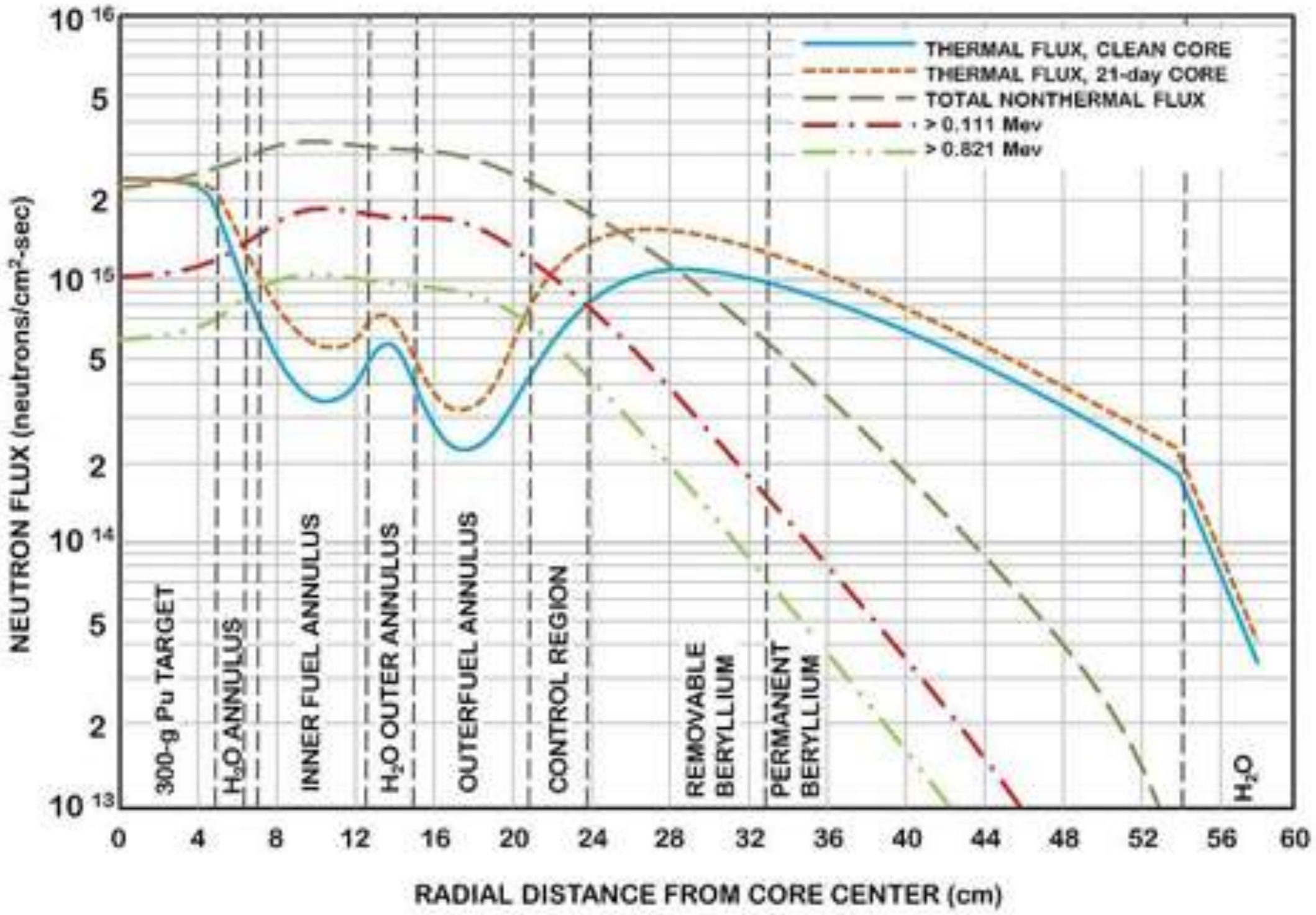

**Figure 26.** Neutron flux as a function of radial distance from the core centerline.

# Appendix D.14: Ohio University, Edwards Accelerator Laboratory

| |
|---|
| **General Description:** University-based facility. |
| **Accelerator:** High current 4.5-MV T-type Pelletron tandem |
| **Beams:** *Neutrons*: 0.5 to 25 MeV<br>*Light particles*: $^{1}$H, $^{2}$H, $^{3}$He, $^{4}$He, Li, B, C beams |
| **Present detector capabilities:** 1-7 detector arrays of NE213 scintillators of 2.5-cm-thick x17.8-cm-diameter or 5-cm-thick x12.7-cm-diameter; lithium glass scintillators; a 10-arm charged particle TOF-E Chamber; and BGO, NaI(Tl), and HPGe gamma detectors. |
| **Research Focus:** nuclear structure, nuclear astrophysics, condensed matter physics, and applied nuclear physics. |
| **Contact person:** Carl Brune (740)-593-1975<br>brune@ohio.edu |

*Prepared by T.N. Massey, C.E. Parker, and C.R. Brune*

## Overview

The Edwards Accelerator Laboratory at Ohio University (OU) was originally constructed with funds supplied by the U.S. Atomic Energy Commission and the State of Ohio. The 4.5-MV tandem van de Graaff accelerator was built and installed by the High Voltage Engineering Company, with the first experiments being performed in 1971. The accelerator has a unique T-shape configuration, with the recently installed Pelletron charging system running perpendicular to the acceleration column, which is designed to support high beam intensities. The laboratory was expanded in 1994, and now includes a vault for the accelerator, two target rooms, a control room, a chemistry room, an electronics shop, an undergraduate teaching laboratory, and offices for students, staff, and faculty.

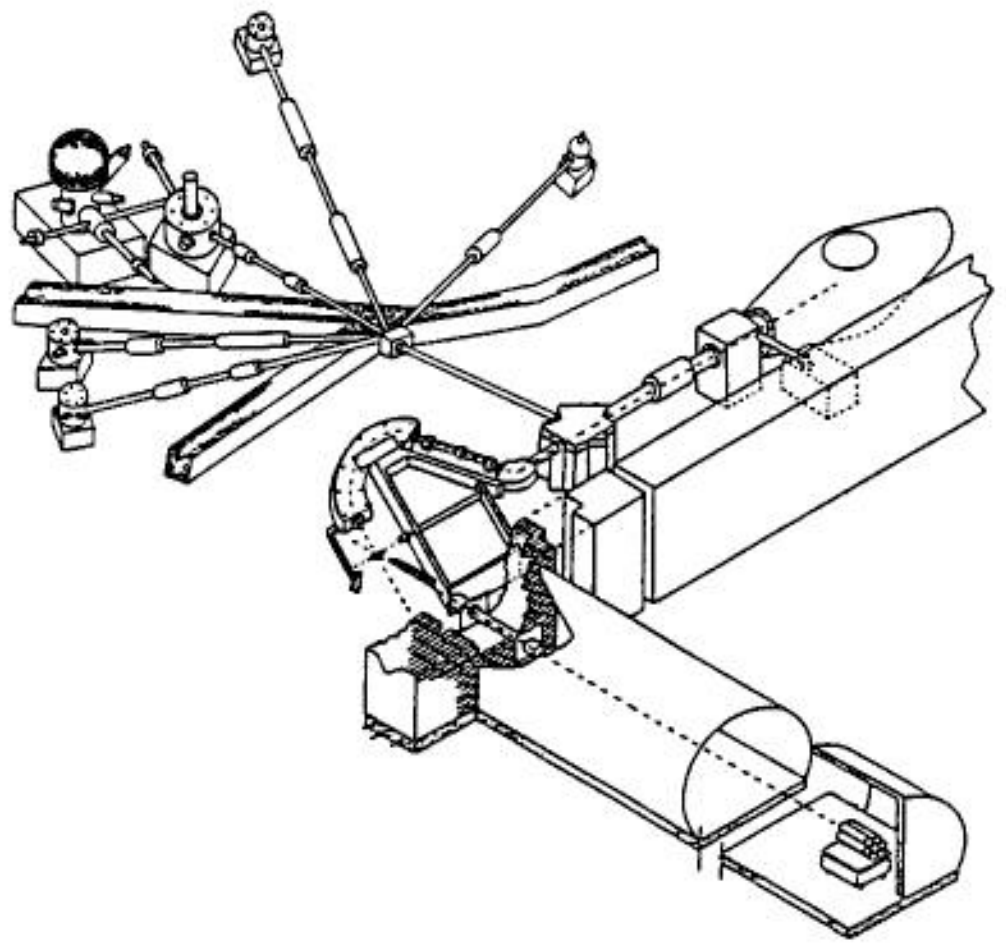

**Figure 27.** Edwards Accelerator Lab Layout.

The 4.5-MV tandem accelerator and beamlines are shown in Figure 27. This machine is presently equipped with a Cs sputter ion source that is used for the production of proton, deuteron, lithium, boron, and carbon beams. The typical maximum beam current available on target for proton and deuteron beams is 10 μA. A duoplasmatron charge-exchange ion source is available for producing $^{3}$He and $^{4}$He beams. For these beams, the typical maximum beam current available on target is 0.5 μA. Pulsing and bunching equipment are capable of achieving 1 ns bursts for proton and deuteron beams, 2.5 ns bursts for $^{3,4}$He beams, and 3 ns bursts for $^{6,7}$Li. The 5 MHz fundamental frequency of our pulsing system leads to 200 ns between pulses. The time between pulses can be increased by discarding pulses using an electronic chopper.

The Edwards Accelerator Laboratory is a unique national facility. The combination of continuous and monoenergetic neutrons together with a well-shielded 30 meter flight path does not exist anywhere else in North America. The beam swinger facility is described in Ref. [1]. This combination of equipment permits measurements with high precision and low background. Several types of neutron detectors are available, including lithium glass, NE213, and fission chambers. The laboratory has the licenses and equipment necessary for utilizing both solid and gaseous tritium targets

## Outside Users

Several groups visit the laboratory each year to conduct experiments. Many outside groups utilize our unique neutron time-of-flight capabilities. The arrangements with outside users may or may not be collaborative. In some cases, outside users may pay for beam time.

## Specific Neutron Sources

The laboratory has both monoenergetic and “white" neutron sources available for measurements and detector calibrations. The available reactions utilizing gas cells include $^{3}$H(p,n), $^{2}$H(d,n), $^{3}$H(d,n), $^{15}$N(p,n), and $^{15}$N(d,n). In total these will cover the energy range of 0.5 to 24 MeV with our available proton and deuteron energies. We also have the capability to rapidly cycle between two gas cells, a technique that is very useful for measuring backgrounds [2]. A summary of the neutron production for these reactions is shown in Figure 28.

For some applications, solid targets are desirable. Lower-energy neutrons can be produced by utilizing the (p,n) reaction on thin metallic $^{7}$Li or titanium tritide targets. We also commonly produce ~15-MeV neutrons via the $^{3}$H(d,n) reaction by bombarding a solid titanium tritide stopping target with a 500-keV deuteron beam (the practical low-energy limit of our accelerator). This configuration generates 2.4 x $10^{7}$ n/sr/μA/s neutrons. In this case, typical beam currents are 1-3 μA, with the beam current being limited by the transmission of the accelerator, which is not optimized for such low-energy beams. We have produced beams up to 25 MeV using a solid tritium target.

## Detector Calibration

For calibration of detectors with a “white" source and the time-of-fight technique, a standard has been developed: neutrons at 120° from the 7.44-MeV deuteron bombardment of a thick aluminum target [3]. This standard has been measured relative to the primary standard of $^{235}$U fission. We also have a low-mass $^{252}$Cf fission chamber that is available for neutron detector calibration [4]. The shape of the neutron energy spectrum is known to 1-2% accuracy from 1 to 8 MeV neutron energy [5].

## Gamma and Charged Particle Capabilities

Gamma-ray detection equipment includes HPGe, BGO, LaBr, and NaI detectors. Charged-particle detection equipment includes a scattering chamber optimized for Rutherford Backscattering and another chamber for time-of-flight measurements with flight paths of up to 2 m. The W.M. Keck Thin Film Analysis Facility consists of an integrated set of UHV chambers that includes PVD and CVD deposition facilities with MeV ion beam analysis (RBS, NRA, ERS, channeling), LEED, and electron spectroscopy (Auger, XPS, UPS).

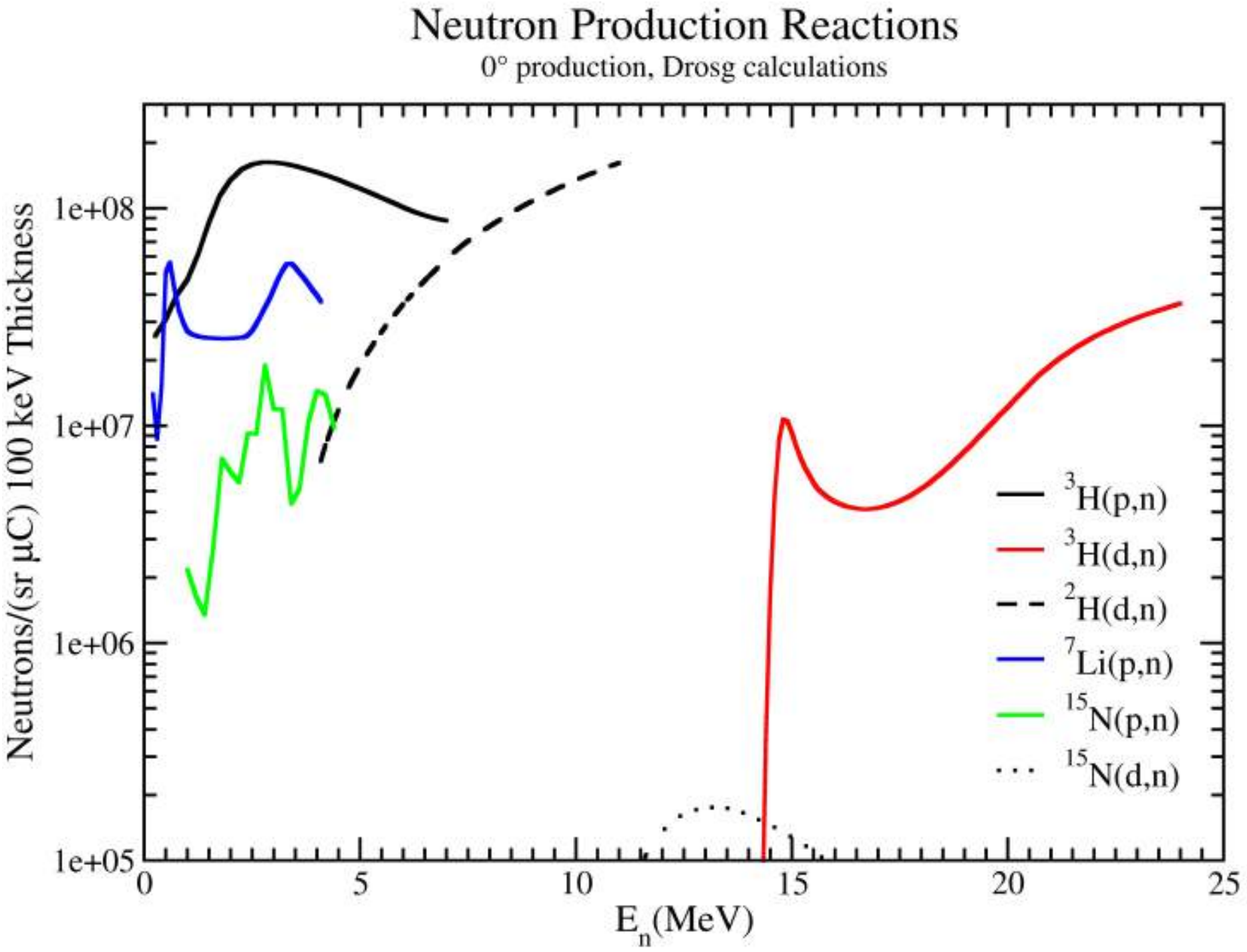

**Figure 28.** The neutron yield of various neutron production reactions is shown based on a thickness equivalent to a 100 keV energy loss in the target. The calculation is for 100 keV energy loss in the pure gas, except for $^{7}$Li, where it is from the pure metal.

## References

1. R. W. Finlay, C. E. Brient, D. E. Carter, A. Marcinkowski, S. Mellema, G. Randers-Pehrson and J. Rapaport, Nucl. Instrum. Methods **198**, 197 (1982).
2. S. M. Grimes, P. Grabmayr, R. W. Finlay, S. L. Graham, G. Randers-Pehrson, and J. Rapaport, Nucl. Instrum. Methods **203**, 269 (1982).
3. T. N. Massey, S. Al-Quraishi, C. E. Brient, J. F. Guillemette, S. M. Grimes, D. K. Jacobs, J. E.
4. O'Donnell, J. Oldendick, and R. Wheeler, Nucl. Sci. Eng. **129**, 175 (1998).
5. N. V. Kornilov, I. Fabry, S. Oberstedt, F.-J. Hambsch, Nucl. Instrum. Methods A **599**, 226 (2009).
6. W. Mannhart, “Status of the Evaluation of the Neutron Spectrum of $^{252}$Cf(sf)," IAEA Consultants' Meeting, 13-15 October 2008.

# Appendix D.15: Rensselaer Polytechnic University, Gaerttner Linear Accelerator Laboratory

| |
|---|
| **General Description**: University based center that specializes in measurements of electron, photon, and neutron induced reactions.<br><br>The center is equipped with variety of neutron production targets and detector setups. The center supports external users for a fee. |
| **Accelerator:** Electron LINAC Electron beam energy: 9-60 MeV Pulse width: 5-5000 ns Repetition rate: 1-400 Hz |
| **Beams:** Neutron beams delivered to several flight path stations from 15-250 m. |
| **Experimental**: Neutron induced reactions; photon induced reactions, medical isotope production research, and radiation damage to electronics |
| **Detectors**: For setups of neutron transmission, capture, scattering, fission<br><br>Includes: organic and inorganic scintillators, ionization chamber, fission chambers, and solid-state detectors. |
| **Contact person:** Prof. Yaron Danon, Gaerttner Linear Accelerator Center, Rensselaer Polytechnic Institute, Troy, NY 12180<br><br>Email: danony@rpi.edu |

*Prepared by Yaron Danon*

The Gaerttner LINAC Center uses a 60 MeV LINAC to produce short pulses of electrons which are used to produce photons and neutrons. Over the years the facility has been used for a range of research topics, including electron, photon, and neutron interactions, neutron photoproduction, medical isotopes, radiation damage, and applied radiation applications.

The principal research focus is on nuclear data, primarily related to neutron interactions. The experimental setup is very flexible, providing multiple setups of neutron production targets and neutron detectors, which are designed to optimize a variety of experiments. More information on the facility and examples are available in references [1] and [2].

The motivation of the nuclear data research is applications in nuclear power generation and criticality safety. The LINAC target room has a large space that enables experiments in proximity to the neutron production target as illustrated in Figure 29. To cover the wide range of neutron energies found in nuclear reactor and other criticality applications, measurement capabilities from thermal to 20 MeV were developed with a focus on the resonance region. The measurement capability matrix is shown in Figure 30 as a function of incident neutron energy.

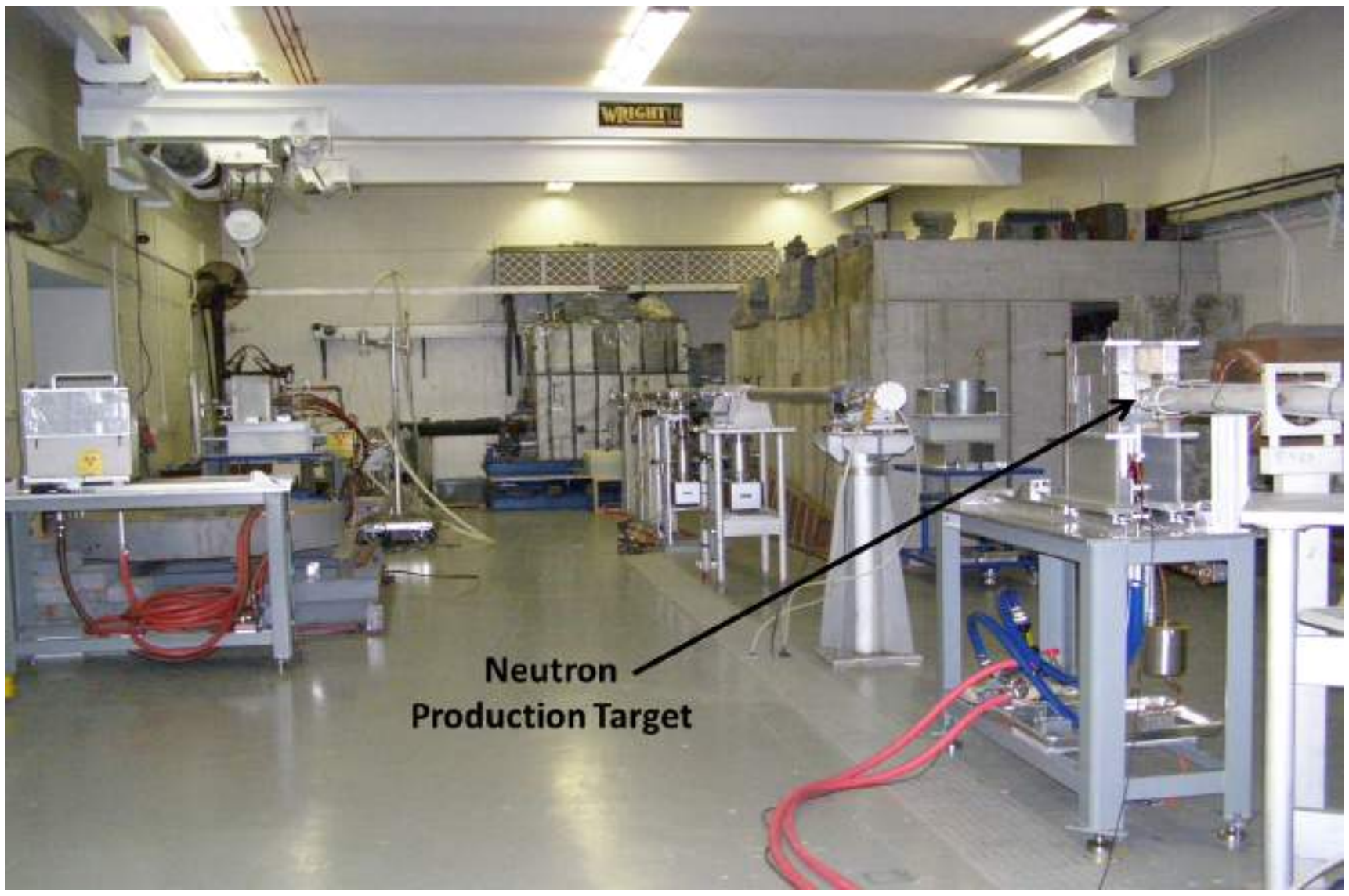

**Figure 29.** The spacious LINAC target room.

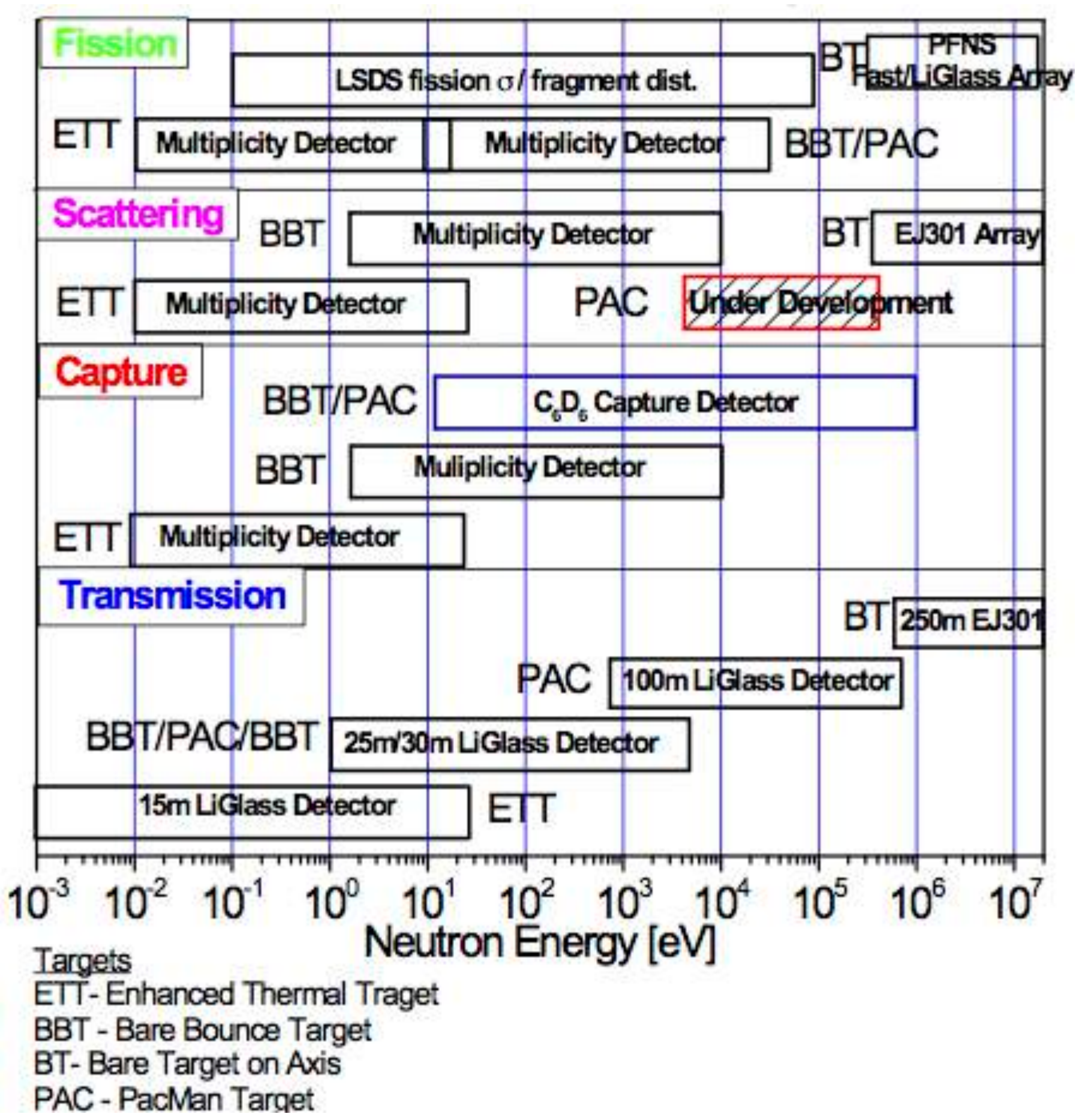

**Figure 30.** The Gaerttner LINAC Center capability matrix for neutron induced reactions measurements. A keV neutron scattering system is under development.

Below we provide short descriptions of current experimental setups.

## Neutron Transmission

Neutron transmission experiments include several setups located at different flight path stations, which use different combinations of neutron production targets and detector types to optimize the measurements for a given incident neutron energy range.

***Thermal neutron transmission*** (0.001-20 eV) uses a Li-Glass detector located at 15 m flight path and neutron production from the Enhanced Thermal Target.

***Epithermal neutron transmission*** (1 eV-10 keV) uses a Li-Glass detector located at 35 m flight path station and neutron production from the Bare Bounce Target.

***Mid Energy neutron transmission*** (5 keV -1 MeV) uses an array of Li-Glass detectors located at the 100 m flight path station and neutron production from the Pacman Target.

***High energy neutron transmission*** (0.4-20 MeV), uses an array of liquid scintillators located at 250 m flight path station and neutron production from the Bare Target.

## Neutron Capture

Currently there are two time-of-flight setups for neutron capture measurements:

***Low and epithermal energy neutron capture*** (0.01 eV – 3 keV), uses the neutron multiplicity detector; an array of 16 NaI gamma detectors surrounding the sample. Located at a 25 m flight path and uses the Enhanced Thermal Target or the Bare Bounce Target.

***Mid energy neutron capture*** (1 eV – 2 MeV) an array of 4 $C_6D_6$ liquid scintillator gamma detectors designed to measure gammas from neutron capture for incident neutron energy in the keV region where neutron scattering reactions dominated. The array is located at a 45 m flight station and uses the Pacman neutron production target.

Figure 31 below is an example of transmission and capture measurements of Re used to generate new resonance parameters [3]. The data were measured using an experimental setup for the thermal region for both transmission and capture measurements.

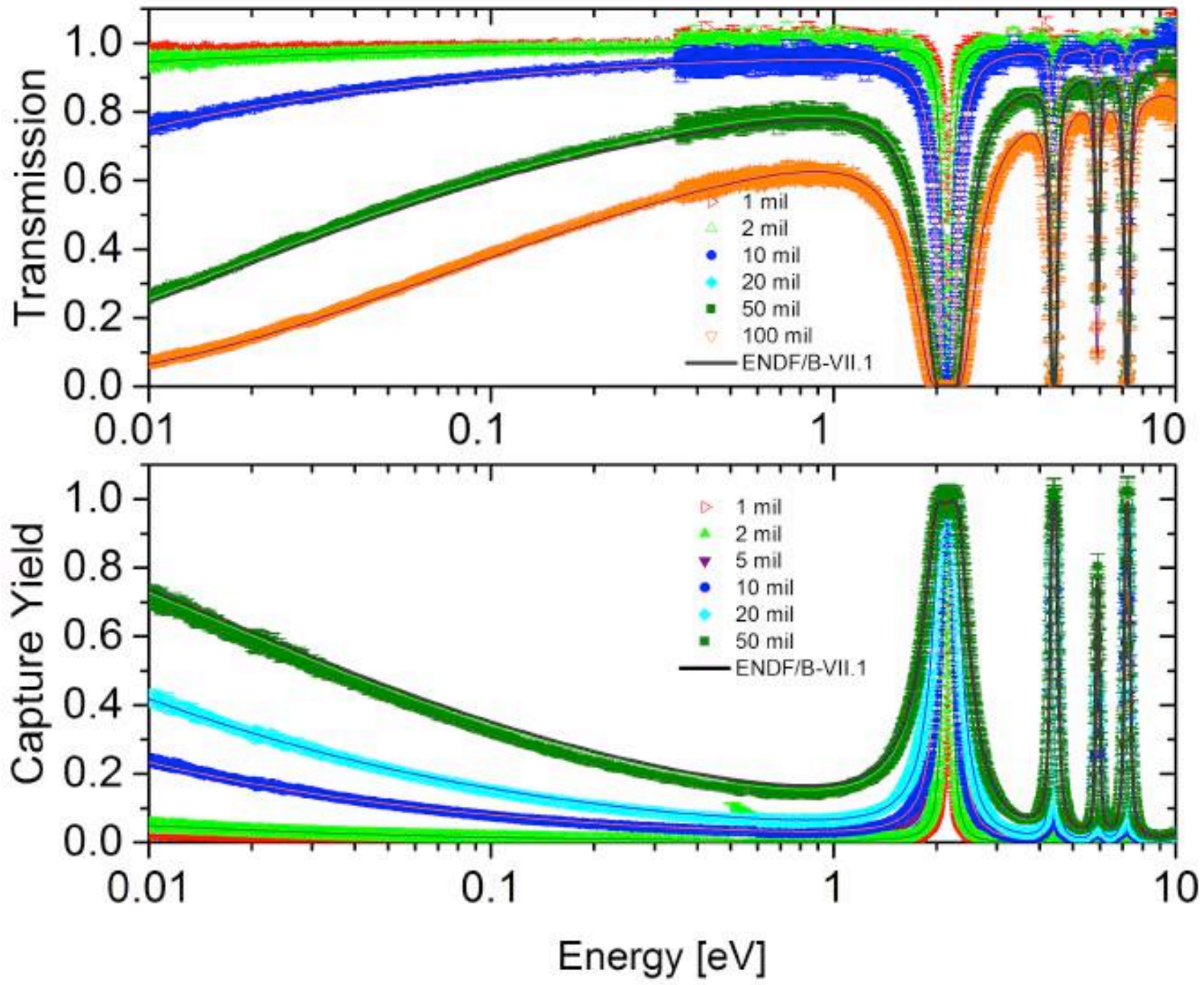

**Figure 31.** An example of resonance transmission (top) and capture (bottom) measurements on Re. The plot also includes curves generated from fitted resonance parameters and the ENDF/B-VII.1 evaluation [3].

## Fast Neutron Scattering

An array of 8 liquid scintillators located at a 30 m flight path station is used for neutron detection. The Bare Target is used for neutron production. The setup is designed to measure neutron scattering in the energy range from 0.5-20 MeV. The detectors use pulse shape analysis to identify photons.

An example of measured neutron backscatter from a $^{238}U$ sample [4] is shown in Figure 32. This measurement benchmarks both the scattering cross section and the angular distribution.

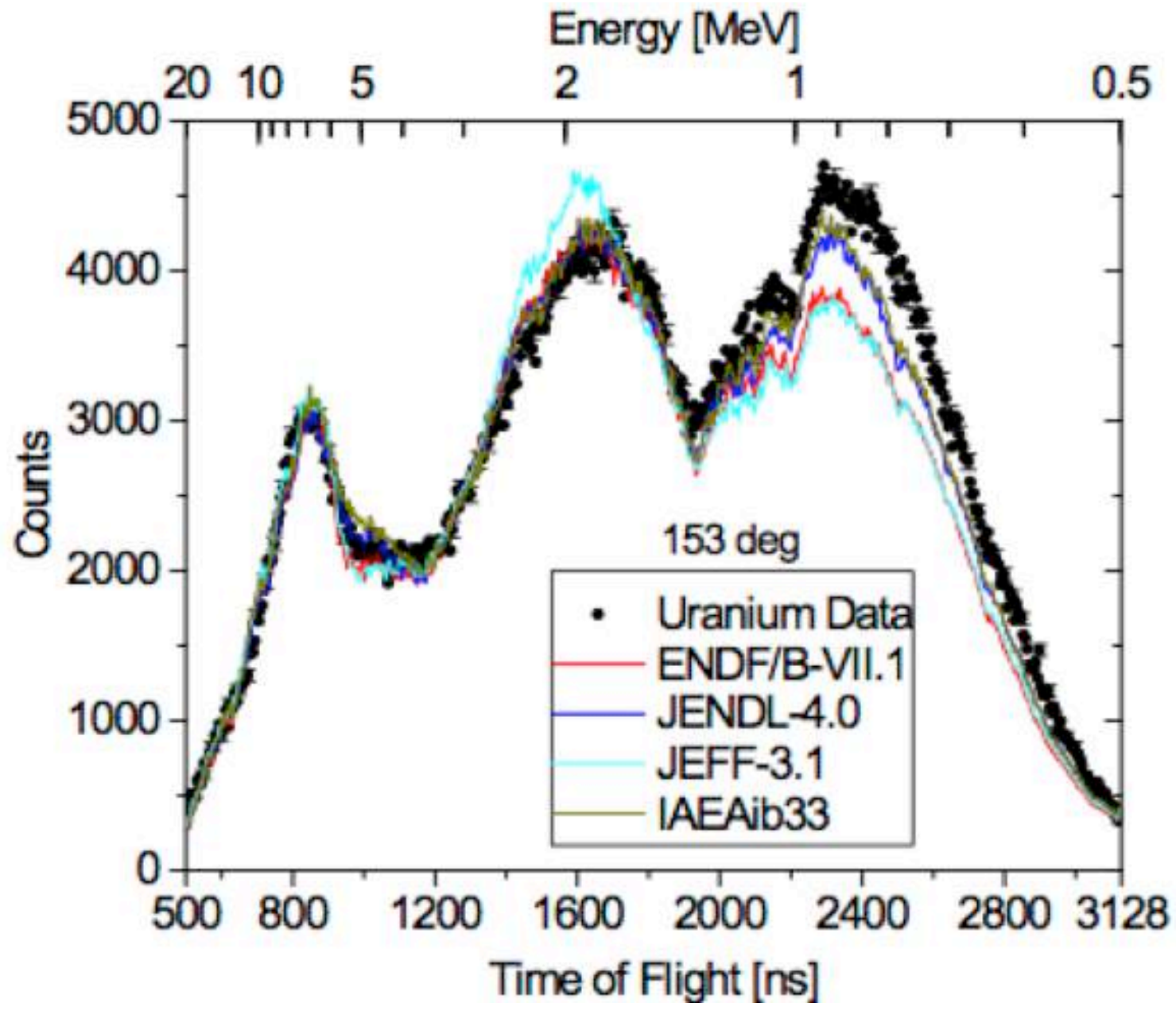

**Figure 32.** Comparison of experimental data and evaluations for neutron scattering to 153° from a $^{238}$U sample [4]. In this case the JENDL 4.0 and IAEA ib33 evaluations perform best.

## Prompt Fission Neutron Spectrum

This setup utilizes the scattering detector array, with the addition of plastic scintillators and 4 large $BaF_2$ gamma detectors. The gamma detectors are used to form a fission tag, enabling a double time-of-flight experiment that is used to measure the prompt fission neutron spectrum as a function of incident neutron energy.

## Lead Slowing Down Spectrometer

The Lead Slowing-Down Spectrometer (LSDS) is a unique setup, in which the LINAC pulse neutron source is located in the center of a 5.83 $m^3$ pure lead cube. The neutron slowing down process results in a very high neutron flux, which enables measurements of very small samples (nanograms) or samples with small cross sections (microbarns). The LSDS was used to measure fission cross sections, fission fragment mass and energy distributions, (n,$\alpha$) and (n,p) cross sections, and capture cross sections.

## References

1. Robert C. Block, Yaron Danon, Frank Gunsing, Robert C. Haight, "Neutron Cross Section Measurements", chapter one in Dan Cacuci (Editor), "Handbook of Nuclear Engineering", Vol. 1, Springer, ISBN: 978-0-387-98150-5, 2010.
2. Y. Danon, L. Liu, E.J. Blain, A.M. Daskalakis, B.J. McDermott, K. Ramic, C.R. Wendorff, D.P. Barry, R.C. Block, B.E. Epping, G. Leinweber, M.J. Rapp, T.J. Donovan, "Neutron Transmission, Capture, and Scattering Measurements at the Gaerttner LINAC Center", Transactions of the American Nuclear Society, Vol. 109, p. 897-900, Washington, D.C., Nov. 10–14, 2013.
3. Epping, Brian E., "Neutron Transmissions, Capture Yields, and Resonance Parameters in the Energy Range of 0.01 eV to 1 keV in Rhenium," MS Thesis, The University of Texas at Austin, Dec. 2013.
4. A.M. Daskalakis, R.M. Bahran, E.J. Blain, B.J. McDermott, S. Piela, Y. Danon, D.P. Barry, G. Leinweber, R.C. Block, M.J. Rapp, R. Capote, A. Trkov, "Quasi-differential neutron scattering from $^{238}$U from 0.5 to 20 MeV", Annals of Nuclear Energy, Vol. 73, Pages 455-464, Nov. 2014.

# Appendix D.16: Texas A&M University, Radiation Effects Facility

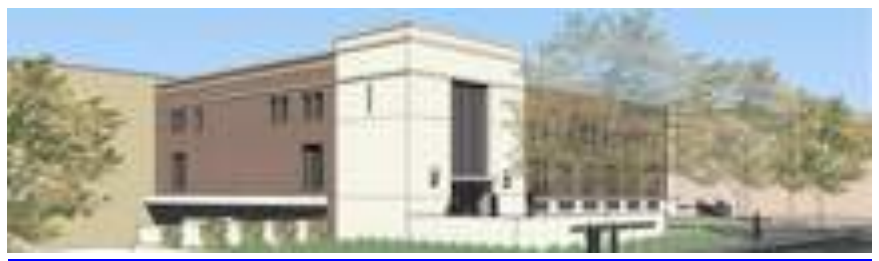

**General Description:** Heavy ions and protons for Single Event Upset (SEU) testing, detector calibration, implantations, basic nuclear physics studies, and any other application utilizing low to medium energy particle beams.

**Accelerators:** K500 Superconducting Cyclotron and K150 (88-in) Cyclotron

**Beams:** The K500 Superconducting Cyclotron produces heavy ion beams between ~3 – 80 MeV/nucleon and proton beams at 30, 40 and 55 MeV. The K150 (88-in) Cyclotron produces heavy ions from ~3 – 15 MeV/nucleon and protons from 10 – 55 MeV. For SEU testing, three beam energy series are provided: 15 MeV/nucleon (He, N, Ne, Ar, Cu, Kr, Ag, Xe, Pr, Ho, Ta, Au), 25 MeV/nucleon (He, N, Ne, Ar, Kr, Ag, Xe) and 40 MeV/nucleon (N, Ne, Ar, Kr).

**Website:** http://cyclotron.tamu.edu/ref/

**Host Location:** Cyclotron Institute, Texas A&M University, College Station, TX.

**Availability:** 24 hours/day, 7 days/week.

**Contact person:** Henry Clark; clark@comp.tamu.edu; 979-845-1411

*Prepared by Henry L Clark*

Since 1994, the Cyclotron Institute's Radiation Effects Facility has provided a convenient and low cost solution to commercial, governmental and educational agencies in need of studying, testing and simulating the effects of ionizing radiation on electronic and biological systems. Starting at just 100 hours/year at inception, the demand for beam time has grown to 3000 hours/year and has remained consistent at this level for several years.

The facility is installed on a dedicated beam line with diagnostic equipment for beam quality and complete dosimetry analysis. As a part of the Cyclotron Institute the facility is fully staffed, including electronic and machine shops that are available to assist with special customer needs. Beam time may be scheduled in 8 hours blocks either consecutively or interleaved with other testing groups.

Testing may be conducted in either our 30" diameter vacuum chamber or with our convenient in-air positioning system. Both provide precise positioning in x, y, and z as well as rotations up to 60 degrees in both theta and roll angle. Our custom-made SEUSS software carries out positioning and dosimetry.

A degrader foil system makes it possible to change beam energy without cyclotron retuning or target rotations.

Our 15 MeV/nucleon (He, N, Ne, Ar, Cu, Kr, Ag, Xe, Pr, Ho, Ta, Au) series allows testing with Linear Energy Transfer (LET) from 1 – 93 MeV/mg/$cm^2$ in Si. Our 25 MeV/nucleon (He, N, Ne, Ar, Kr, Ag, Xe) and 40 MeV/nucleon (N, Ne, Ar, Kr) series offer heavy ions for long range testing from 286 μm to 2.3 mm in Si. Typical beam time changes are between 30 minutes to 1 hour.

The beam flux is adjustable between 1E1 – 2E7 ions/$cm^2$/sec. A higher flux of 1E10 protons/$cm^2$/sec is obtainable from the K150 cyclotron. The beam spot size is selectable between 0.1 – 2 in. in diameter. Beam uniformity is typically better than 90%.

The beam uniformity and dosimetry are determined by an array of five plastic scintillators coupled to photo multiplier tubes. These scintillators are located in the diagnostic chamber adjacent to and upstream from the target area. The control software determines beam uniformity, axial gain, and beam flux (in particles/$cm^2$/sec), based on scintillator count rates. The results are displayed and updated once per second.

# Appendix D.17: Triangle Universities Nuclear Laboratory

**General Description:** Three-university facility, DOE Center of Excellence. Primary research activities range from nuclear weak interaction to nuclear strong interaction physics. The four TUNL accelerator facilities LENA, LEBAF, FN-Tandem, and HIGS provide light-ion, neutron- and gamma-ray beams covering a large range of energies and operating characteristics.

**TUNL:** TUNL is the world's most versatile facility for providing mono-energetic neutron beams for neutron-induced cross-section measurements (elastic and inelastic scattering, radiative capture, fission *etc*.) in the 1 to 30 MeV energy range.

**Beam time cost:** $150/hour.

**HIGS:** "High-Intensity Gamma-ray Source" based on Compton backscattering of FEL photons from relativistic electrons to produce monoenergetic and tunable $\gamma$-ray beams in the 1.5 to 100 MeV energy range. HIGS consists of a 160 MeV Linac, a 1.2 GeV booster synchrotron, a 1.2 GeV electron storage ring equipped with undulator magnets to provide linearly and circularly polarized FEL photons. Gamma-ray energy spread adjustable through collimation. Typical collimator size: 3/4" dia., resulting in $\Delta E/E \sim 3\%$ and between $10^7$ and $3 \times 10^8$ $\gamma$-rays per second, dependent on $\gamma$-ray energy and FEL mirror quality. Highest flux in the 10 to 15 MeV $\gamma$-ray energy range. HIGS is the world's most intense accelerator-driven $\gamma$-ray source, with $10^3$ /(eV s).

**Beam time cost:** $1000/hour.

**LENA:** Proton accelerator facility for "Low-Energy Nuclear Astrophysics" consisting of a 200 kV high-current ECR ion source and a 1 MV Van de Graaff accelerator.

**ECR:** $I_{max}$=3 mA dc and 200 μA pulsed.

**Van de Graaff:** $I_{max}$=250 μA

**Beam time cost:** $150/hour.

**LEBAF:** "Low-Energy Beam Accelerator

| |
|---|
| Facility" utilizing the Atomic Beam Polarized Ion Source for delivering polarized (or unpolarized) hydrogen or deuterium beams, which can be accelerated from 60 keV to 680 keV using a 200 kV mini-tandem in conjunction with a scattering chamber, operated at 200 kV.<br><br>$I_{max}$=50 μA of positive ions with energies between 60 and 120 keV and $I_{max}$=10 μA for negative ions at the higher energies.<br><br>**Beam time cost:** $150/hour. |
| **FN-Tandem:** 10 MV tandem accelerator with ion sources to accelerate p, d, $^3$He and $^4$He ions. Pulsed beam operation (1.5 to 3 ns time resolution) at 2.5 MHz or reduced repetition rate.<br><br>• $I_{max}$=10 μA dc and 1 μA pulsed for protons and deuterons and<br>• $I_{max}$=2 μA dc & 0.2 μA pulsed for $^3$He and $^4$He.<br>• Polarized proton and deuteron beam intensities: $I_{max}$=2 μA dc. |
| **Secondary beams:** Mono-energetic or quasi mono-energetic neutrons in the 0.1 MeV to 35 MeV neutron energy range using the reactions $^7$Li(p,n)$^7$Be, $^3$H(p,n)$^3$He, $^2$H(d,n)$^3$He and $^3$H(d,n)$^4$He with neutron fluxes up to $10^8$ n/(cm$^2$ sec) at 1 cm distance from the neutron source in dc operation and up to 3 x $10^7$ n/(cm$^2$ sec) in pulsed mode operation.<br><br>Collimated neutron beam with adjustable cross sectional area (up to 6 cm in diameter) and $10^4$ n/(cm$^2$ sec) in the 4 to 20 MeV neutron energy |
| Experiments in support of fundamental physics applications include neutron-induced background reactions relevant to neutrino-less double-beta decay and dark-matter searches. |
| **Contact person:** C.R. Howell |

*Prepared by Werner Tornow*

An important part of the applied research program is conducted in collaboration with scientists from LANL and LLNL and focuses on neutron- and gamma-ray induced reactions on actinide nuclei, especially fission and nuclear forensics, but also has a strong component in support of ongoing research to better understand the complicated physics governing the inertial confinement DT fusion plasma at the National Ignition Facility (NIF) at LLNL.

Other studies focus on plant growth under elevated $CO_2$ concentrations using $^{13}C$ as a marker, and on Rutherford backscattering measurements to identify trace elements absorbed in filters used in water treatment facilities.

Standard charged-particle and gamma-ray detectors as well as sophisticated fast neutron detectors, including the neutron time-of-flight spectrometer shown in Figure 36 are part of the detector pool available at TUNL. An Enge split-pole spectrometer is available for special applications.

Recent nuclear physics applications at the tandem laboratory included neutron-induced fission product yield measurements on $^{235}U$, $^{238}U$ and $^{239}Pu$ between 0.5 and 15 MeV, and cross section measurements involving the reactions $^{235}U(n,n'\gamma)$, $^{238}U(n,n'\gamma)$, $^{241}Am(n,2n)^{240}Am$, $^{181}Ta(n,2n)^{180}Ta$, $^{124,136}Xe(n,2n)^{123,135}Xe$ and neutron capture on a number of nuclei, including $^{124,136}Xe(n,\gamma)^{125,137}Xe$,

Recent nuclear physics applications at HIGS concentrated on $^{241}Am(\gamma,n)^{240}Am$, $^{235}U(\gamma,\gamma')^{235}U$, $^{238}U(\gamma,\gamma')^{238}U$, $^{239}Pu(\gamma,\gamma')^{239}Pu$, $^{240}Pu(\gamma,\gamma')^{240}Pu$ and $^{235}U(\gamma,f)$, $^{238}U(\gamma,f)$, and $^{239}Pu(\gamma,f)$.

## References

1. W. Tornow, Nuclear Physics News International, Vol. 11 (4), 6 (2001).
2. H.R. Weller, M.W. Ahmed, H. Gao, W. Tornow, Y.K. Wu, M. Gai, R. Miskimen, Progress in Particle and Nuclear Physic 62, 257 (2009)

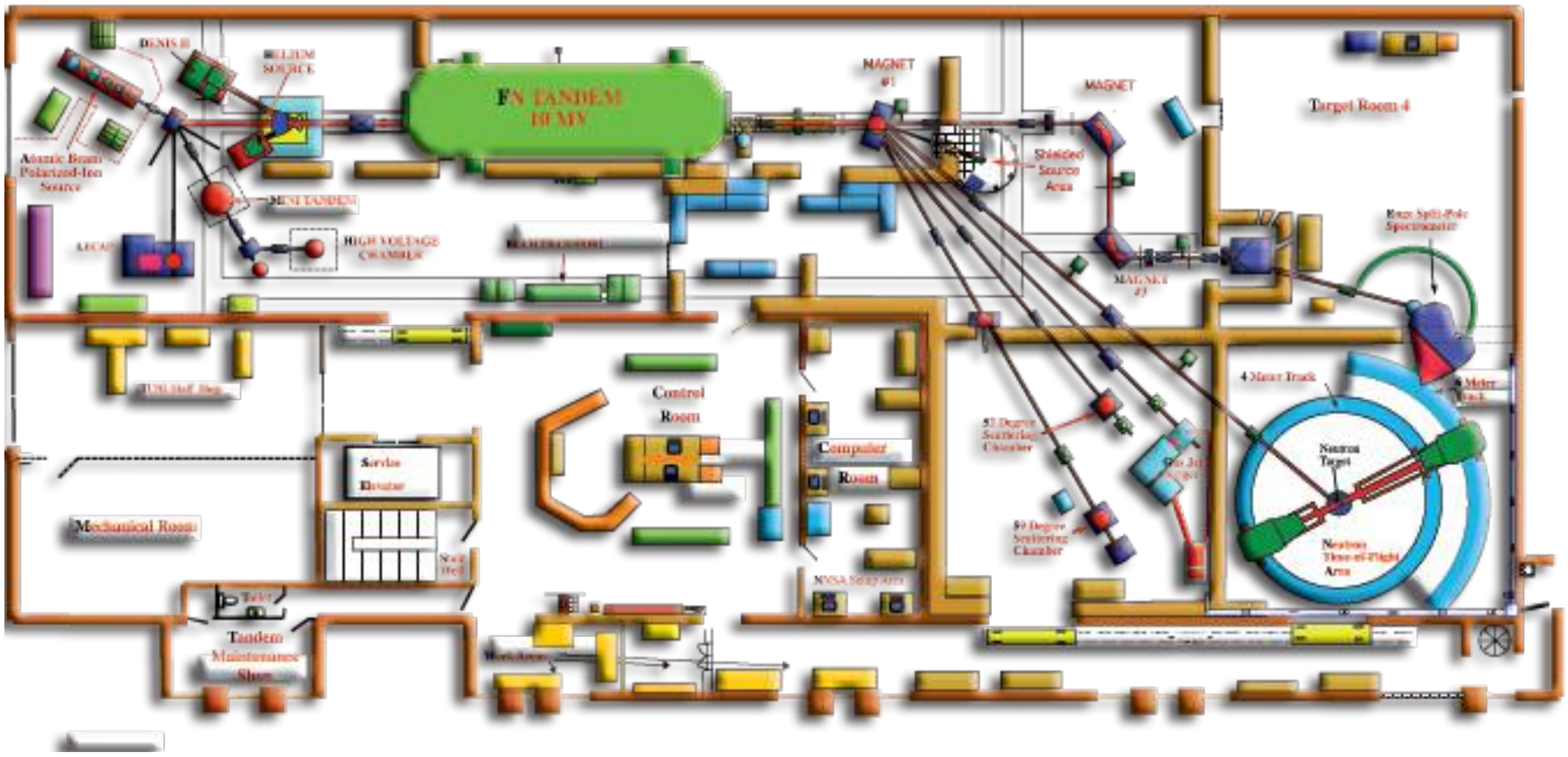

**Figure 33.** Tandem accelerator laboratory floor plan.

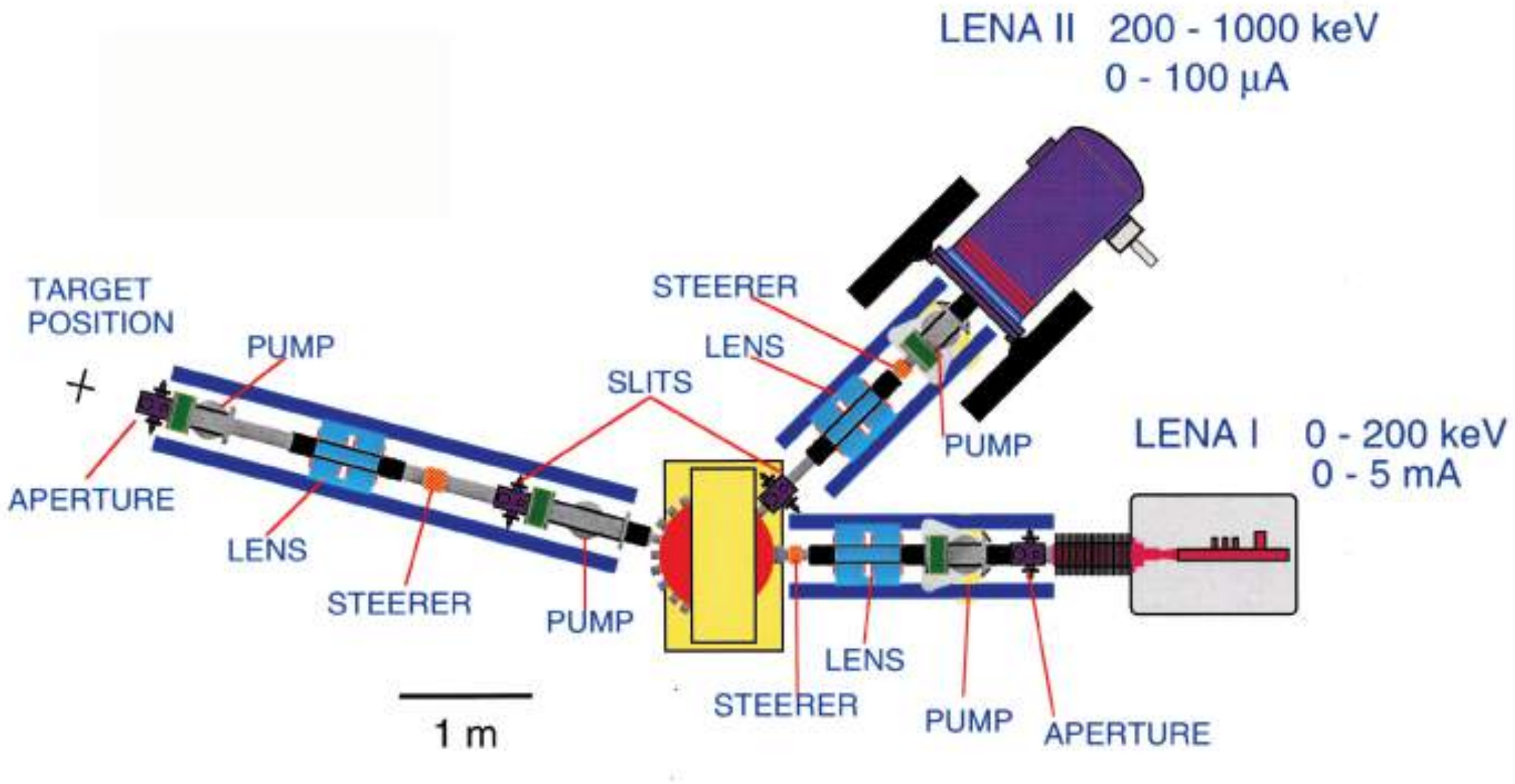

**Figure 34.** Laboratory for Nuclear Astrophysics.

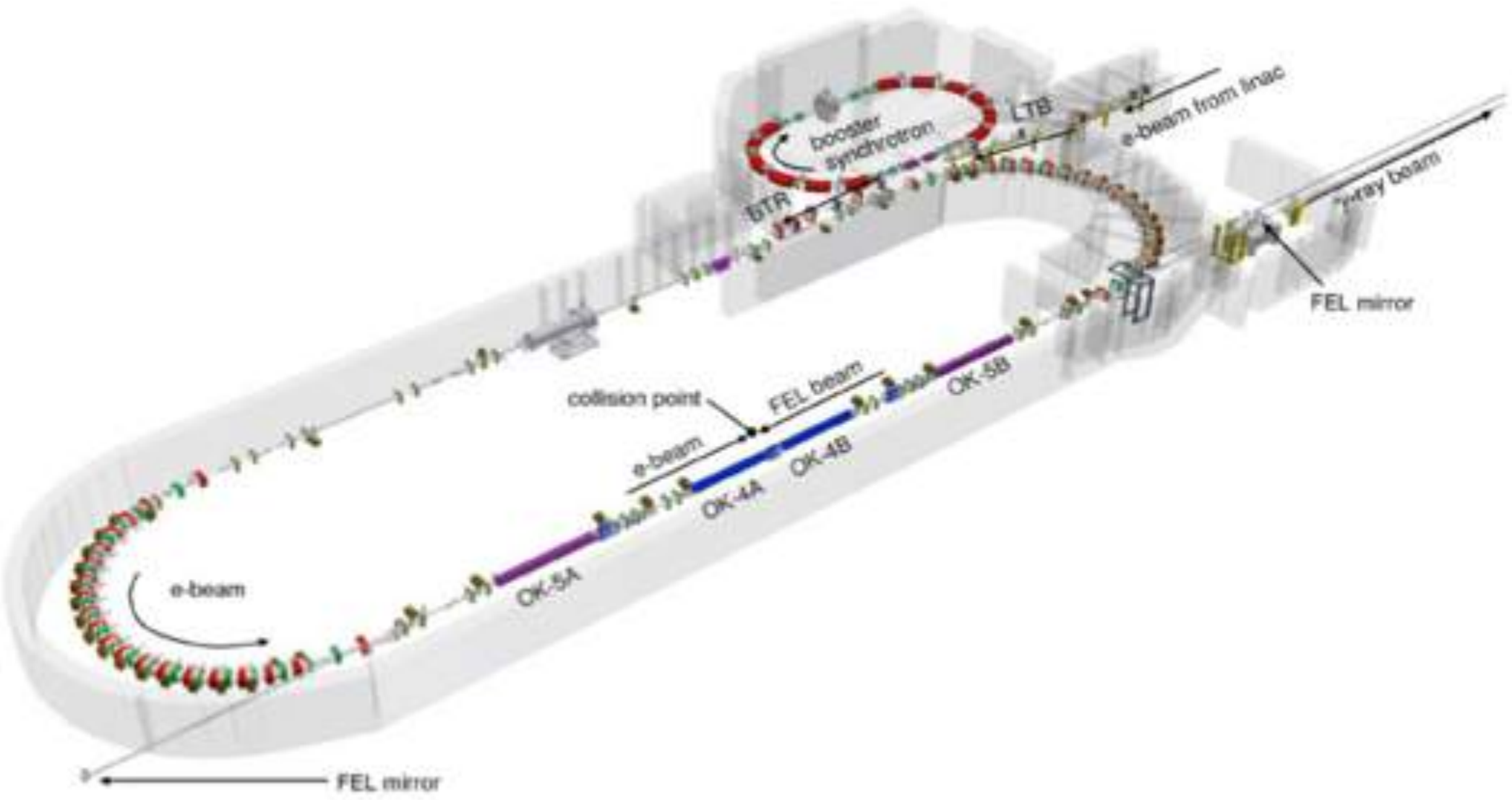

**Figure 35.** High-Intensity Gamma-ray Source (HIGS) facility for mono-energetic γ-ray production.

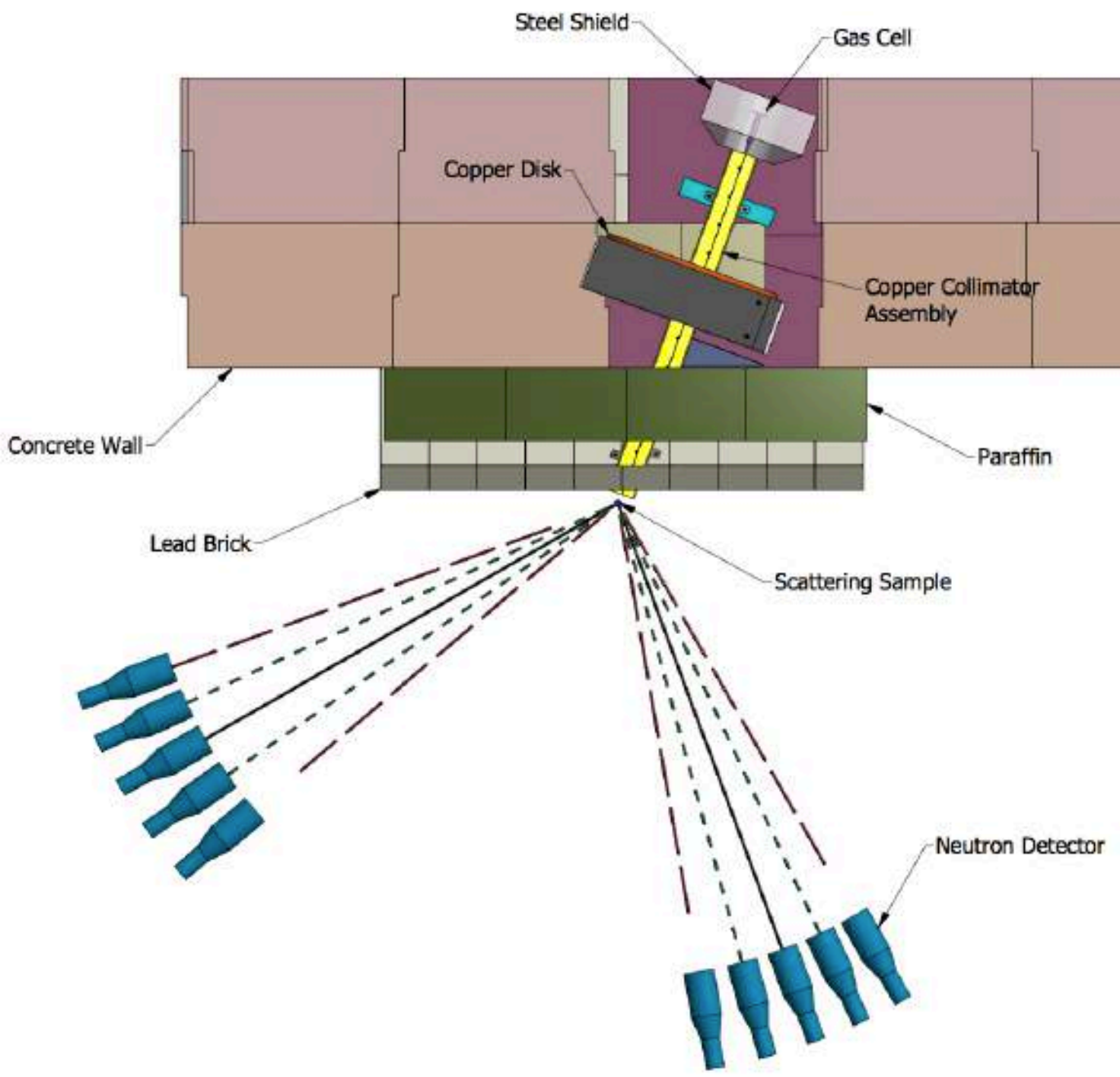

**Figure 36.** Neutron Time-of-Flight spectrometer.

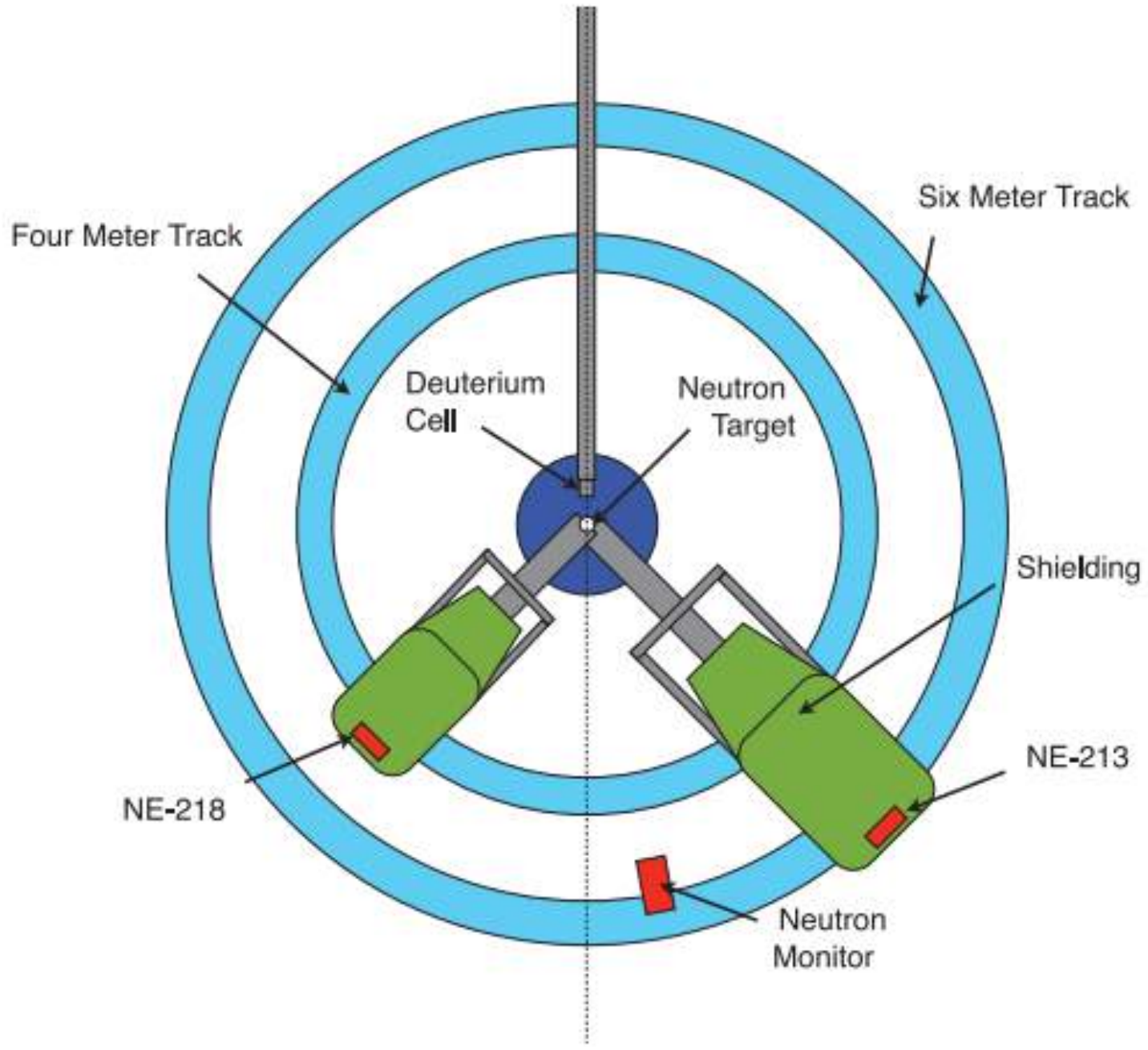

**Figure 37.** Shielded neutron source area for collimated neutron beams.